\documentclass[5p,twocolumn,sort&compress,times,preprint]{elsarticle}
\usepackage[utf8]{inputenc}
\usepackage{bm}
\usepackage{amsmath}
\usepackage[pdftex]{hyperref}
\hypersetup{
 setpagesize=false,
 bookmarksnumbered=true,%
 bookmarksopen=true,%
 colorlinks=true,%
 linkcolor=blue,
 citecolor=red,
}
\usepackage{color}
\usepackage{here}
\usepackage{cases}
\newcommand{\bG}{\mbox{\bf G}}
\newcommand{\bq}{\mbox{\bf q}}
\newcommand{\br}{\mbox{\bf r}}
\newcommand{\bR}{\mbox{\bf R}}

\journal{Computer Physics Communications}

\begin{document}


\begin{frontmatter}

\title{RESPACK: An {\it ab initio} tool for derivation of effective low-energy model of material}

\author[a]{Kazuma Nakamura\corref{author}}
\cortext[author] {Corresponding author.\\\textit{E-mail address:} kazuma@msn.kyutech.ac.jp}
\address[a]{
Quantum Physics Section, Department of Basic Sciences, , Kyushu Institute of Technology,1-1 Sensui-cho, Tobata, Kitakyushu, Fukuoka, 804-8550, Japan
}
\author[b]{Yoshihide Yoshimoto}
\address[b]{
Department of Computer Science, The University of Tokyo,7-3-1 Hongo, Bunkyo-ku, Tokyo 113-0033, Japan
}
\author[c]{Yusuke Nomura}
\address[c]{ 
RIKEN Center for Emergent Matter Science, 2-1 Hirosawa, Wako, Saitama 351-0198, Japan 
}
\author[d]{Terumasa Tadano}
\address[d]{
Research Center for Magnetic and Spintronic Materials, National Institute for Materials Science, Tsukuba 305-0047, Japan
}
\author[e]{Mitsuaki Kawamura} 
\address[e]{
Institute for Solid State Physics, 
The University of Tokyo, Kashiwa 277-8581, Japan
}
\author[f]{Taichi Kosugi} 
\address[f]{
Laboratory for Materials and Structures, Institute of Innovative Research, Tokyo Institute of Technology, Yokohama 226-8503, Japan
}
\author[e]{Kazuyoshi Yoshimi}
\author[e]{Takahiro Misawa} 
\author[e]{Yuichi Motoyama} 

\date{}

\begin{abstract}
RESPACK is a first-principles calculation software for evaluating the interaction parameters of materials and is able to calculate maximally localized Wannier functions, response functions based on the random phase approximation and related optical properties, and frequency-dependent electronic interaction parameters. RESPACK receives its input data from a band-calculation code using norm-conserving pseudopotentials with plane-wave basis sets. Automatic generation scripts that convert the band-structure results to the RESPACK inputs are prepared for xTAPP and {\sc Quantum ESPRESSO}. An input file for specifying the RESPACK calculation conditions is designed pursuing simplicity and is given in the Fortran namelist format. RESPACK supports hybrid parallelization using OpenMP and MPI and can treat large systems including a few hundred atoms in the calculation cell. 
\end{abstract}
\end{frontmatter}
\begin{small}
\noindent
{\bf PROGRAM SUMMARY} \\ 

\noindent
{\em Program Title:} RESPACK 
\\
{\em Program summary URL:}\\ 
https://sites.google.com/view/kazuma7k6r 
\\ 
{\em Licensing provisions:} GNU General Public Licence v3.0  
\\
{\em Programming language:} {\sc Fortran}, {\sc Python} 
\\
{\em External routines:} {\sc LAPACK}, {\sc BLAS}, {\sc MPI} 
\\
{\em Computer:} Any architecture with Fortran 90 compiler 
\\
{\em Operating system:} GNU/Linux 
\\ 
{\em Has the code been vectorised or parallelized?}: Yes 
\\ 
{\em Nature of problem:} 
{\it Ab initio} calculations for maximally localized Wannier function, response function with random-phase approximation, and matrix-element evaluations of frequency-dependent screened direct and exchange interactions. With this code, an effective low-energy model of materials is derived from first principles. 
\\
{\em Solution method:}
Our method is based on {\it ab initio} many-body perturbation calculation and the maximally localized Wannier function calculation. The program employs the plane-wave basis set, and evaluations of matrix elements are performed with the fast Fourier transformation. The generalized tetrahedron method is used for the Brillouin Zone integral.  
\\
{\em Additional comments including Restrictions and Unusual features:}
RESPACK supports xTAPP and {\sc Quantum ESPRESSO} packages, and automatic generation scripts for converting the band-calculation results to the RESPACK inputs are prepared for these software. The current RESPACK only supports band-calculation codes using norm-conserving pseudopotentials with plane-wave basis sets. RESPACK supports hybrid parallelization using OpenMP and MPI to treat large systems in which a few hundred atoms are contained in unit cell.  
\\
\end{small}

\section{Introduction} \label{sec_intro}
First-principles calculations based on density functional theory (DFT)~\cite{Hohenverg_1964,Kohn_1965} have currently been established and are widely used not only by theoretical researchers but also by experimental researchers. The users only input the crystal structure, and can easily evaluate the electronic and structural properties of materials. Density-functional calculations based on the local density approximation (LDA) are attractive due to its reasonable accuracy and low computational cost. Thanks to the recent advances in computer powers, it has successfully been applied to large-scale systems~\cite{Iwata_2010,Hine_2009,Duy_2014-1,Duy_2014-2,Bowler_2006}. There are also many efforts to predict new materials by performing a huge number of DFT calculations~\cite{vandeWalle_2002,Zhang_2013}. On the other hand, it is well known that the LDA often fails to describe strongly-correlated electron systems~\cite{Imada_2010}. Low-energy excitation and quantum fluctuations due to local interactions (on the order of several eV) dominate the low-energy properties of strongly correlated electron systems; the LDA cannot describe such quantum fluctuations. Therefore, the development of {\it ab initio} methods for strongly correlated materials has been an active research subject~
\cite{Nakamura_2008,Nakamura_2009,Nakamura_2010,Nakamura_2016,cRPA_Miyake_2009,cRPA_Wehling_2011,cRPA_Nomura_2012,cRPA_Nomura_2013,cRPA_Ersoy_2012,cRPA_Vaugier_2012,cRPA_Hansmann_2013,cRPA_Nilsson_2013,cRPA_Amadon_2014,cRPA_Yamaji_2014,cRPA_Okamoto_2014,cRPA_Kim_2016,cRPA_Hirayama_2018,cRPA_Hirayama_2019,cRPA_Tadano_2019,cRPA_Nomura_arXiv,cRPA_Hirayama_arXiv}. 

A well-known procedure for this attempt is the first-principle effective-model approach~\cite{Imada_2010,Aryasetiawan_2004}. In this method, an effective model of a strongly correlated electron system is derived from first principles, and then the resulting effective model is numerically analyzed. Applying the highly accurate solvers which are able to evaluate electronic correlation accurately (exact diagonalization~\cite{Kawamura_2017}, dynamical mean field theory~\cite{Georges_1996,Kotliar_2006}, many-variable variational Monte Carlo method~\cite{Tahara_2008_1,Tahara_2008_2,Misawa_2019}, configuration interaction method~\cite{Nakamura_CI_2008}, etc.) to the {\it ab initio} derived effective model enables quantitative understanding of real strongly correlated materials. It is important to establish a reliable {\it ab initio} derivation method for the effective models, and many studies have been performed for this purpose~\cite{Nakamura_2008,Nakamura_2009,Nakamura_2010,Nakamura_2016,cRPA_Miyake_2009,cRPA_Wehling_2011,cRPA_Nomura_2012,cRPA_Nomura_2013,cRPA_Ersoy_2012,cRPA_Vaugier_2012,cRPA_Hansmann_2013,cRPA_Nilsson_2013,cRPA_Amadon_2014,cRPA_Yamaji_2014,cRPA_Okamoto_2014,cRPA_Kim_2016,cRPA_Hirayama_2018,cRPA_Hirayama_2019,cRPA_Tadano_2019,cRPA_Nomura_arXiv,cRPA_Hirayama_arXiv}. So far, development of the program was often closed at the laboratory level, and therefore, in order to spread the first-principle effective-model approach, software development and release are necessary.

In this paper, we introduce a software RESPACK~\cite{RESPACK_URL} which contains a program group (maximally localized Wannier function and {\it ab initio} many body perturbation calculation) for deriving an effective low-energy model from first principles. 
The present paper focuses on a feature of the derivation tool for the effective model, but RESPACK has currently been extended to include the GW calculation, the spin-orbit interaction, and the electron-lattice coupling evaluation, which will be reported in the future. The present paper is organized as follows: In Section~\ref{sec_method}, we describe the methodological background. Some program details are given in Section~\ref{sec_technical}. We give in Section~\ref{sec_calculation_flow} calculation procedure of RESPACK, and demonstrate in Section~\ref{sec_install} how to install and compile of the source codes. In Section~\ref{sec_benchmark}, we show a quantitative check for derived effective-model parameters of a cubic perovskite oxide SrVO$_3$. A summary is given in Section~\ref{sec_conclusion}. We give in \ref{sec_input} input details for the RESPACK calculation. In \ref{sec_tr}, we introduce a utility tool for transfer analysis after the RESPACK calculations. 

\section{Theoretical background} \label{sec_method}

\subsection{Effective low-energy model}
We consider the derivation of the following extended Hubbard model within the two-center integrals written as 
\begin{eqnarray}
\mathcal{H}
&=& \sum_{\sigma} \sum_{{\bm R} {\bm R'}} \sum_{ij}  
  t_{i {\bf R} j {\bf R}'} 
                   a_{i {\bm R}}^{\sigma \dagger} 
                   a_{j {\bm R'}}^{\sigma}   \nonumber \\
&+& \frac{1}{2} \sum_{\sigma \rho} \sum_{{\bm R} {\bm R'}} \sum_{ij} 
\biggl\{ U_{i {\bf R} j {\bf R}'} 
      a_{i {\bm R}}^{\sigma \dagger} 
      a_{j {\bm R'}}^{\rho \dagger}
      a_{j {\bm R'}}^{\rho} 
      a_{i {\bm R}}^{\sigma} \nonumber \\ 
&+& J_{i {\bf R} j {\bf R}'} 
\bigl(a_{i {\bm R}}^{\sigma \dagger} 
      a_{j {\bm R'}}^{\rho \dagger}
      a_{i {\bm R}}^{\rho} 
      a_{j {\bm R'}}^{\sigma}
    + a_{i {\bm R}}^{\sigma \dagger} 
      a_{i {\bm R}}^{\rho \dagger}
      a_{j {\bm R'}}^{\rho} 
      a_{j {\bm R'}}^{\sigma}\bigr) \biggr\}, 
\label{eq:H}                
\end{eqnarray}
where $a_{i {\bm R}}^{\sigma \dagger}$ and $a_{i {\bm R}}^{\sigma}$ are creation and annihilation operators, respectively, of an electron with spin $\sigma$ in the $i$th Wannier orbital in the lattice $\bR$. In this expression, the Wannier orbital is taken to be real. $t_{i {\bf R} j {\bf R}'}$ is a transfer integral defined as 
\begin{eqnarray}
\hspace{-0.4cm}t_{i {\bf R} j {\bf R}'}
= \langle \phi_{i {\bf R}} | \mathcal{H}_0 |  \phi_{j {\bf R}'} \rangle 
= \int_V d\br \phi_{i {\bf R}}^{*} (\br)  \mathcal{H}_0 (\br) \phi_{j {\bf R}'} (\br') 
\label{eq:t}
\end{eqnarray}
with $|\phi_{i {\bf R}} \rangle =a_{i {\bm R}}^{\dagger}|0\rangle$ and the diagonal term ($i=j$, ${\bf R=R}'$) being onsite energy. $\mathcal{H}_0$ in Eq.~(\ref{eq:t}) is the one-body part of $\mathcal{H}$ and is often taken to be the Kohn-Sham (KS) Hamiltonian $\mathcal{H}_{KS}$. The integral in Eq.~(\ref{eq:t}) is taken over the crystal volume $V$. 

Effective direct-Coulomb and exchange integrals are expressed respectively as
\begin{eqnarray}
\hspace{-0.8cm}U_{i {\bf R} j {\bf R}'}(\omega) 
\!\!\!\!&=&\!\!\!\!\langle \phi_{i {\bf R}} \phi_{i {\bf R}} | W(\omega)| \phi_{j {\bf R}'} \phi_{j {\bf R}'} \rangle \nonumber \\
\!\!\!\!&=&\!\!\!\!\!\!\!\int_V\!\!\!d\br\!\!\int_V\!\!\!d\br'\!
  \phi_{i {\bf R}}^{*}(\br) \phi_{i {\bf R}}(\br) 
  W(\br,\br', \omega) \phi_{j {\bf R}'}^{*}(\br') \phi_{j {\bf R}'}(\br') 
\label{eq:U}
\end{eqnarray} 
and 
\begin{eqnarray}
\hspace{-0.8cm}J_{i {\bf R} j {\bf R}'}(\omega)   
\!\!\!\!&=&\!\!\!\! \langle \phi_{i {\bf R}} \phi_{j {\bf R}'} | W(\omega) | \phi_{j {\bf R}'} \phi_{i {\bf R}} \rangle \nonumber \\
\!\!\!\!&=&\!\!\!\!\!\!\!\int_V\!\!\!d\br\!\!\int_V\!\!\!d\br'\!
  \phi_{i {\bf R}}^{*}(\br) \phi_{j {\bf R}'}(\br) 
  W(\br,\br',\omega) \phi_{j {\bf R}'}^{*}(\br') \phi_{i {\bf R}}(\br')
\label{eq:J}
\end{eqnarray}
with $W(\br,\br',\omega)$ being the frequency-dependent screened Coulomb interaction. In practical calculations, we evaluate $W(\br,\br',\omega)$ based on random phase approximation (RPA) with imposing the constraint to the polarization function (see Section.~\ref{method:chiqw}). 

The static limit of the screened direct-Coulomb $U_{i{\bf R}j{\bf R}'}(\omega)$ and exchange $J_{i{\bf R}j{\bf R}'}(\omega)$ integrals gives interaction parameters in the Hamiltonian ${\mathcal H}$, which are given by  
\begin{eqnarray}
 U_{i {\bf R} j {\bf R}'}
= \lim_{\omega \to 0} U_{i {\bf R} j {\bf R}'}(\omega) 
\label{eq:static-U}
\end{eqnarray} 
and 
\begin{eqnarray}
J_{i {\bf R} j {\bf R}'} 
= \lim_{\omega \to 0} J_{i {\bf R} j {\bf R}'} (\omega), 
\label{eq:static-J}
\end{eqnarray}
respectively. $t_{i {\bf R} j {\bf R}'}$, $U_{i{\bf R}j{\bf R}'}(\omega)$, and $J_{i{\bf R}j{\bf R}'}(\omega)$ have the lattice translational symmetry of   
\begin{eqnarray}
 t_{i {\bf 0} j {\bf R'-R}}&=&t_{i {\bf R} j {\bf R'}}, \label{eq:tt} \\ 
 U_{i {\bf 0} j {\bf R'-R}}(\omega)&=&U_{i {\bf R} j {\bf R'}}(\omega), \label{eq:tU} \\ 
 J_{i {\bf 0} j {\bf R'-R}}(\omega)&=&J_{i {\bf R} j {\bf R'}}(\omega), \label{eq:tJ}
\end{eqnarray} 
where we used $\phi_{i{\bf R}}({\bf r})=\phi_{i{\bf 0}}({\bf r}-{\bf R})$. In this paper, we focus on an {\it ab initio} derivation of these parameters.

\subsection{Wannier function} \label{method:wannier}
The calculation of the Wannier function follows the algorithm for the maximally-localized Wannier function~\cite{Marzari_1997,Souza_2001}. The $i$th Wannier function of the lattice {\bf R} is defined as 
\begin{eqnarray}
\phi_{i{\bf R}}({\bf r})=
\frac{1}{\sqrt{N_k}}\sum_{{\bf k}}^{N_k} 
\sum_{\alpha=N_s^{\bf k}+1}^{N_s^{\bf k}+N_b^{\bf k}}
U_{\alpha i}^{{\bf k}} \psi_{\alpha{\bf k}}({\bf r})e^{-i{\bf k}\cdot{\bf R}}, 
\label{wir}
\end{eqnarray} 
where ${\bf k}$ is a wave vector in the first Brillouin zone and $N_k$ is the total number of the Monkhorst-Pack $k$ mesh. 
The Wannier function is constructed from the $N_b^{\bf k}$ KS bands [from $(N_s^{\bf k}+1)$-th to ($N_s^{\bf k}+N_b^{\bf k}$)-th bands]. $N_s^{\bf k}$ and $N_b^{\bf k}$ are determined from the energy-window information. 
$U_{\alpha i}^{{\bf k}}$ is a matrix that transforms the $\alpha$th Bloch wave function into the $i$th Wannier function. 
The $\alpha$th Bloch wave function is defined as 
\begin{eqnarray}
\psi_{\alpha{\bf k}}({\bf r})
=\frac{1}{\sqrt{N_k}} \sum_{{\bf G}}^{N_{G}^{\psi}} 
C_{{\bf G}\alpha}({\bf k})\frac{1}{\sqrt{\Omega}} e^{i({\bf k+G})\cdot{\bf r}},  
\label{pak} 
\end{eqnarray} 
where $\bG$ is a reciprocal lattice vector, and $N_G^{\psi}$ is the total number of the plane waves used for the expansion of the wave function, which is determined by the cutoff energy $E_{cut}^{\psi}$ from the inequality $\frac{1}{2}|{\bf k+G}|^2\le E_{cut}^{\psi}$. $\Omega$ is the volume of the unit cell. Both the Wannier function and the Bloch function are normalized for the crystal volume $V=N_k\Omega$. $C_{{\bf G}\alpha}({\bf k})$ is the expansion coefficient of the plane wave $e^{i({\bf k+G})\cdot{\bf r}}/\sqrt{\Omega}$. By inserting Eq.~(\ref{pak}) into Eq.~(\ref{wir}), the Wannier function at the home cell ({\bf R} = {\bf 0}) is written as
\begin{eqnarray}
\phi_{i{\bf 0}}({\bf r})
=\frac{1}{N_k}
\sum_{{\bf k}}^{N_k}
\sum_{{\bf G}}^{N_G^{\psi}} 
\tilde{C}_{{\bf G}i}({\bf k})
\frac{1}{\sqrt{\Omega}}
e^{i({\bf k+G})\cdot{\bf r}}
\label{wi0}
\end{eqnarray}
with 
\begin{eqnarray} 
\tilde{C}_{{\bf G}i}({\bf k})
=\sum_{\alpha=N_s^{{\bf k}}+1}^{N_s^{{\bf k}}+N_b^{{\bf k}}} 
 C_{{\bf G}\alpha}({\bf k}) U_{\alpha i}^{{\bf k}}
\label{cikg}.  
\end{eqnarray}
Here, $\tilde{C}_{{\bf G}i}({\bf k})$ is the expansion coefficient of the plane wave for the Wannier function. The center of the $i$th Wannier orbital at the lattice ${\bf R}$ is defined as 
\begin{eqnarray}
  \langle {\bf r} \rangle_{i{\bf R}} 
= \langle \phi_{i{\bf R}} | {\bf r} | \phi_{i{\bf R}} \rangle 
= \langle \phi_{i{\bf 0}} | {\bf r} | \phi_{i{\bf 0}} \rangle + {\bf R}. 
\label{wannier-center} 
\end{eqnarray}
Similarly, the spread of the Wannier orbital is defined as 
\begin{eqnarray}
 S_{i{\bf R}}
=\langle r^2 \rangle_{i{\bf R}} - |\langle \mathbf{r} \rangle_{i{\bf R}}|^2 
=\langle r^2 \rangle_{i{\bf 0}} - |\langle \mathbf{r} \rangle_{i{\bf 0}}|^2
\label{wannier-spread}
\end{eqnarray}
with $\langle r^2 \rangle_{i{\bf R}} = \langle \phi_{i{\bf R}} | r^2 | \phi_{i{\bf R}} \rangle$.

\subsection{Response function with random phase approximation} \label{method:chiqw}
In the RPA and constrained RPA, the polarization function in the plane-wave basis is written as   
\begin{eqnarray}
\hspace{-0.5cm}\chi_{{\bf GG'}}({\bf q},\omega)
\!\!\!\!\!&=&\!\!\!\!\!\frac{2}{N_k} \sum_{{\bf k}}^{N_k} \sum^{unocc}_{\alpha} \sum^{occ}_{\beta}
\Bigl(1-T_{\alpha{\bf k+q}} T_{\beta{\bf k}} \Bigr) \nonumber \\
\!\!\!\!\!&\times&\!\!\!\!\! M_{\alpha\beta}^{{\bf G}}({\bf k+q,k})
M_{\alpha\beta}^{{\bf G'}}({\bf k+q,k})^{*}
X_{\alpha{\bf k+q},\beta{\bf k}}(\omega) 
\label{eq:chi} 
\end{eqnarray} 
with 
\begin{eqnarray} 
M_{\alpha\beta}^{{\bf G}}({\bf k+q,k}) 
= \langle \psi_{\alpha {\bf k+q}} | e^{i({\bf q+G})\cdot{\bf r}} |\psi_{\beta {\bf k}} \rangle
\label{ism} 
\end{eqnarray} 
and 
\begin{eqnarray}
\hspace{-0.8cm}
X_{\alpha{\bf k+q},\beta{\bf k}}(\omega)
=\frac{1}{\omega\!-\!E_{\alpha{\bf k+q}}\!+\!E_{\beta{\bf k}}\!+\!i\delta} 
\!-\!\frac{1}{\omega\!+\!E_{\alpha{\bf k+q}}\!-\!E_{\beta{\bf k}}\!-\!i\delta}. 
\label{Xakqbk}
\end{eqnarray}
Here, {\bf q} is a wave vector in the first Brillouin zone, $\omega$ is frequency, and indices $\alpha$ and $\beta$ specify the unoccupied and occupied bands, respectively. The interstate matrix $M_{\alpha\beta}^{{\bf G}}({\bf k+q,k})$ is evaluated using the fast Fourier transformation technique. $E_{\alpha{\bf k}}$ and $\delta$ in Eq.~(\ref{Xakqbk}) are the energy of the Bloch state and the  broadening factor, respectively. The quantity $X_{\alpha{\bf k+q},\beta{\bf k}}(\omega)$ is calculated with the generalized tetrahedron technique~\cite{Fujiwara_2003,Nohara_2009} as 
\begin{eqnarray}
X_{\alpha{\bf k+q},\beta{\bf k}}(\omega)
\sim x({\bf k}; \alpha, \beta, {\bf q}, \omega, \delta, \delta_a, \delta_r), 
\label{tetrahedron} 
\end{eqnarray}
where $\delta_a$ and $\delta_r$ in Eq.~(\ref{tetrahedron}) are parameters to judge an energy degeneracy of $E_{\alpha{\bf k+q}}$ and $E_{\beta{\bf k}}$. The tetrahedron routine returns a value of $x({\bf k})$ in Eq.~(\ref{tetrahedron}), and the \{$x({\bf k})$\} data are used for the Brillouin-zone integral. $T_{\alpha{\bf k}}$ in Eq.~(\ref{eq:chi}) is the transition probability from the Wannier states \{$|\phi_{i {\bf R}}\rangle$\} to the Bloch state $|\psi_{\alpha{\bf k}}\rangle$, which is calculated as 
\begin{eqnarray}
T_{\alpha {\bf k}} = \sum_{i}^{N_w} U_{\alpha i}^{{\bf k}} U_{\alpha i}^{{\bf k}*}, 
\label{tak}
\end{eqnarray} 
where $N_w$ is the total number of the Wannier orbitals and $U_{\alpha i}^{{\bf k}}$ is defined in Eq.~(\ref{wir}). The quantity $T_{\alpha{\bf k}}$ is introduced to calculate constrained  polarization~\cite{Sasioglu_2011}, and, in the usual RPA, the $T_{\alpha{\bf k}}$ are set to zero.   

The symmetric dielectric function~\cite{Hyvertsen_chi_1987} can be written with using the polarization function as 
\begin{eqnarray}
\epsilon_{{\bf G}{\bf G}'}(\bq,\omega) 
= \delta_{{\bf G}{\bf G}'} 
- \frac{4\pi}{\Omega}\frac{1}{|{\bf q+G}|} \chi_{{\bf G}{\bf G}'}(\bq,\omega) \frac{1}{|{\bf q+G'}|}. 
\label{epsGG} 
\end{eqnarray}
When $|{\bf q+G}|$ and $|{\bf q+G'}|$ are large, the contribution from the second term in the right hand side is negligible and  $\epsilon_{{\bf G}{\bf G}'}(\bq,\omega) \approx \delta_{{\bf G}{\bf G}'}$. Therefore, the matrix $\chi_{{\bf GG'}}({\bf q},\omega)$ is restricted to $N_G^{\epsilon}\times N_G^{\epsilon}$, where $N_G^{\epsilon}$ ($ < N_G^{\psi}$) is the number of the plane waves used to expand the polarization function, which is determined by the cutoff energy $E_{cut}^{\epsilon}$ as an inequality $\frac{1}{2}|{\bf q+G}|^2\le E_{cut}^{\epsilon}$. For $E_{cut}^{\epsilon}\!<\!\frac{1}{2}|\bq\!+\!\bG|^2\!<\!E_{cut}^{\psi}$, $\epsilon_{{\bf G}{\bf G}'}(\bq,\omega)$ is assumed to be  $\epsilon_{{\bf G}{\bf G}'}(\bq,\omega) = \delta_{{\bf G}{\bf G}'}$.

In the calculation of $\epsilon_{{\bf G}{\bf G}'}(\bq,\omega)$, we note on our special treatment of the head component which corresponds to the ${\bf G}={\bf G}'={\bf 0}$ component in the ${\bf q}\to{\bf 0}$ limit. In the usual RPA case, we calculate the following~\cite{Draxl_2006}
\begin{eqnarray}
\lim_{{\bf q}\to{\bf 0}} \epsilon_{{\bf 00}}({\bf q},\omega) 
= 1-\frac{4\pi}{\Omega} \frac{\partial^2\chi_{{\bf 00}}({\bf q},\omega)}{\partial q^2}
-\frac{(\omega_{pl}^{\mu\nu})^2}{\omega(\omega+i\delta)}.  
\label{eps00} 
\end{eqnarray}
Here, the last term results from the intraband transition, and $\delta$ is the broadening factor introduced in Eq.~(\ref{Xakqbk}). $\omega_{pl}$ is the bare plasma frequency calculated via the Fermi-surface integral as 
\begin{eqnarray}
\omega_{pl}^{\mu\nu}
=\frac{1}{N_k} \sum_{{\bf k}}^{N_k} \sum_{\alpha}   
\bigl(1-(T_{\alpha{\bf k}})^2\bigr) p_{\alpha\alpha{\bf k}}^{\mu} p_{\alpha\alpha{\bf k}}^{\nu}
\delta(E_{\alpha{\bf k}}-E_F), 
\label{plasma} 
\end{eqnarray} 
where $E_F$ is the Fermi energy determined in the DFT band calculation, and $p_{\alpha\alpha{\bf k}}^{\nu}$ is the diagonal element of the transition-moment matrix with respect to the bands as  
\begin{eqnarray}
p_{\alpha\beta{\bf k}}^{\mu}
&=&-i \Bigl\langle \psi_{\alpha{\bf k}} \Bigl| \frac{\partial}{\partial x_{\mu}}+[V_{NL},x_{\mu}] 
\Bigr|\psi_{\beta{\bf k}} \Bigr\rangle \nonumber \\ 
&\sim& -i \Bigl\langle \psi_{\alpha{\bf k}} \Bigl| \frac{\partial}{\partial x_{\mu}} 
\Bigr|\psi_{\beta{\bf k}} \Bigr\rangle   
\label{momentum} 
\end{eqnarray} 
with $x_{\mu}$ being the Cartesian coordinate. On the above evaluation, we ignore the contribution from the non-local part of the pseudopotential, $V_{NL}$. 
We note that this neglect is not so serious in the evaluation of the effective interaction, because the effective interaction is written in terms of the sum over the $q$ points [see Eq.~(\ref{eq:W})]; in this case, the contribution from the ${\bf q}=0$ to the effective interaction becomes small relatively. On the other hand, in the optical response, the non-local pseudopotential contribution may manifest itself as a significant effect, especially for the transition metals~\cite{Marini2001}. This is because the optical properties are completely the ${\bf q}=0$ quantity [see Eqs.~(\ref{eq:epsilonM}), (\ref{eq:eels}), (\ref{eq:optical-conductivity}), and (\ref{eq:reflectance})].

For the head-component calculation in the constrained RPA, we calculate the following 
\begin{eqnarray}
\lim_{{\bf q} \to 0}\epsilon_{{\bf 00}}({\bf q},\omega)
=1-\frac{4\pi}{\Omega}\frac{\partial^2 \chi_{{\bf 00}}({\bf q},\omega)}{\partial q^2},  
\label{eps00-cRPA} 
\end{eqnarray}
which is obtained by dropping the last term in the right hand side of Eq.~(\ref{eps00}). The Wannier functions are constructed to include the low-energy bands near the Fermi level, and thus, $T_{\alpha{\bf k}}=1$ is expected and the bare plasma frequency $\omega_{pl}^{\mu\nu}$ in Eq.~(\ref{plasma}) becomes zero. This is why we dropped the last term in the right hand side of Eq.~(\ref{eps00}). The second derivative of the polarization function with respect to the wavenumber $q$ in Eqs.~(\ref{eps00}) or (\ref{eps00-cRPA}) can be calculated analytically, as has been done for insulators~\cite{Hyvertsen_chi_1987}.   

The optical properties such as the macroscopic dielectric function $\epsilon_{{\rm M}}(\omega)$, the electronic energy loss spectrum (EELS) $L(\omega)$, the real part of the optical conductivity $\sigma(\omega)$, and the reflectance spectrum $R(\omega)$ are also calculated from the inverse of the matrix $\epsilon_{{\bf G} {\bf G}'}(\bq,\omega)$ in Eqs.(\ref{epsGG}) and (\ref{eps00}) or (\ref{eps00-cRPA}) as 
\begin{eqnarray}
\epsilon_{{\rm M}}(\omega)
 =\lim_{{\bf q}\to 0} \frac{1}{\epsilon_{{\bf 0}{\bf 0}}^{-1}(\bq,\omega)},  
 \label{eq:epsilonM} 
\end{eqnarray}
\begin{eqnarray}
 L(\omega)
 =-{\rm Im}\lim_{{\bf q}\to 0}\epsilon_{{\bf 0}{\bf 0}}^{-1}(\bq,\omega), 
 \label{eq:eels} 
\end{eqnarray}
\begin{eqnarray}
 {\rm Re} \bigl[ \sigma(\omega) \bigr]
 =\frac{\omega}{4\pi} {\rm Im} \lim_{{\bf q}\to 0} \frac{1}{\epsilon_{{\bf 0}{\bf 0}}^{-1}(\bq,\omega)},
 \label{eq:optical-conductivity} 
\end{eqnarray}
and
\begin{eqnarray}
 R(\omega)=\Biggl|\frac{1-\sqrt{\lim_{{\bf q}\to 0}\epsilon_{{\bf 0}{\bf 0}}^{-1}(\bq,\omega)}}
                       {1+\sqrt{\lim_{{\bf q}\to 0}\epsilon_{{\bf 0}{\bf 0}}^{-1}(\bq,\omega)}}\Biggr|,
\label{eq:reflectance} 
\end{eqnarray}
respectively. 

\subsection{Direct-Coulomb and exchange integrals}\label{method:calc_int}
The evaluations of the interaction integrals $U_{i {\bf 0} j {\bf R}'}(\omega)$ in Eq.~(\ref{eq:tU}) and $J_{i {\bf 0} j {\bf R}'}(\omega)$ in Eq.~(\ref{eq:tJ}) proceed as follows: First, the screened Coulomb interaction is written in the reciprocal space by using the Fourier transform as  
\begin{eqnarray}
\hspace{-0.8cm}
W(\br,\br',\omega)
=\frac{1}{N_q}\sum_{{\bf q}}^{N_q} \sum_{{\bf G},{\bf G}'}^{N_G^{\psi}} 
 e^{i ({\bf q}+{\bf G})\cdot {\bf r}} 
 W_{{\bf G} {\bf G}'} (\bq,\omega) 
 e^{-i ({\bf q}+{\bf G}')\cdot {\bf r'}}. 
 \label{eq:W}
\end{eqnarray} 
Note that the $q$-grid is the same as the $k$-grid; thus, $N_q=N_k$. $W_{{\bf G} {\bf G}'} (\bq,\omega)$ in the right-hand side is written with using the inverse dielectric matrix as follows: 
\begin{eqnarray}
\hspace{-0.8cm} 
W_{{\bf G} {\bf G}'} (\bq,\omega) =   
 \left\{
  \begin{array}{ll} 
  \frac{4\pi}{\Omega} 
  \frac{\epsilon_{{\bf G} {\bf G}'}^{-1}(\bq,\omega)}{|\bq+\bG||\bq+\bG'|}, 
  &\mbox{$\frac{1}{2}|\bq+\bG|^2<E_{cut}^{\epsilon}$,} \\
  \frac{4\pi}{\Omega} 
  \frac{1}{|\bq+\bG|^2}, 
  &\mbox{$E_{cut}^{\epsilon}<\frac{1}{2}|\bq+\bG|^2<E_{cut}^{\psi}$.} 
  \end{array}
 \right.
 \label{eq:Wq} 
\end{eqnarray}
Note that the inverse dielectric matrix has off-diagonal elements in the first $N_G^{\epsilon}\times N_G^{\epsilon}$ block and becomes diagonal matrix with unity in the area beyond this block. By inserting Eqs.~(\ref{eq:W}) and (\ref{eq:Wq}) into Eq.~(\ref{eq:U}) and noting the lattice-translational symmetry [Eqs.~(\ref{eq:tU}) and (\ref{eq:tJ})], we obtain the form of 
\begin{eqnarray}
U_{i {\bf 0} j {\bf R}} (\omega)\!\!\!\!&=&\!\!\!\! 
 \frac{4\pi}{\Omega N_q} 
 \sum_{{\bf q}}^{N_q} \sum_{{\bf G},{\bf G}'}^{N_G^{\epsilon}} 
  e^{-i {\bf q}\cdot {\bf R}} 
  \rho_{i {\bf q}}(\bG) 
  \epsilon_{{\bf G} {\bf G}'}^{-1}(\bq,\omega) 
  \rho_{j {\bf q}}^{*}(\bG') \nonumber \\  
 &+&\!\!\!\!\frac{4\pi}{\Omega N_q} \sum_{{\bf q}}^{N_q} \sum_{{\bf G}}^{(N_G^{\epsilon}+1:N_G^{\psi})}
  e^{-i {\bf q} \cdot{\bf R}} 
  \rho_{i {\bf q}}(\bG) 
  \rho_{j {\bf q}}^{*}(\bG)
\label{eq:matrix_U}
\end{eqnarray}
with 
\begin{eqnarray}
   \rho_{i {\bf q}}(\bG) 
 = \frac{1}{|\bq+\bG|N_k} 
    \sum_{{\bf k}}^{N_k} \langle \tilde{\phi}_{i {\bf k}+{\bf q}} | 
    e^{i ({\bf q}+{\bf G})\cdot {\bf r}} | \tilde{\phi}_{j {\bf k}} \rangle 
    \label{rhoiqG} 
\end{eqnarray}
and $|\tilde{\phi}_{i {\bf k}}\rangle = \sum_{{\bf R}}^{N_R} |\phi_{i {\bf R}} \rangle e^{i{\bf k}\cdot{\bf R}}$ with $N_R$ being the total number of the lattices in the system. The divergence of Eq.~(\ref{rhoiqG}) in ${\bf q}\to{\bf 0}$ with {\bf G}={\bf G}'={\bf 0} is removed by following the prescription of Ref.~\cite{Hyvertsen_GW_1986}.   

The dependence of the static ($\omega=0$) direct-Coulomb integral on the distance between the two Wannier functions is evaluated via 
\begin{eqnarray}
U(r_{ij{\bf R}})=U_{i{\bf 0}j{\bf R}}(0),
\label{WvsR} 
\end{eqnarray}
where $r_{ij{\bf R}}$ is the distance between the two Wannier centers,  
\begin{eqnarray}
r_{ij{\bf R}}
=\bigl|\langle{\bf r}\rangle_{j{\bf R}}-\langle{\bf r}\rangle_{i{\bf 0}} \bigr|
=\bigl|{\bf R}+\langle{\bf r}\rangle_{j{\bf 0}}-\langle{\bf r}\rangle_{i{\bf 0}}\bigr|. 
\label{wannier-distance} 
\end{eqnarray}

Matrix elements of the bare (or unscreened) Coulomb interaction,  
$U_{i {\bf 0} j {\bf R}}^{{\rm bare}} = \langle \phi_{i {\bf 0}} \phi_{i {\bf 0}} | v | \phi_{j {\bf R}} \phi_{j {\bf R}} \rangle$, are calculated with replacing $\epsilon_{{\bf G} {\bf G}'}^{-1}(\bq,\omega)$ of Eq.~(\ref{eq:matrix_U}) by $\delta_{{\bf G} {\bf G}'}$ as 
\begin{eqnarray}
 U_{i {\bf 0} j {\bf R}}^{{\rm bare}} 
 =\frac{4\pi}{\Omega N_q} \sum_{{\bf q}}^{N_q} 
  \sum_{{\bf G}}^{N_G^{\psi}}
  e^{-i {\bf q}\cdot {\bf R}} 
  \rho_{i {\bf q}}(\bG) 
  \rho_{j {\bf q}}^{*}(\bG).
\label{eq:matrix_V}
\end{eqnarray}
The dependence of $U_{i{\bf 0}j{\bf R}}^{{\rm bare}}$ on the distance between two Wannier functions is obtained from   
\begin{eqnarray}
U^{\rm bare}(r_{ij{\bf R}})=U_{i{\bf 0}j{\bf R}}^{{\rm bare}}.
\label{VvsR}
\end{eqnarray}

The parallel argument can be applied to the derivation of the screened exchange integrals in Eq.~(\ref{eq:tJ}). The result is 
\begin{eqnarray}
J_{i {\bf 0} j {\bf R}} (\omega)\!\!\!\!&=&\!\!\!\! 
 \frac{4\pi}{\Omega N_q} 
 \sum_{{\bf q}}^{N_q} 
 \sum_{{\bf G},{\bf G}'}^{N_G^{\epsilon}} 
 \rho_{ij {\bf R}{\bf q}}(\bG) 
 \epsilon_{{\bf G} {\bf G}'}^{-1}(\bq,\omega) 
 \rho_{ij {\bf R}{\bf q}}^{*}(\bG') \nonumber \\ 
 &+&\!\!\!\!\frac{4\pi}{\Omega N_q} 
 \sum_{{\bf q}}^{N_q} 
 \sum_{{\bf G}}^{(N_G^{\epsilon}+1:N_G^{\psi})}
 \rho_{ij {\bf R}{\bf q}}(\bG) 
 \rho_{ij {\bf R}{\bf q}}^{*}(\bG)
 \label{eq:matrix_J}
\end{eqnarray}
with
\begin{eqnarray}
 \rho_{ij {\bf R} {\bf q}} (\bG)
 =\frac{1}{ N_k | \bq+\bG | } 
  \sum_{{\bf k}}^{N_k} e^{-i {\bf k}\cdot {\bf R}}  
  \langle \tilde{\phi}_{i {\bf k}+{\bf q}} | e^{i ({\bf q}+{\bf G})\cdot {\bf r}} | 
  \tilde{\phi}_{j {\bf k}} \rangle. 
\end{eqnarray}
The bare exchange integral $J_{i {\bf 0} j {\bf R}}^{{\rm bare}} = \langle \phi_{i {\bf 0}} \phi_{j {\bf 0}} | v | \phi_{j {\bf R}} \phi_{i {\bf R}} \rangle$ is given as 
\begin{eqnarray}
 J_{i {\bf 0} j {\bf R}}^{{\rm bare}} 
 = \frac{4\pi}{\Omega N_q} 
 \sum_{{\bf q}}^{N_q} 
 \sum_{{\bf G}}^{N_G^{\epsilon}}
 \rho_{ij {\bf R}{\bf q}}(\bG) 
 \rho_{ij {\bf R}{\bf q}}^{*}(\bG). 
 \label{eq:matrix_X}
\end{eqnarray}

\section{Some technical aspects} \label{sec_technical}

\subsection{Symmetry}\label{sec_symmetry} 
RESPACK makes use of the space group symmetries of an input crystal structure and requires irreducible data obtained from a band calculation.  
The Bloch function at a reducible $k$ point, ${\bf k}$, is calculated from its irreducible part via     
\begin{eqnarray}
   \psi_{\alpha{\bf k}}({\bf r}) 
 =\hat{T}\hat{R} \psi_{\alpha{\bf k}^*}({\bf r}) 
 =\psi_{\alpha{\bf k}^*}({\bm R}^{-1}({\bf r}-{\bm T})),  
\label{red-wavefunction}
\end{eqnarray}
where ${\bf k}^*$ is the corresponding irreducible $k$ point. $\hat{R}$ and $\hat{T}$ are the operators of a rotation and a fractional translation, respectively, which are represented by the 3$\times$3 matrix ${\bm R}$ and three-dimensional vector ${\bm T}$. Equations~(\ref{pak}) and (\ref{red-wavefunction}) leads to 
\begin{eqnarray}
\psi_{\alpha{\bf k}}({\bf r}) 
\!\!\!\!\!&=&\!\!\!\!\!
  \frac{1}{\sqrt{N_k}} 
  \sum_{{\bf G}^*}^{N_G^{\psi}} 
  C_{{\bf G}^*\alpha}({\bf k}^*)  
  \frac{1}{\sqrt{\Omega}} e^{i({\bf k}^*+{\bf G}^*) \cdot ({\bm R}^{-1}({\bf r}-{\bm T}))} \nonumber \\ 
&\sim&\!\!\!\!\!\! 
  \frac{1}{\sqrt{N_k}}  
  \sum_{{\bf G}^*}^{N_G^{\psi}} 
  C_{{\bf G}^*\alpha}({\bf k}^*)
  e^{-i ({\bm R}^{-1})^{t} {\bf G}^* \cdot {\bm T} }  
  \frac{e^{ i ({\bm R}^{-1})^{t} ({\bf k}^*+{\bf G}^*) \cdot {\bf r}}}{\sqrt{\Omega}} 
\label{wavefunction}
\end{eqnarray}
with ${\bf G}^*$ being reciprocal lattice vector for expansion of the wave function at the irreducible $k$ point. In the above expression, we remove the global phase which does not depend on ${\bf G}^*$.
By comparing Eq.~(\ref{wavefunction}) with Eq.~(\ref{pak}), we find the following relations:  
\begin{eqnarray}
 {\bf k}&=&({\bm R}^{-1})^{t} {\bf k}^* + {\bf \Delta}_{rw}, \label{k-and-kirr} \\
 {\bf G}&=&({\bm R}^{-1})^{t} {\bf G}^* - {\bf \Delta}_{rw}, \label{G-and-Girr} \\ 
 C_{{\bf G}\alpha}({\bf k}) 
 &=&C_{{\bf G}^*\alpha}({\bf k}^*) e^{-i ({\bm R}^{-1})^{t} {\bf G}^* \cdot {\bm T}}. \label{coef-G}
\end{eqnarray}
Here, ${\bf \Delta}_{rw}$ is a rewind vector which is introduced to pull back the rotated ${\bf k}^*$ vector to the first Brillouin zone. From Eq.~(\ref{G-and-Girr}), ${\bf G}^*$ and ${\bf G}$ have the following relationship  
\begin{eqnarray}
{\bf G}^*={\bm R}^{t}({\bf G}+{\bf \Delta}_{rw}), 
\label{G-and-Gstar}
\end{eqnarray}
and with Eqs.~(\ref{coef-G}) and (\ref{G-and-Gstar}), we obtain  
\begin{eqnarray}
   C_{{\bf G}\alpha}({\bf k}) 
  =C_{{\bm R}^{t}({\bf G}+{\bf \Delta}_{rw})\alpha}({\bf k}^*)  
  e^{-i({\bf G}+{\bf \Delta}_{rw}) \cdot {\bm T}}.
\end{eqnarray}
More specifically, in the code, we treat ${\bf S}=({\bm R}^{-1})^{t}$ instead of ${\bm R}$, so the following expression is practically implemented 
\begin{eqnarray}
   C_{{\bf G}\alpha}({\bf k}) 
  =C_{{\bf S}^{-1}({\bf G}+{\bf \Delta}_{rw})\alpha}({\bf k}^*)  
  e^{-i({\bf G}+{\bf \Delta}_{rw}) \cdot {\bm T}}.
\end{eqnarray}
Similarly, the inverse dielectric matrix at a reducible $q$ point is generated from the irreducible one as follows: 
\begin{eqnarray}
\epsilon^{-1}_{\bf G,G'}({\bf q},\omega)
\!=\!\epsilon^{-1}_{ {\bf S}^{-1}({\bf G}+{\bf \Delta}_{rw}), {\bf S}^{-1}({\bf G'}+{\bf \Delta}'_{rw}) }
({\bf q}^*\!,\!\omega) e^{-i({\bf G}-{\bf G'})\cdot{\bm T}}. 
\label{symeps} 
\end{eqnarray}

\subsection{Frequency grid}\label{sec_frequency_grid} 
\begin{figure}[htpb] 
\begin{center} 
\includegraphics[width=0.4\textwidth]{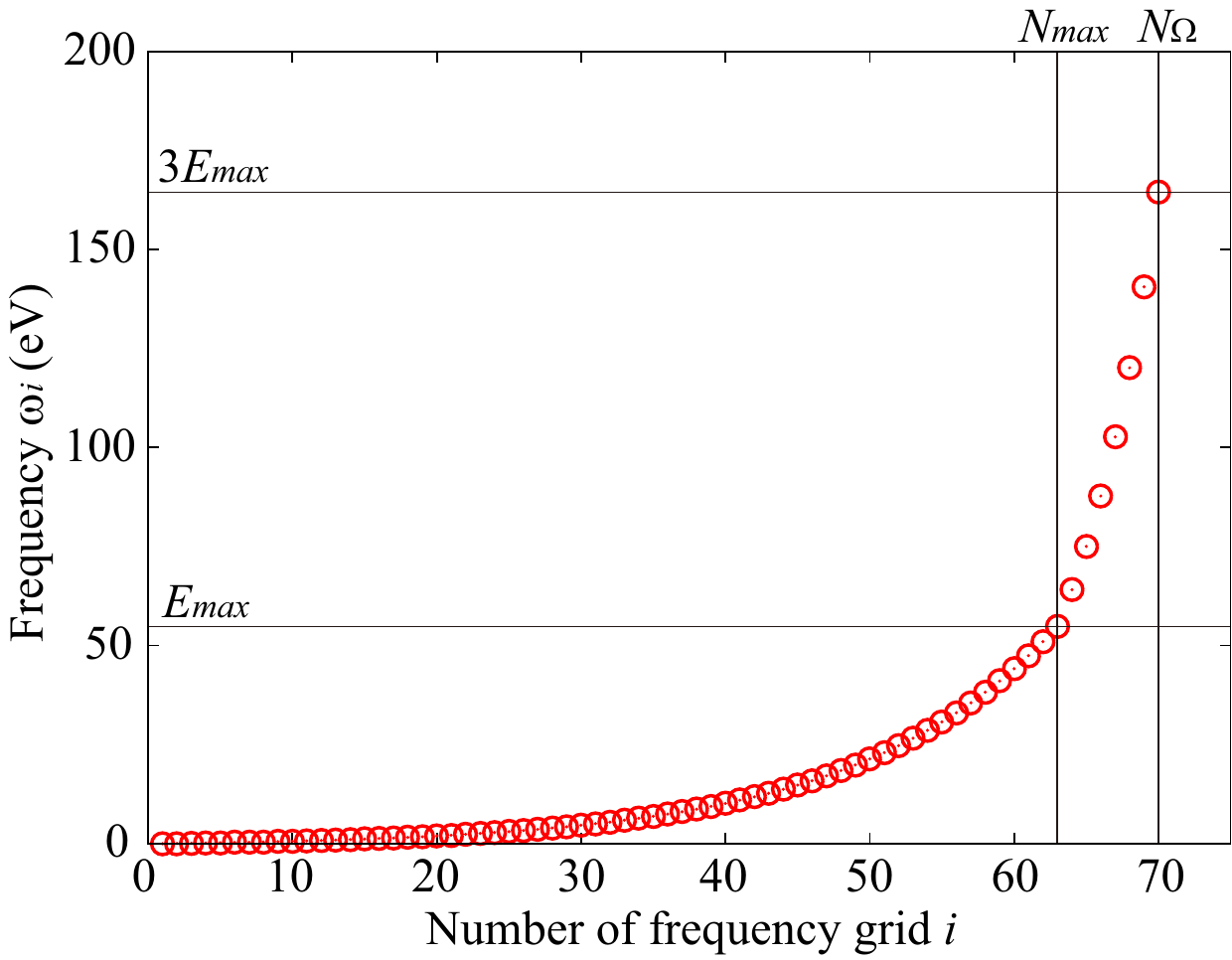}
\end{center} 
\vspace{-0.5cm} 
\caption{Frequency grid generated with Eq.~(\ref{eq:grid}).} 
\label{fig-wgrid}
\end{figure}

The frequency grid of the polarization function is generated as a logarithmic grid:  
\begin{eqnarray}
\hspace{-0.8cm} \omega_i=  
 \left\{
  \begin{array}{ll} 
  \frac{\Delta\omega}{s-1}\biggl( \exp\bigl[ (i-1)\ln s \bigr]-1\biggr), 
  &\mbox{$i=1, \ldots, N_{max}$,} \\
  E_{max}\exp\biggl[ \ln3 \frac{i-N_{max}}{N_{\omega}-N_{max}}\biggr], 
  &\mbox{$i=N_{max}+, \ldots, N_{\Omega}$.} 
  \end{array}
 \right. 
 \label{eq:grid} 
\end{eqnarray}
Here, $\omega_i$ is the $i$th frequency, $E_{max}=\max(\{E_{\alpha{\bf k}}\})-\min(\{E_{\alpha{\bf k}}\})$, $N_{\Omega}$ is the total number of the frequency grids, and $N_{max}$ is the total number of the frequency grids in the frequency range $0 \le \omega_i \le E_{max}$. By default, $N_{\Omega}$ and $N_{max}$ are set to 70 and $(9N_{\Omega})/10$. The parameter $s$ in Eq.~(\ref{eq:grid}) is determined by solving the equation $s^{N_{max}-1}+\frac{E_{max}}{\Delta\omega}(s-1)=1$. The resulting grids satisfy the following boundary conditions: (i) $\omega_1=0$, (ii) $\omega_2=\Delta\omega$ with $\Delta\omega$ being 0.05 eV by default, (iii) $\omega_{N_{max}}=E_{max}$, and (iv) $\omega_{N_{\Omega}}=3E_{max}$. An example of the generated grid is shown in Fig.~\ref{fig-wgrid}. 

\subsection{Interpolation treatment}\label{sec_interpolation_treatment} 
Using the resulting transfer data $t_{i{\bf 0}j{\bf R}}$ in Sec.~\ref{method:wannier}, the one-body Hamiltonian matrix at an arbitrary $k$-point ${\bf k}'$ is calculated as 
\begin{eqnarray}
H_{ij}({\bf k'})
=\sum_{{\bf R}}^{N_R} 
t_{i{\bf 0}j{\bf R}} w_{{\bf R}} 
e^{+i{\bf k}'\cdot{\bf R}}. 
\label{Hijk} 
\end{eqnarray}
Here, ${\bf k}'$ is the $k$ point employed in the band dispersion or the $k$ point used in the Monkhorst-Pack mesh for the density of state calculation:
\begin{numcases}
 {\{{\bf k'}\}=\!\!}   
  \!\!\!\{{\bf k}_{{\rm disp}}\},&\hspace{-0.6cm}\text{for band-dispersion calculation,}  \label{k-disp-grid} \\
  \!\!\!\{{\bf k}_{{\rm MP  }}\},&\hspace{-0.6cm}\text{for density-of-state calculation.}
  \label{k-dos-grid} 
\end{numcases}
Also, $w_{{\bf R}}$ in Eq.~\ref{Hijk} is a weight factor at the lattice ${\bf R}$, which is introduced to avoid the double counting of the transfer at the boundary edge of the system with the periodic boundary condition. Note that $w_{{\bf R}}$ satisfies the following sum rule $\sum_{{\bf R}}w_{{\bf R}}=N_k$. 

By diagonalizing the matrix $H_{ij}({\bf k'})$, 
\begin{eqnarray}
\sum_{j}^{N_w}H_{ij}({\bf k'})
 C_{j\alpha}({\bf k'})
= C_{i\alpha}({\bf k'}) 
 \epsilon_{\alpha{\bf k'}},   
\end{eqnarray}
we obtain eigenvectors \{$C_{j\alpha}({\bf k'})$\} and eigenvalues $\epsilon_{\alpha{\bf k'}}$. With $\epsilon_{\alpha{\bf k'}}$, we can calculate the density of state as 
\begin{eqnarray}
\rho(\omega)=\sum_{i}^{N_w} \rho_i(\omega)  
\end{eqnarray}
with 
\begin{eqnarray}
\hspace{-0.3cm} 
\rho_i(\omega)
=\frac{2}{N_{k'}}
\sum_{\bf k'}^{N_{k'}}
\sum_{\alpha}^{N_w}
|C_{i\alpha}({\bf k'})|^2 
\frac{1}{\pi} 
{\rm Im} \frac{1}{\omega-\epsilon_{\alpha{\bf k'}}-i\delta}  
\end{eqnarray}
being the partial density of state associated with the Wannier orbital $\phi_{i{\bf 0}}$. The factor of 2 comes from the sum over spin degrees of freedom. The Brillouin Zone integral is performed with the generalized tetrahedron technique~\cite{Fujiwara_2003,Nohara_2009}. 

As a similar quantity, a density matrix is calculated as follows:
\begin{eqnarray}
 D_{ij}({\bf R})
 &=&\frac{2}{N_k}
 \sum_{{\bf k}}^{N_{k'}} 
 \sum_{\alpha}^{N_w}
  C_{i\alpha}({\bf k'}) C_{j\alpha}^*({\bf k'}) e^{i{\bf k}\cdot{\bf R}} \nonumber \\ 
  &\times& \frac{1}{\pi} 
  \int_{-\infty}^{E_F}\frac{1}{\omega-\epsilon_{\alpha{\bf k}}-i\delta} d\omega. 
\end{eqnarray}
This is convenient to monitor occupancy of each Wannier orbital or bond order between the Wannier orbitals. The Fermi surface is also calculated as constant energy surface $F({\bf k}')$  
\begin{eqnarray}
 F({\bf k}')\ {\rm to\ satisfy}\ \epsilon_{\alpha{\bf k}'}=E_F. 
\end{eqnarray}
The output can be visualized by software {\sc Fermisurfer}~\cite{fermisurfer}. 

\subsection{Parallel calculation}\label{sec_parallel_calculation}  
The polarization function can be calculated in parallel. There are two parallelization levels; one over the irreducible $q$ points and the other over band pairs. First, let us consider the parallel calculation over the band pairs. To see this treatment, we rewrite the polarization function [Eq.~(\ref{eq:chi})] as follows:  
\begin{eqnarray}
 \chi_{{\bf GG'}}({\bf q},\omega) =\sum_{\alpha}^{N_{vir}} \sum_{\beta}^{N_{occ}} 
 \chi_{{\bf GG'}}^{\alpha\beta}({\bf q},\omega).   
\end{eqnarray}
The band sums are divided, and each $\chi_{{\bf GG'}}^{\alpha\beta}({\bf q},\omega)$ can be calculated by an independent MPI process. Now, we write this process as follows: 
\begin{eqnarray}
 \hspace{-0.3cm} \chi_{{\bf GG'}}({\bf q},\omega) 
\! \! \!  &=& \! \! \! 
 \sum_{n}^{N_{{\rm MPI}}} \sum_{m}^{N_{{\rm MPI}}} \biggl\{
  \sum_{\alpha_m}^{N_{vir}^{m}} 
  \sum_{\beta_n}^{N_{occ}^{n}} 
  \chi_{{\bf GG'}}^{\alpha_m \beta_n}({\bf q},\omega) \biggr\}.  
\end{eqnarray}
Here, $n$ and $m$ specify an index of an MPI process. 
The occupied-state and virtual-state data \{$\beta$\} and \{$\alpha$\} are divided into $N_{\rm MPI}$ processes; (\{$\beta_1$\}, ..., \{$\beta_{N_{\rm MPI}}$\}) and (\{$\alpha_1$\}, ..., \{$\alpha_{N_{\rm MPI}}$\}). The divided virtual-state data are interchanged among the MPI processes to compute the partial-sum contribution to the polarization function. {\tt MPI\_SENDRECV} routine is used for this data interchange.  

\begin{figure}[htpb] 
\begin{center} 
\includegraphics[width=0.475\textwidth]{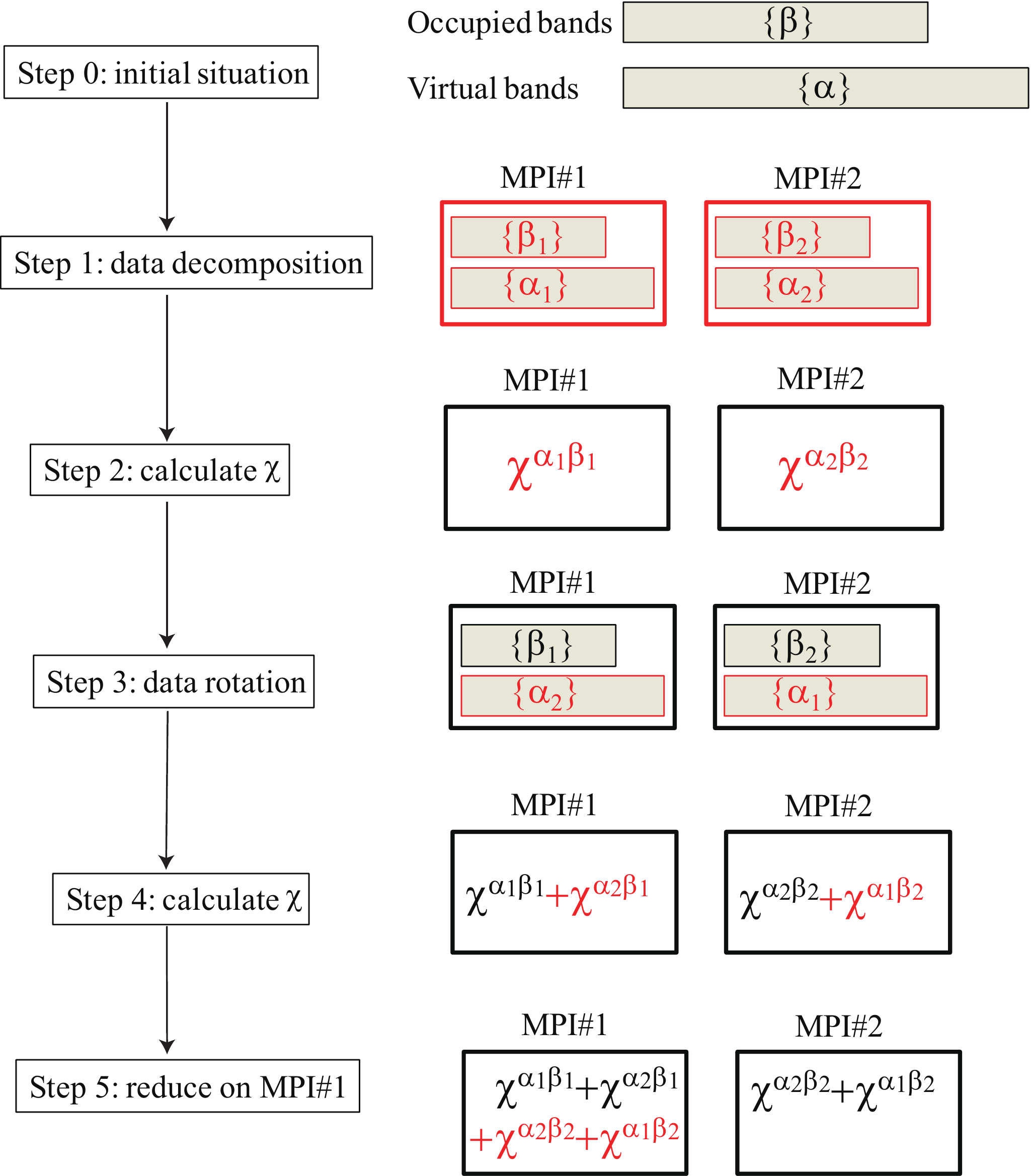}
\end{center} 
\vspace{-0.5cm} 
\caption{Practical procedure for an MPI calculation of polarization function. On the right side, work images are displayed, and the work at each step is highlighted in red. In this example, the total number of the MPI processes is 2. The band data are divided into two and accommodated in each MPI process (step 1). After performing polarization calculations in each MPI process (step 2), only virtual data are interchanged among the MPI processes (step 3). The polarization calculations are performed again (step 4). Finally, the partial-sum data of the polarization functions stored in each MPI process are collected in the master node (step 5).
} 
\label{fig-MPI}
\end{figure}

Figure~\ref{fig-MPI} is a practical procedure for the case of $N_{{\rm MPI}}=2$. The occupied-state and virtual-stat data are divided into two, and each data are stored in each MPI process (Step 1). After performing the polarization calculations in each MPI process (Step 2), only the virtual-state data \{$\alpha_1$\} and \{$\alpha_2$\} are interchanged between the two MPI processes (Step 3). Then, the polarization calculation is performed again (Step 4). 
Finally, the data stored in each MPI process is collected in the master process (Step 5).

\begin{figure}[h] 
\centering
\includegraphics[width=0.475\textwidth]{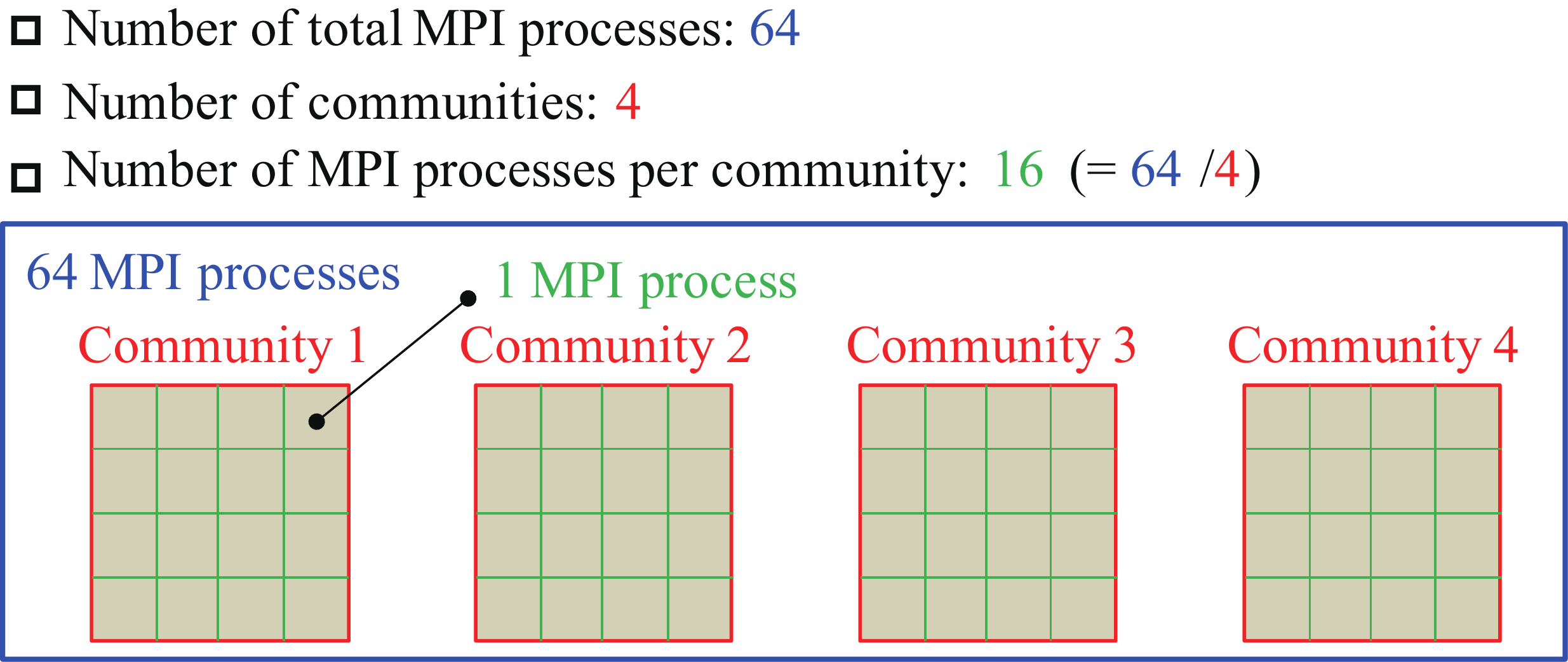}
\caption{Schematic diagram of a parallel computation on $q$ points. In this example, the total number of MPI processes is 64. All the MPI processes are divided into 4 MPI communities with setting the $q$-point parallel degree to 4. Hence, each MPI community consists of 16 MPI processes which are allocated to the parallel calculation for the band pair.}
\label{qcommunity}
\end{figure}

A parallel calculation over the $q$ points is more trivial. Consider the case where the total number of the MPI processes is 64, and the number of irreducible $q$ points is 4. 
We first divide all the 64 MPI processes into 4 communities, and thus each community consists of 16 MPI processes. One community performs the polarization calculation of one $q$ point. Figure.~\ref{qcommunity} is a schematic figure showing this procedure. {\tt MPI\_COMM\_SPLIT} routine is used for splitting to the communities. The 16 MPI processes in each community are assigned to perform the parallel calculation over the band pairs mentioned above. 

\section{Calculation flow} \label{sec_calculation_flow}
\begin{figure}[htpb] 
\begin{center} 
\includegraphics[width=0.475\textwidth]{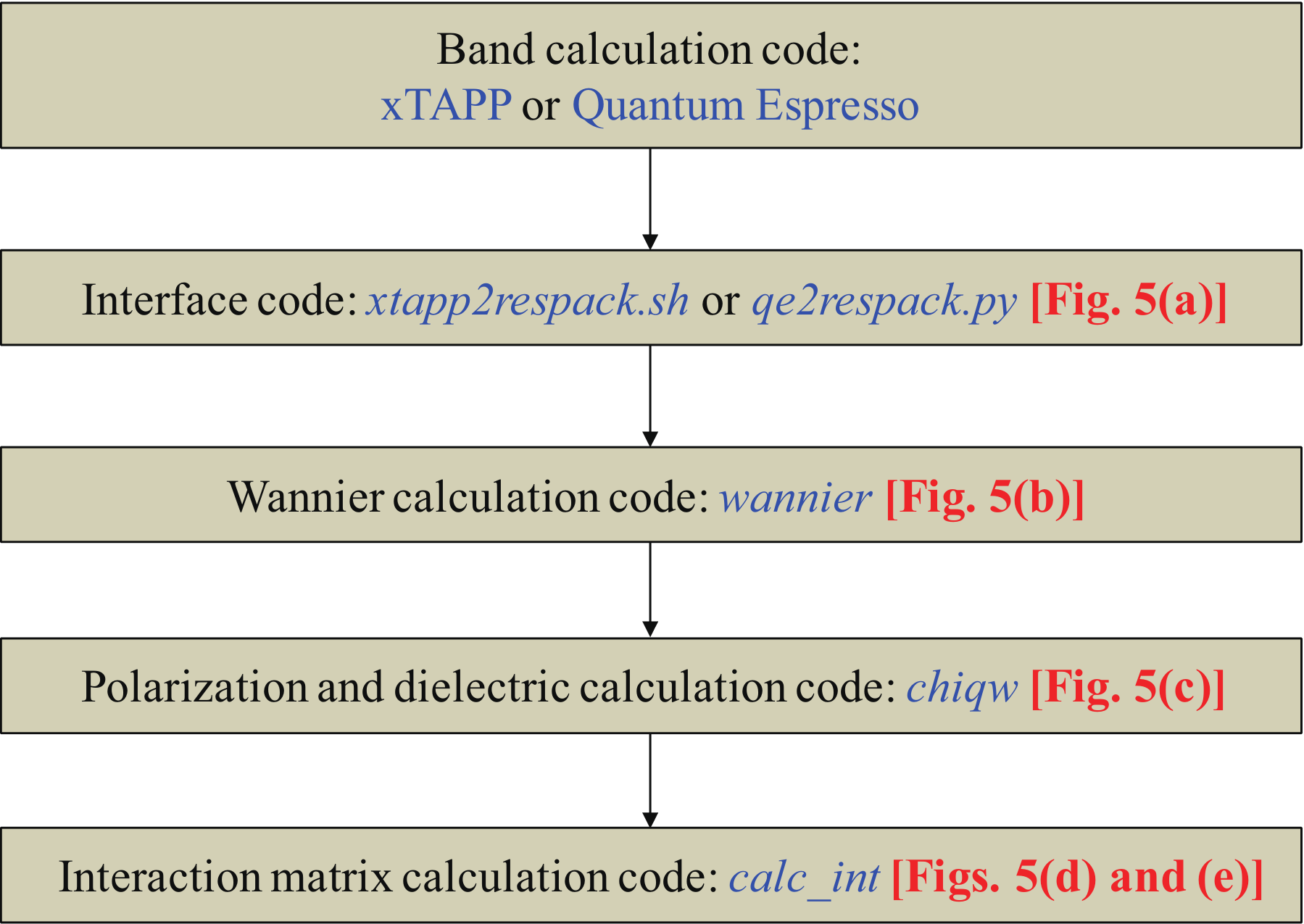}
\end{center} 
\vspace{-0.3cm} 
\caption{Flow diagram of an {\it ab initio} derivation for an effective low-energy model in Eq.~(\ref{eq:H}) with RESPACK.  After the band calculations with xTAPP or {\sc Quantum Espresso}, via an interface code, inputs of the RESPACK calculations are prepared. Lower three codes ({\it wannier}, {\it chqw}, and {\it calc\_int}) compose the main part of RESPACK to derive the effective model.} 
\label{flow-respack}
\end{figure}

We next describe the practical procedure of a RESPACK calculation~\cite{RESPACK_URL}. Figure~\ref{flow-respack} shows an overall flow diagram of calculation processes; first, we perform band-structure calculations with xTAPP~\cite{Yamauchi_1996} or {\sc Quantum Espresso}~\cite{QE-2009,QE-2017}. Next, with using the interface script, we convert the band-structure results to the inputs of RESPACK. Then, with the obtained band-calculation data and an input file that specifies the RESPACK-calculation condition, we perform the Wannier-function calculation. We call this calculation $wannier$. Then, we calculate the polarization and dielectric functions, and this calculation is called $chiqw$. In the constrained RPA, the polarization process is restricted by using the information of the Wannier function [see Eqs.~(\ref{eq:chi}) and (\ref{tak})], so one has to perform the {\it wannier} calculation before the {\it chiqw} calculation. Lastly, with the {\it wannier} and {\it chiqw} outputs, we evaluate the matrix elements of the screened interaction. This calculation is called $calc\_int$. In the following subsections, we describe details.  

\subsection{Preparation for RESPACK}  
We show in Fig.~\ref{flow}(a) a preparation process from band-structure calculation with xTAPP or {\sc Quantum Espresso} to RESPACK.  In RESPACK, interface scripts that convert outputs of the band calculation to inputs for RESPACK are prepared for these two codes. In the case of xTAPP, {\tt xtapp2respack.sh} generates a directory {\tt dir-wfn}, in which the following 9 files are created. 
\vspace{3mm}\hrule
\begin{enumerate}
\item {\tt dat.bandcalc} (band calculation information)
\item {\tt dat.sample-k} (sample $k$ points)
\item {\tt dat.symmetry} (symmetry operations)
\item {\tt dat.lattice} (lattice vectors)
\item {\tt dat.eigenvalue} (energy eigenvalues)
\item {\tt dat.nkm} (number of reciprocal lattice vectors)
\item {\tt dat.wfn} (wave functions)
\item {\tt dat.kg} (reciprocal lattice vectors)
\item {\tt dat.atom\_position} (atomic positions)
\end{enumerate}
\hrule\vspace{3mm}
For the format of each file, see the manual~\cite{RESPACK_URL}. In the case of {\sc Quantum ESPRESSO}, {\tt qe2respack.py} is the generation script. After this process, RESPACK calculations are performed with the data in {\tt dir-wfn}.
%

\subsection{Wannier calculation}  
Figure~\ref{flow}(b) is a flow diagram of $wannier$. With the data in {\tt dir-wfn} and an input {\tt input.in} that describes conditions of the {\it wannier} calculation, the calculation is performed with an executable file {\tt calc\_wannier}. Details of {\tt input.in} are described in ~\ref{sec_input}. After the calculation, two directories {\tt dir-wan} and {\tt dir-model} are generated, in which the calculation results are saved. For details of the generated output files, see the descriptions in Fig.~\ref{flow}(b).

\subsection{Chiqw calculation}  
We next show in Fig.~\ref{flow}(c) a flow diagram of {\it chiqw}. This code calculates the polarization and dielectric functions. With the data in {\tt dir-wfn} and {\tt dir-wan} and the input file {\tt input.in}, the calculation is performed with an executable file {\tt calc\_chiqw}. After the calculation, a directory {\tt dir-eps} is generated and, under this  directory, subdirectories {\tt q001}, {\tt q002}, ... {\tt qNirr} are generated, where {\tt 001}, {\tt 002}, and {\tt Nirr} are the numbers of the irreducible $q$ points. The calculation results of every $q$ points are saved in each subdirectory.  

\subsection{Calc\_int calculation}
\begin{figure*}[htpb] 
\begin{center} 
\includegraphics[width=0.88\textwidth]{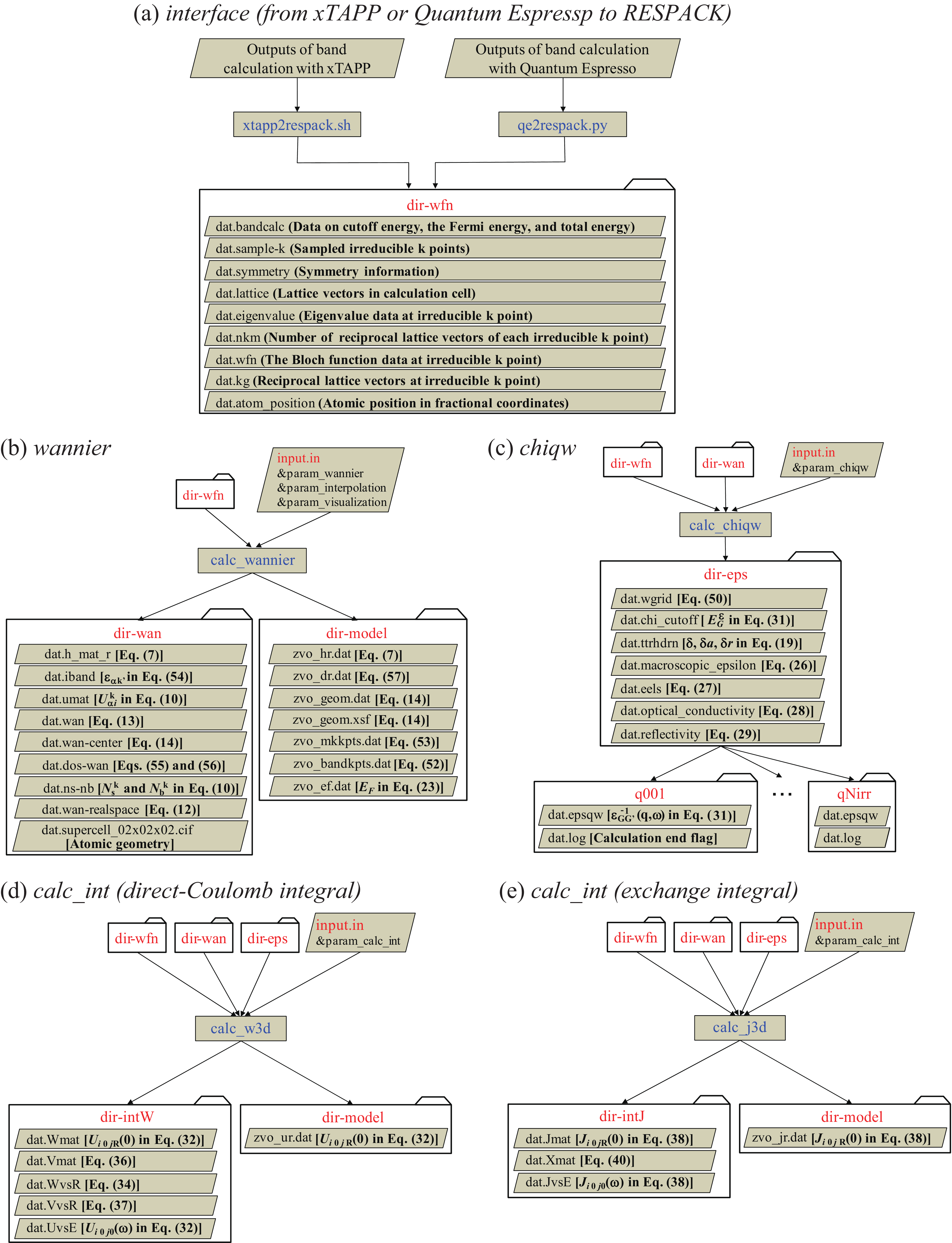}
\end{center} 
\vspace{-0.5cm} 
\caption{Flow diagram of RESPACK. (a) Preparation process from DFT codes of xTAPP or {\sc Quantum Espresso} to RESPACK. The band-calculation results are converted to inputs of RESPACK with an interface program ({\tt xtapp2respack.sh} for xTAPP and {\tt qe2respack.py} for {\sc Quantum Espresso}). Nine files are generated in the directory {\tt dir-wfn}. (b) {\it wannier} calculation: With the data in {\tt dir-wfn} and {\tt input.in}, the $wanner$ calculation is performed with an executable file {\tt calc\_wannier}. The calculation results are stored in the directories {\tt dir-wan} and {\tt dir-model}. (c) {\it chiqw} calculation: With the data in {\tt dir-wfn} and {\tt dir-wan} and {\tt input.in}, the $chiqw$ calculation is performed with an executable file {\tt calc\_chiqw}. The calculation results are saved in the directories {\tt dir-eps} and subdirectories {\tt q001}, {\tt q002}, ... {\tt qNirr}. See text for details. (d) and (e) {\it calc\_int} calculation [(d) direct-Coulomb integral and (e) exchange integral]. With the data in {\tt dir-wfn}, {\tt dir-wan}, and {\tt dir-eps} and {\tt input.in}, the $calc\_int$ calculation is performed with an executable file {\tt calc\_w3d} for the direct-Coulomb integrals and {\tt calc\_j3d} for the exchange integrals. The calculation results are stored in the directories {\tt dir-intW}, {\tt dir-intJ}, and {\tt dir-model}.} 
\label{flow}
\end{figure*}

Figures~\ref{flow}(d) and (e) show flow diagrams for the direct-Coulomb-integral and exchange-integral calculations, respectively. A common namelist {\tt \&param\_calc\_int} described in {\tt input.in} can be used for the two calculation programs (see Appendix~\ref{param_calc_int}). Executable files are {\tt calc\_w3d} for the direct-Coulomb integral and {\tt calc\_j3d} for the exchange integral. The calculation results are saved in the directories {\tt dir-intW}, {\tt dir-intJ}, and {\tt dir-model}.

\section{Installation instructions}\label{sec_install}

\subsection{Download source files and compile}
The source code of RESPACK can be obtained from the official website https://sites.google.com/view/kazuma7k6r. A gzipped tar file {\tt RESPACK.tar.gz} contains everything necessary for installation. 
When moving to the directory {\tt src}, one finds three source directories: {\tt calc\_int}, {\tt chiqw}, {\tt wannier}. {\tt Makefile} is prepared in each source directory, and one executes the {\tt make} command to compile these source codes. In the case of {\tt chiqw}, the work so far is as follows: 
\vspace{3mm}\hrule\vspace{3mm}
{\tt > tar -zxvf RESPACK.tar.gz}

{\tt > cd RESPACK/src/chiqw/}

{\tt > make} 
\vspace{3mm}\hrule\vspace{3mm}
\noindent 
Here, {\tt >} is a prompt character. After {\tt make}, an executable file {\tt calc\_chiqw} is generated. This procedure is the same for the other programs.

\subsection{Compile using cmake}
RESPACK can also be compiled using {\tt CMake}. In {\tt CMake}, one needs to make a temporary directory for compilation, and executes the {\tt cmake} and {\tt make} commands from that directory as follows: 

\vspace{3mm}\hrule\vspace{3mm} 
{\tt > tar -zxvf RESPACK.tar.gz}

{\tt > cd RESPACK}

{\tt > mkdir build}

{\tt > cd build} 

{\tt > cmake -DCONFIG=gcc} 

\hspace{0.2cm}{\tt -DCMAKE\_INSTALL\_PREFIX=PATH\_TO\_INSTALL ../}

{\tt > make} 

{\tt > make install} 

\vspace{3mm}\hrule\vspace{3mm}
\noindent
If {\tt make} is successful, executable files are generated in each directory under {\tt RESPACK/build/src.} By executing {\tt make install}, the executable files will be installed in the {\tt bin} directory under the directory specified by the {\tt -DCMAKE\_INSTALL\_PREFIX} option. If the {\tt -DCMAKE\_INSTALL\_PREFIX} option is omitted, it will be installed under {\tt /usr/local/bin}. The {\tt -DCONFIG} option is used for reading the {\tt CMake} configuration files stored in the {\tt RESPACK/config} directory. For {\tt -DCONFIG}, the following options are available:
\begin{itemize}
    \item {\tt intel}: Intel compiler 
    \item {\tt gcc}: GNU compiler
\end{itemize}
If one wants to execute the {\tt cmake} command again, it is recommended that one deletes the temporary directory {\tt build} and restart from scratch, because the previous settings may remain. 
%
%
%

\section{Benchmark} \label{sec_benchmark}

\begin{table*}[htpb] 
\caption{Input file {\tt input.in} for a RESPACK calculation. This input describes a derivation of parameters specifying an effective low-energy model for the $t_{2g}$ band of SrVO$_3$. {\tt \&param\_wannier}, {\tt \&param\_interpolation}, and {\tt \&param\_visualization} specify the namelists for the {\it wannier} calculation [see Fig.~\ref{flow}(b) and details are described in \ref{param-wannier}, \ref{param-interpolation}, and \ref{param-visalization}]. {\tt \&param\_chiqw} is the namelist for the $chiqw$ calculation [Fig.~\ref{flow}(c) and \ref{param-chiqw}]. {\tt \&param\_calc\_int} is the namelist for the $calc\_int$ calculation [Figs.~\ref{flow}(d) and (e) and \ref{param-calcint}]. A brief description of each variable is given in the right of {\tt !}. Default values are shown with a bold font in parentheses. A variable that does not include a bold-font value is a required variable.} 
\begin{center} 
\scalebox{1.0}{
\begin{tabular}{l} \hline \hline  \\ [-10pt] 
 {\tt \&param\_wannier}  \\
 {\tt N\_wannier=3,} ! Number of the Wannier functions you want to calculate \\
 {\tt Lower\_energy\_window=6.50, } ! Lower bound of energy window \\   
 {\tt Upper\_energy\_window=9.70, } ! Upper bound of energy window \\   
 {\tt N\_initial\_guess=3,} ! Number of initial guesses \\
 / \\ 
 {\small {\tt dxy 0.50 0.50  0.50  0.50 ! vec\_ini(1)\%orb vec\_ini(1)\%a vec\_ini(1)\%x vec\_ini(1)\%y vec\_ini(1)\%z}} \\
 {\small {\tt dyz 0.50 0.50  0.50  0.50 ! vec\_ini(2)\%orb vec\_ini(2)\%a vec\_ini(2)\%x vec\_ini(2)\%y vec\_ini(2)\%z}} \\
 {\small {\tt dzx 0.50 0.50  0.50  0.50 ! vec\_ini(3)\%orb vec\_ini(3)\%a vec\_ini(3)\%x vec\_ini(3)\%y vec\_ini(3)\%z}} \\
 \\ 
 {\tt \&param\_interpolation} \\   
 {\tt N\_sym\_points=5,} ! Number of symmetric $k$ points in calculation lines for band dispersion\\
 / \\
 {\small {\tt 0.50 0.50 0.50 ! SK\_sym\_pts(1,1) SK\_sym\_pts(2,1) SK\_sym\_pts(3,1)}}: R\\ 
 {\small {\tt 0.00 0.00 0.00 ! SK\_sym\_pts(1,2) SK\_sym\_pts(2,2) SK\_sym\_pts(3,2)}}: Gamma \\ 
 {\small {\tt 0.50 0.00 0.00 ! SK\_sym\_pts(1,3) SK\_sym\_pts(2,3) SK\_sym\_pts(3,3)}}: X \\ 
 {\small {\tt 0.50 0.50 0.00 ! SK\_sym\_pts(1,4) SK\_sym\_pts(2,4) SK\_sym\_pts(3,4)}}: M \\ 
 {\small {\tt 0.00 0.00 0.00 ! SK\_sym\_pts(1,5) SK\_sym\_pts(2,5) SK\_sym\_pts(3,5)}}: Gamma \\ 
 \\
 {\tt \&param\_visualization} \\   
 {\tt Flg\_vis\_wannier=1,} ! Calculate realspace Wannier function (do not: 0, do: 1) {\bf (0)} \\
 / \\
 \\ 
 {\tt \&param\_chiqw} \\
 {\tt Ecut\_for\_eps=10.0,} ! Cutoff energy for polarization function in Rydberg unit {\bf (1/10 of wave-function cutoff)} \\
 {\tt Num\_freq\_grid=70,} ! Number of frequency grid {\bf (70)} \\
 {\tt Green\_func\_delt=0.1,} ! Smearing value used in tetrahedron calculation (eV) {\bf (0.1 eV)} \\ 
 {\tt MPI\_num\_qcomm=1,} ! Degree of parallelism for $q$-point parallel calculation {\bf (1)}\\ 
 {\tt Flg\_cRPA=1,} ! Flag for constrained RPA or usual RPA (usual RPA: 0, constrained RPA: 1) {\bf (0)} \\
 / \\ 
 \\ 
 {\tt \&param\_calc\_int} \\  
 {\tt Calc\_ifreq=1,} ! Number of frequency to output {\bf (1)} \\
 / \\ 
 \hline \hline
\end{tabular} 
}
\end{center}
\label{input-SrVO3} 
\end{table*} 

\subsection{Typical outputs} \label{typical-output} 
In this section, we show benchmark results for a $t_{2g}$-model derivation of perovskite oxide SrVO$_3$ with a simple cubic structure having a lattice constant of 3.8425 \AA. Density functional calculations with plane-wave basis sets were performed using the xTAPP code~\cite{Yamauchi_1996}, where the norm-conserving  pseudopotential~\cite{Kleinman_1982,Troullier_1991} and the generalized gradient approximation to the exchange correlation energy were employed~\cite{Perdew_1996}. The calculation condition is set to $8 \times 8 \times 8$ $k$-point sampling, 100-Ry wavefunction cutoff, and 400-Ry charge-density cutoff. The 50 bands are considered for the polarization function, which corresponds to considering the excitation from the Fermi level to 35 eV. The numbers of the doubly-occupied, partially-occupied, and unoccupied bands are 12, 3, and 35, respectively.  

Table~\ref{input-SrVO3} is an input file {\tt input.in} for RESPACK calculations. Details of {\tt input.in} are described in \ref{sec_input}. We construct the $t_{2g}$-type Wannier functions from the low-energy bands near the Fermi level. The calculations are performed with both of the constrained RPA and usual RPA to show the difference between these two. The cutoff for the polarization function is set to 10 Ry, and the broadening factor $\delta$ of the generalized tetrahedron calculation is set to 0.1 eV.

\begin{figure}[htpb] 
\begin{center} 
\includegraphics[width=0.47\textwidth]{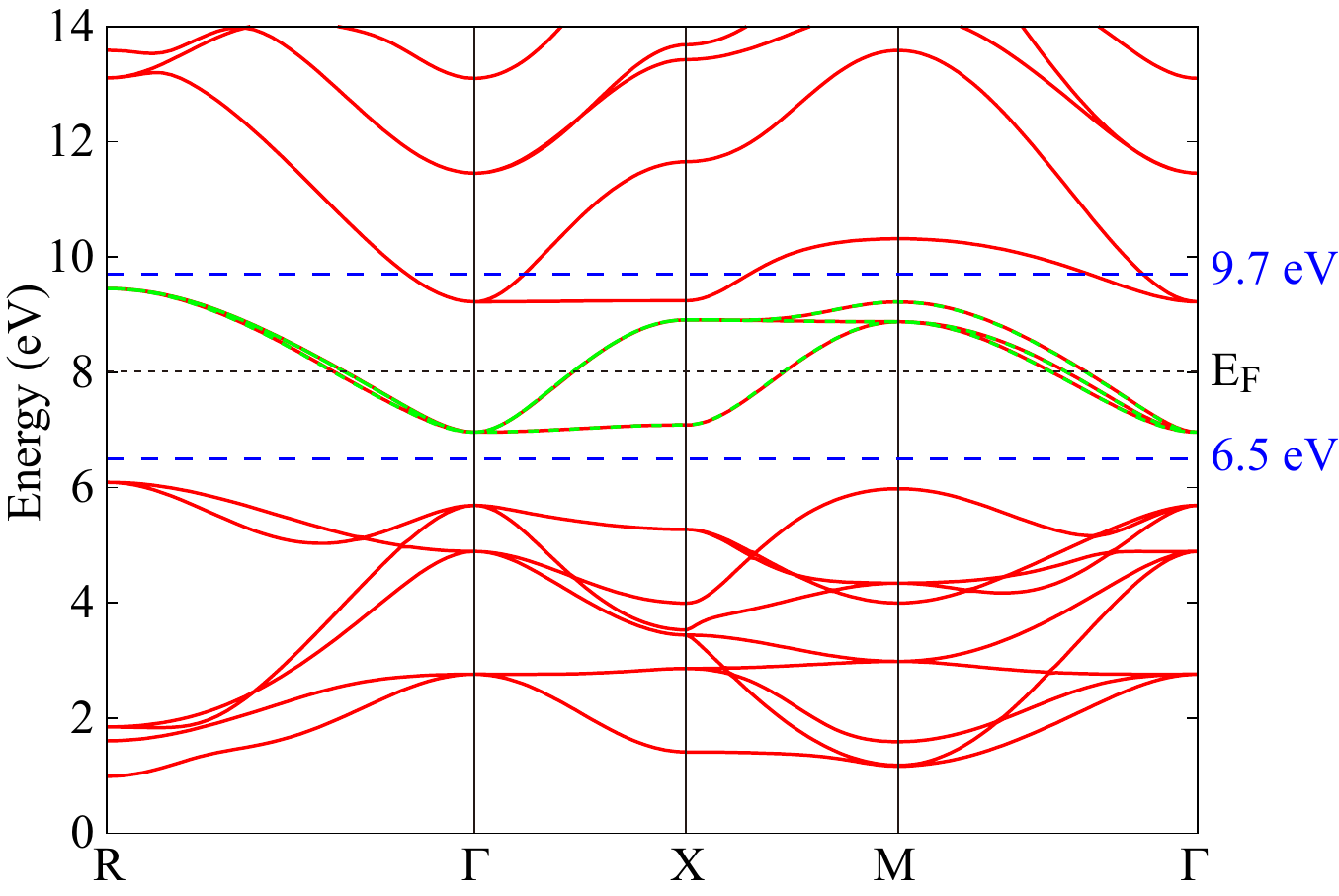}
\end{center} 
\vspace{-0.5cm} 
\caption{Calculated band structure for SrVO$_3$. Red-solid curve is DFT band dispersion by xTAPP and green-dashed curve is the Wannier-interpolated $t_{2g}$-band dispersion with RESPACK-$wannier$ code. A region between the two blue-dashed horizontal lines indicates the energy window region, and the band structure data in this region is used for constructing the Wannier functions. The dotted horizontal line is the Fermi level as 8.02 eV. Dispersions are plotted along the high-symmetry points, where R=$\frac{1}{2}{\bm a}^*+\frac{1}{2}{\bm b}^*+\frac{1}{2}{\bm c}^*$, $\Gamma$=${\bm 0}$, X=$\frac{1}{2}{\bm a}^*$, and M=$\frac{1}{2}{\bm a}^*+\frac{1}{2}{\bm b}^*$ with ${\bm a}^*, {\bm b}^*$, and ${\bm c}^*$ being basis vectors of reciprocal lattice, respectively. In the simple cubic primitive lattice, ${\bm a}^*=\frac{2\pi}{a}\hat{x}$, ${\bm b}^*=\frac{2\pi}{a}\hat{y}$, and ${\bm c}^*=\frac{2\pi}{a}\hat{z}$ with $\hat{x}$, $\hat{y}$, and $\hat{z}$ being unit vectors of the Cartesian coordinates. Also, $a$ is a lattice constant of 3.8425 \AA.} 
\label{band-SrVO3}
\end{figure}

\begin{figure}[htpb] 
\begin{center} 
\includegraphics[width=0.4\textwidth]{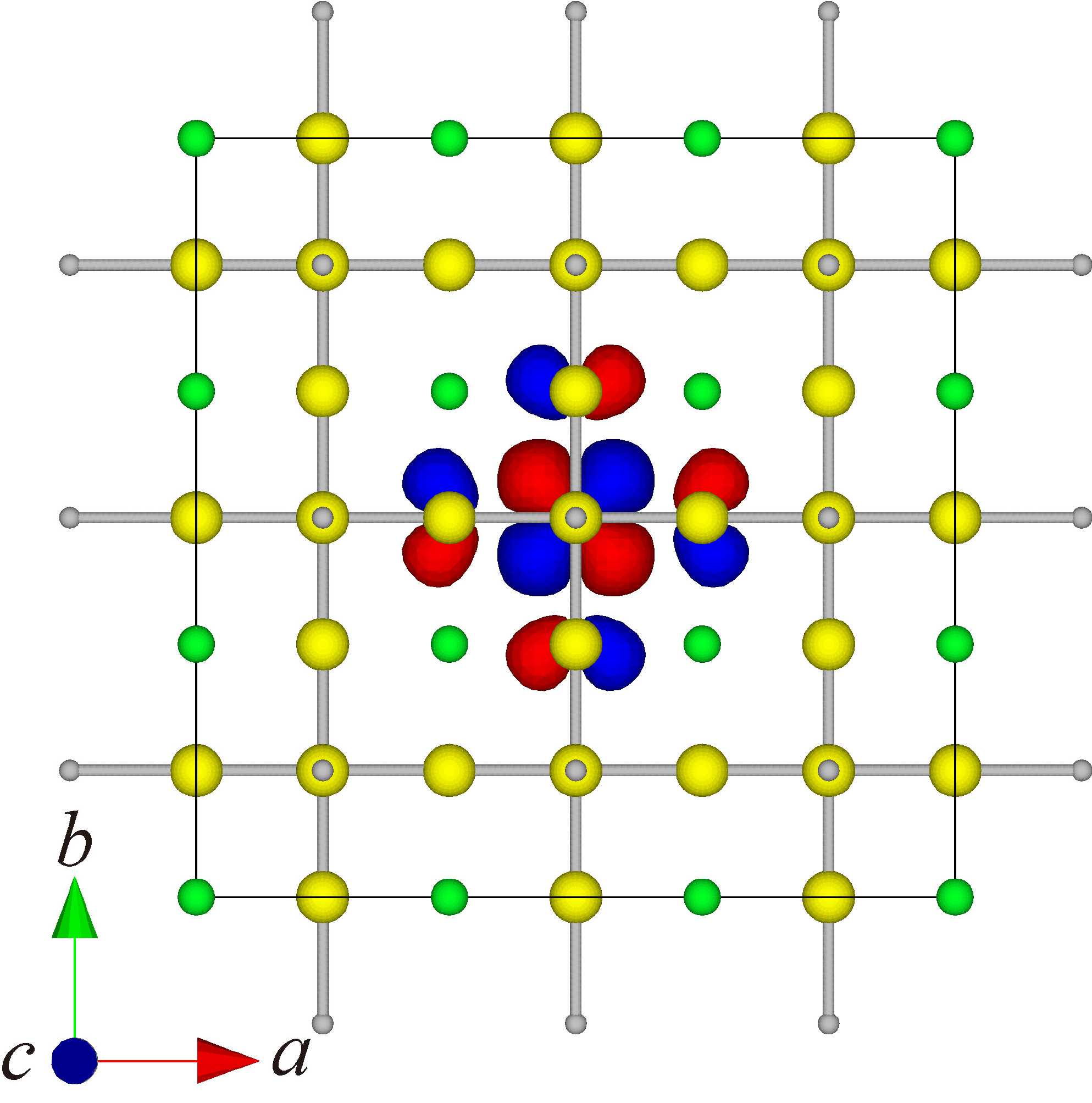}
\end{center} 
\vspace{-0.5cm} 
\caption{Calculated realspace $d_{xy}$-type Wannier function of SrVO$_3$ with the RESPACK-$wannier$ code (drawn by VESTA~\cite{Momma_2011}). Blue and red contour plots indicate positive- and negative-value region of the Wannier function. Large-yellow, middle-green, and small-gray spheres are oxygen, strontium, and vanadium atoms, respectively. In the simple cubic primitive lattice, ${\bm a}=a\hat{x}$, ${\bm b}=a\hat{y}$ and ${\bm c}=a\hat{z}$.} 
\label{realspace-wannier-SrVO3}
\end{figure}

Figure~\ref{band-SrVO3} is a comparison between the original KS band (red-solid curves) and the Wannier-interpolated band (green-dashed curves). A region between the two blue-dashed horizontal lines indicates the energy window used to construct the $t_{2g}$-type Wannier functions. We also show in Fig.~\ref{realspace-wannier-SrVO3} the calculated $d_{xy}$ Wannier function in realspace.

\begin{figure}[htpb] 
\begin{center} 
\includegraphics[width=0.4\textwidth]{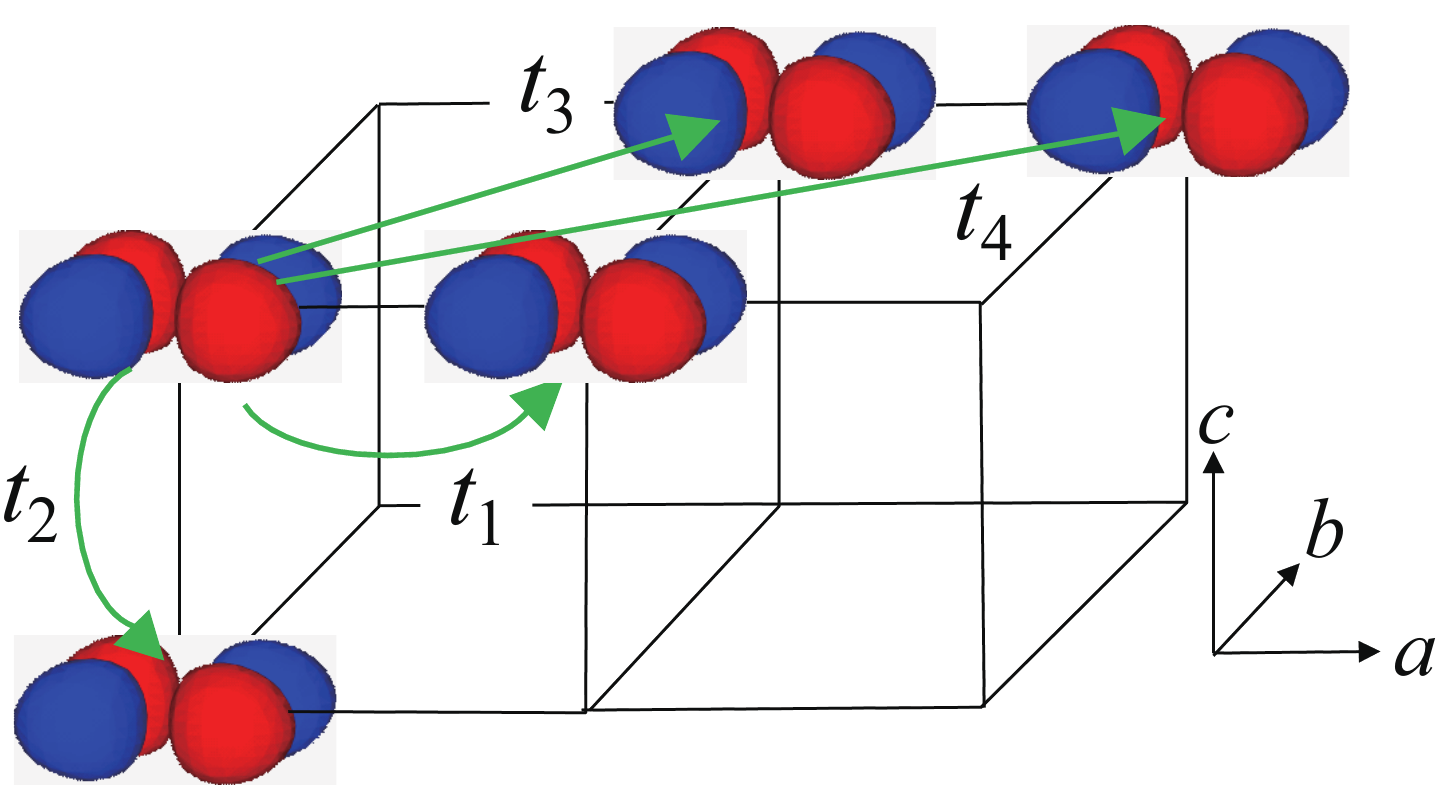}
\end{center} 
\vspace{-0.5cm} 
\caption{Schematic figure for transfer integrals. In this figure, we depict the $d_{xy}$-type Wannier orbital. Equivalent transfers exist for the $d_{yz}$- and $d_{zx}$-type Wannier orbitals.} 
\label{transfer_network}
\end{figure}

We show in Table~\ref{param_transfer} important transfers for the $t_{2g}$ band, where the definition of $t_1$, $t_2$, $t_3$, and $t_4$ are illustrated in Fig.~\ref{transfer_network}. In this figure, we depict the $d_{xy}$ Wannier function as an example. Since the lattice of the system is simple cubic, there exist equivalent transfers for the $d_{yz}$ and $d_{zx}$ orbitals. RESPACK provides a utility code that searches the equivalent transfers, which is described in \ref{sec_tr} in more detail. We note that the original band structure in Fig.~\ref{band-SrVO3} are well reproduced by these four transfers.
\begin{table}[htpb] 
\caption{Derived transfer parameters of a $t_{2g}$ model of SrVO$_3$ as the Wannier matrix elements for the Kohn-Sham Hamiltonian in Eq.~(\ref{eq:t}), where we show main 4 transfer integrals. Definition for $t_1$, $t_2$, $t_3$, and $t_4$ are given in Fig.~\ref{transfer_network}. These four transfers successfully reproduce the original $t_{2g}$-band structure in Fig.~\ref{band-SrVO3}. The unit of transfer integral is eV.}  
\begin{center} 
\scalebox{1.0}{
\begin{tabular}{c@{\ \ \ }c@{\ \ \ }c@{\ \ \ }c} 
\hline \hline  \\ [-10pt] 
  $t_1$    & $t_2$    & $t_3$    & $t_4$    \\
  \hline \\ [-10pt]
  $-0.259$ & $-0.026$ & $-0.086$ & $-0.012$ \\  
  \hline \hline 
\end{tabular}
} 
\end{center}
\label{param_transfer} 
\end{table} 

\begin{figure}[htpb] 
\begin{center} 
\includegraphics[width=0.37\textwidth]{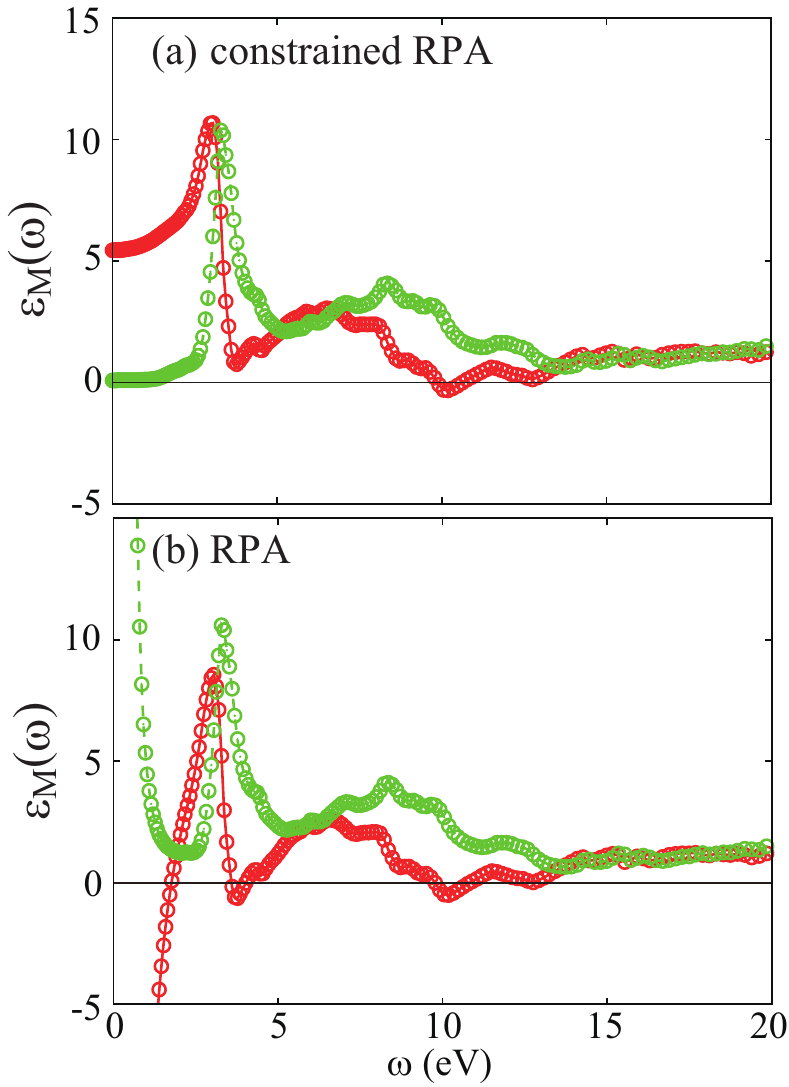}
\end{center} 
\vspace{-0.5cm} 
\caption{Calculated macroscopic dielectric function in Eq.~(\ref{eq:epsilonM}) of SrVO$_3$ with the RESPACK-$chiqw$ code: (a) constrained RPA and (b) usual RPA. Circles represent calculation values, and the red-solid and green-dashed curves describe the real and imaginary parts of the spectra, respectively.} 
\label{epsilonM}
\end{figure}

Figure~\ref{epsilonM} shows calculated macroscopic dielectric functions [Eq.~(\ref{eq:epsilonM})] with the RESPACK-$chiqw$ code. Panels (a) and (b) describe the constrained RPA and usual RPA results, respectively. Red-solid and green-dashed curves describe the real and imaginary parts, respectively, and circles represent calculation values. The difference between the constrained RPA and usual RPA spectra is appreciable in the low-energy excitation region less than 2-3 eV. In the constrained RPA, a metallic charge excitation is excluded by the polarization constraint described in Sec.~\ref{method:chiqw}, and then the real part of $\epsilon_{{\rm M}}(\omega)$ converges to the finite value in the $\omega\to 0$ limit, while, in the usual RPA, the real part of $\epsilon_{{\rm M}}(\omega)$ diverges negatively due to the metallic charge excitation~\cite{about-Re-eM-w-0}, thus leading to the Drude behavior of the imaginary part of $\epsilon_{{\rm M}}(\omega)$. 

It should be noted here that the present spectra neglect the transition moment contributed from the commutation relation between the non-local pseudopotential and electronic position, $[V_{{\rm NL}}, {\bf r}]$, in Eq.~(\ref{momentum}). For transition metals, this contribution is known to affect the spectral property in the low-energy excitation region~\cite{Marini2001}. In the present SrVO$_3$, we checked that this contribution is not significant. Support for this contribution is a future issue in the RESPACK project.

\begin{figure}[htpb] 
\begin{center} 
\includegraphics[width=0.37\textwidth]{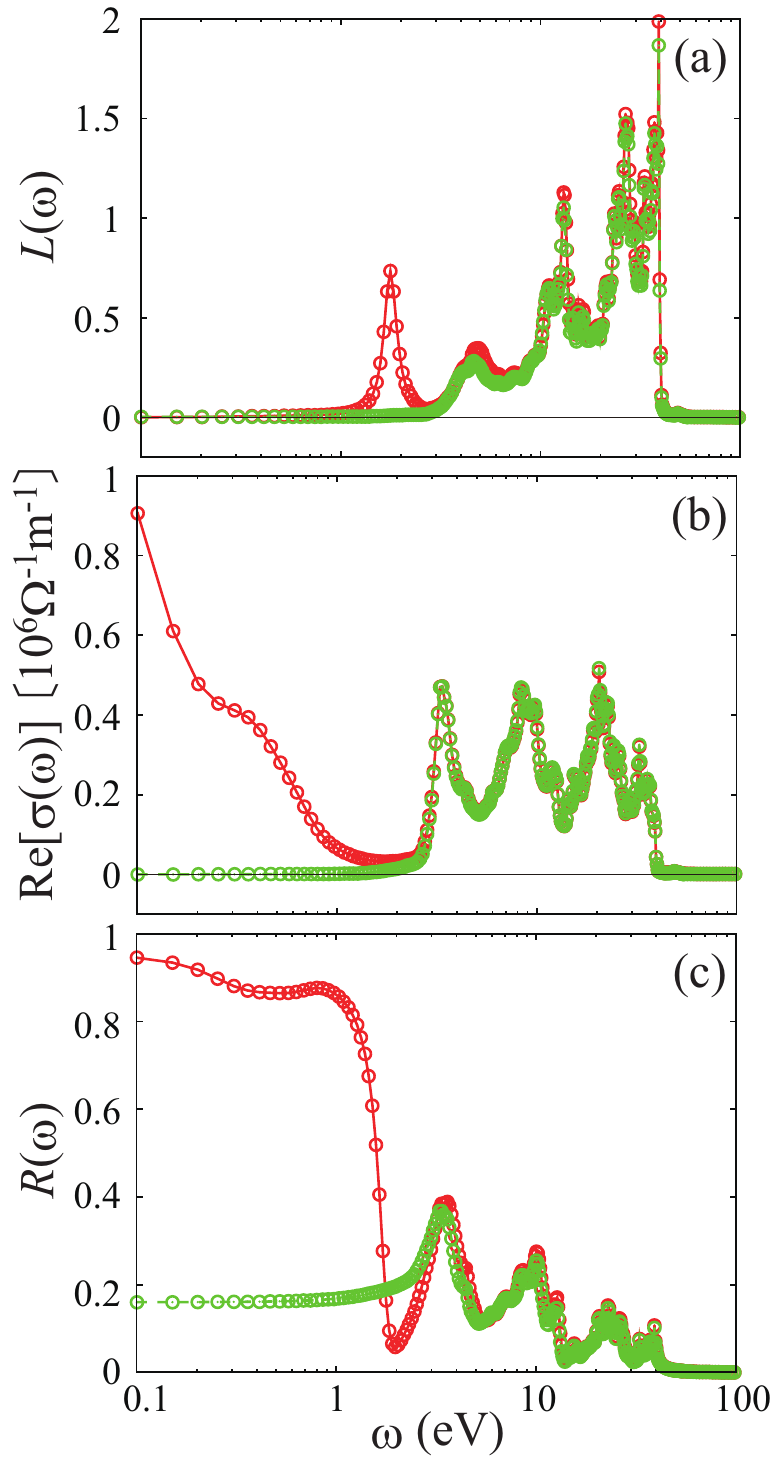}
\end{center} 
\vspace{-0.5cm} 
\caption{Calculated optical properties of SrVO$_3$ with the RESPACK-$chiqw$ code. (a) EELS $L(\omega)$ in Eq.~(\ref{eq:eels}), (b) the real part of the optical conductivity $\sigma(\omega)$ in Eq.~(\ref{eq:optical-conductivity}), (c) reflectance spectrum $R(\omega)$ in Eq.~(\ref{eq:reflectance}). Circles represent calculation values, and the red-solid and green-dashed curves describe the usual RPA and the constrained RPA results, respectively.} \label{eels-sigma-ref}
\end{figure}

We next show in Fig.~\ref{eels-sigma-ref} other optical properties calculated with the usual RPA and constrained RPA. 
Panels (a), (b), and (c) display EELS $L(\omega)$ [Eq.~(\ref{eq:eels})], the real part of the optical conductivity $\sigma(\omega)$ [Eq.~(\ref{eq:optical-conductivity})], and the reflectance spectrum $R(\omega)$ [Eq.~(\ref{eq:reflectance})], respectively. Red-solid and green-dashed curves represent the results based on the usual RPA and constrained RPA, respectively. Circles denote the calculation values. There is a difference between the constrained RPA and usual RPA in the EELS around the low-energy excitation region less than 2 eV; in the usual RPA case, an additional peak appears in $L(\omega)$, which is due to the low-energy plasmon excitation in the $t_{2g}$ band~\cite{Nakamura_2016,Makino-1998}. This plasma excitation is also observed in the reflectance spectrum $R(\omega)$; we see a sharp drop from around 1 to 0 in the RPA reflectance spectrum. On the optical conductivity, the spectral trend is basically the same as the macroscopic dielectric function $\epsilon_{{\rm M}}(\omega)$; the usual RPA spectrum exhibits the Drude behavior characteristic of a metallic system in the low-excitation region less than 2 eV, while the constrained RPA spectrum has no intensity in this frequency region, which is a characteristic aspect of the insulating system. 

\begin{figure}[htpb] 
\begin{center} 
\includegraphics[width=0.42\textwidth]{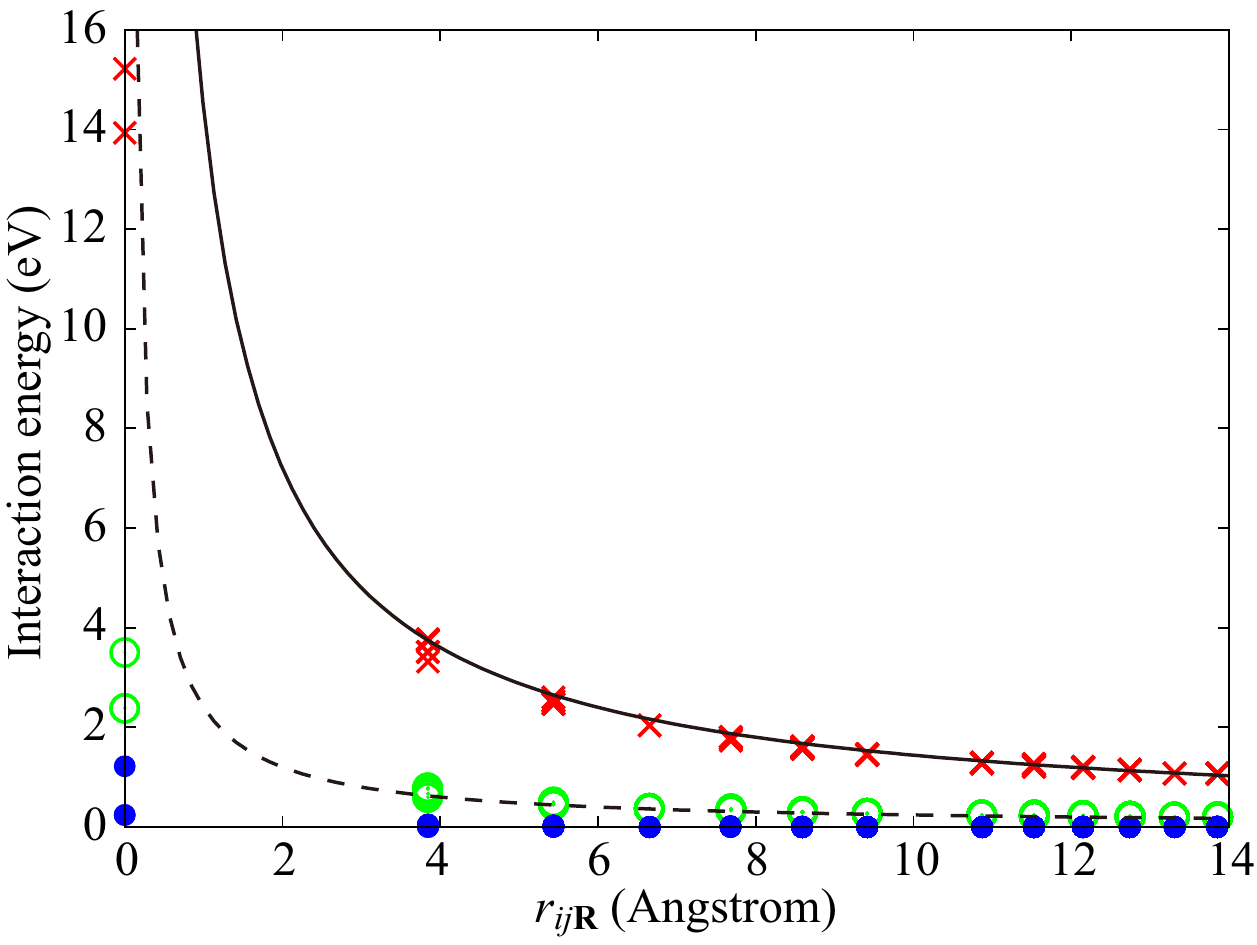}
\end{center} 
\vspace{-0.5cm} 
\caption{Dependence of direct-Coulomb integral on distance between $t_{2g}$ Wannier orbitals of SrVO$_3$ [Eqs.~(\ref{WvsR}) and (\ref{VvsR})], calculated with RESPACK-$calc\_int$ code. Red crosses, green-open circles, and blue dots represent the bare, constrained-RPA, and usual RPA results, respectively. Solid and dashed curves are $1/r$ and $1/(\epsilon_0 r)$ with $\epsilon_0=6$, respectively.} 
\label{WvsR-SrVO3}
\end{figure}

We next show in Fig.~\ref{WvsR-SrVO3} a distance dependence of the static ($\omega=0$) direct-Coulomb integral [Eq.~(\ref{WvsR}) for screened interaction and Eq.~(\ref{VvsR}) for bare interaction], where the distance between the Wannier functions are defined by Eq.~(\ref{wannier-distance}). Red crosses, green-open circles, and blue dots represent the bare, constrained-RPA, and RPA results, respectively. Solid and dashed curves are $1/r$ and $1/(\epsilon_0 r)$ with $\epsilon_0=6$, respectively. 

\begin{figure}[htpb] 
\begin{center} 
\includegraphics[width=0.37\textwidth]{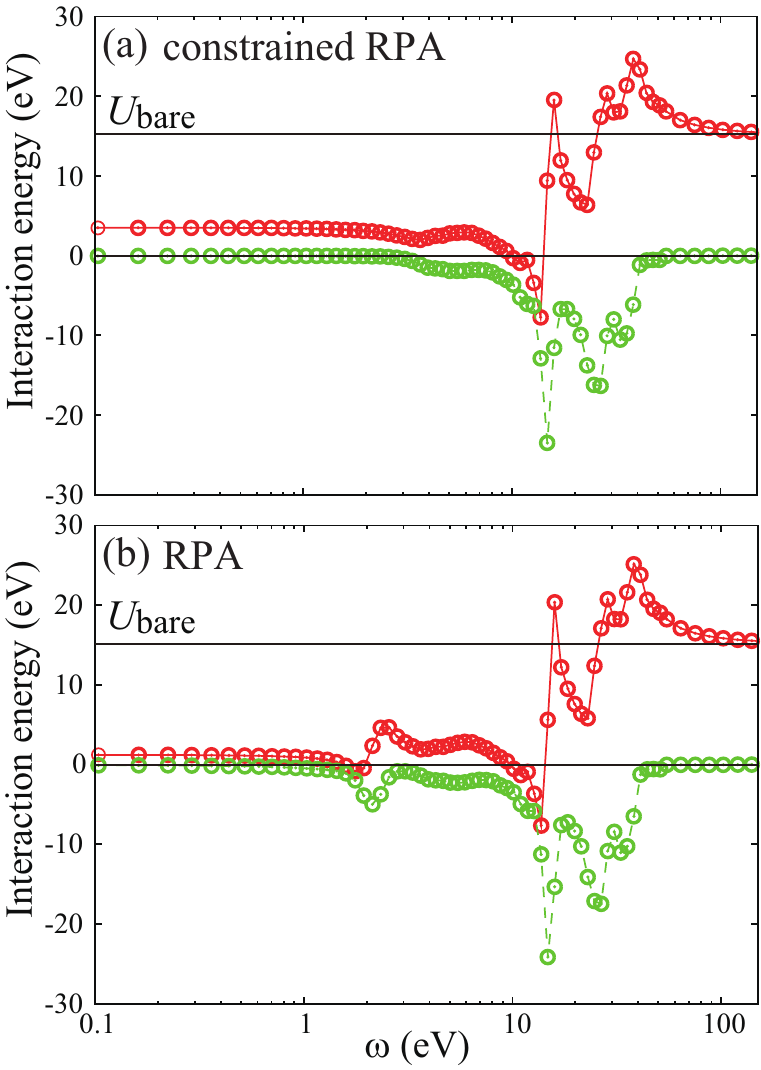}
\end{center} 
\vspace{-0.5cm} 
\caption{Frequency dependence of onsite direct-Coulomb integral of SrVO$_3$ [$U_{d_{xy}{\bf 0}, d_{xy}{\bf 0}}(\omega)$ in Eq.~(\ref{eq:matrix_U})], calculated with RESPACK-$calc\_int$ code. (a) constrained RPA and (b) usual RPA. Circles represent calculation values, and the red-solid and green-dashed curves describe the real and imaginary parts of the spectra, respectively.} 
\label{UvsE-SrVO3}
\end{figure}

Figure~\ref{UvsE-SrVO3} shows a frequency dependence of the onsite direct-Coulomb integral [$U_{1{\bf 0}1{\bf 0}}(\omega)$ in Eq.~(\ref{eq:matrix_U}) with orbital index 1 denoting $d_{xy}$-type orbital]. Panels (a) and (b) represent the constrained RPA and usual RPA results, respectively. Red-solid and green-dashed curves describe the real and imaginary parts, respectively. As well as the optical data, the difference between the constrained RPA and usual RPA occurs in the low-energy excitation region below 2 eV due to the low-energy plasmon excitation considered in the usual RPA calculation. 

\begin{figure}[htpb] 
\begin{center} 
\includegraphics[width=0.37\textwidth]{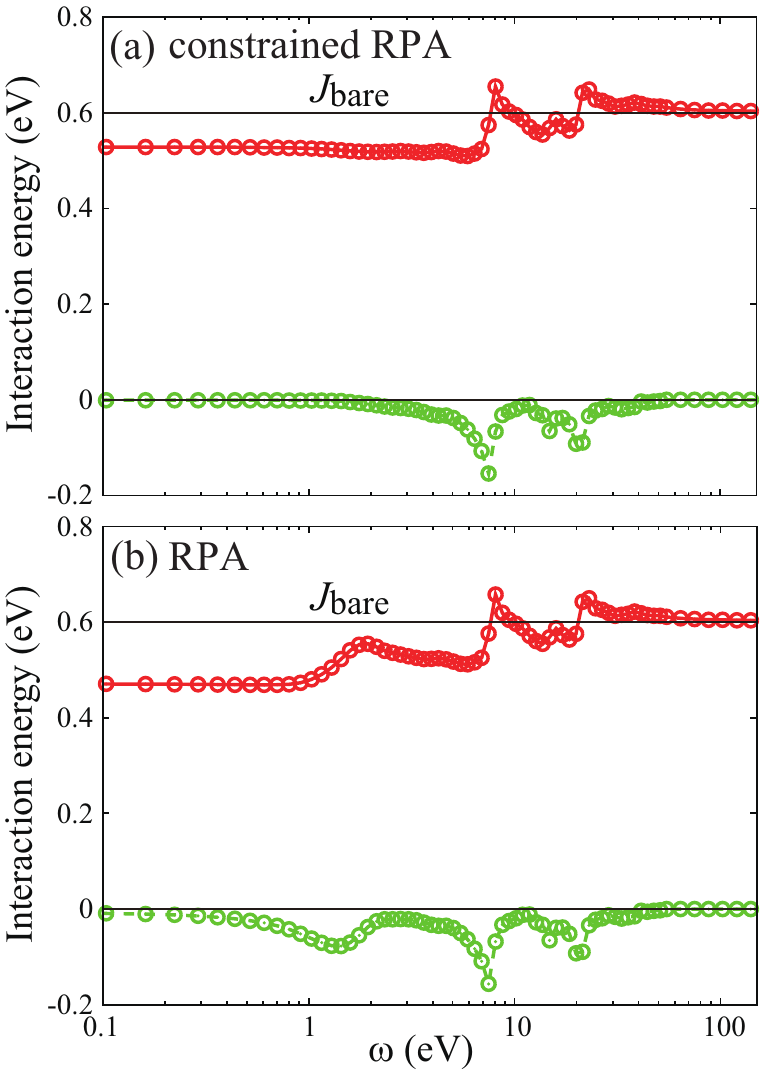}
\end{center} 
\vspace{-0.5cm} 
\caption{Frequency dependence of onsite exchange integral of SrVO$_3$ [$J_{d_{xy}{\bf 0}, d_{yz}{\bf 0}}(\omega)$ in Eq.~(\ref{eq:matrix_J})], calculated with RESPACK-$calc\_int$ code. (a) constrained RPA and (b) usual RPA. Circles represent calculation values, and the red-solid and green-dashed curves describe the real and imaginary parts of the spectra, respectively.} 
\label{JvsE-SrVO3}
\end{figure}

Figure ~\ref{JvsE-SrVO3} is a frequency dependence of the onsite exchange integral. [$J_{1{\bf 0}2{\bf 0}}(\omega)$ in Eq.~(\ref{eq:matrix_J}) with orbital indices 1 and 2 denoting $d_{xy}$-type and $d_{yz}$-type orbitals, respectively]. In exchange integral, differences between the constrained RPA (a) and usual RPA (b) can also be observed around $\omega\sim 2$ eV. 

\subsection{Convergence check to calculation conditions} \label{convergence}  
We next show a convergence behavior of the calculated interaction parameters of the $t_{2g}$ model of SrVO$_3$ with respect to the various computational conditions. Table~\ref{k-dependence} shows a convergence of the static interaction parameters with increasing the sampling $k$-point density, where we list the static onsite-intraorbital interaction $U$, onsite-interorbital $U'$, onsite-exchage $J$, and nearest-neighbor $V$ interactions averaged over orbitals. We see that the convergence is achieved around $8\times8\times8$.  
\begin{table}[htpb] 
\caption{Dependence of the static ($\omega$=0) interaction parameters of the $t_{2g}$ model of SrVO$_3$ on the sampling $k$-point density. The interaction parameters with the bare (unscreened), constrained RPA (cRPA), and usual RPA are listed. $V=(1/N_w)^2\sum_{i,j=1}^{N_w} U_{i {\bf 0}j {\bf R}}$ is the orbital-averaged value of the nearest-neighbor interactions with ${\bf R}=(1,0,0)$. Other conditions are as follows: The total number of the bands $N_{band}=50$, the cutoff energy for the polarization function $E_{cut}^{\epsilon}=10$ Ry, and the broadening factor introduced in the $chiqw$ calculation $\delta=0.1$ eV. The unit of the interaction parameter is eV.} 
\begin{center} 
\scalebox{0.72}{
\begin{tabular}{@{\!}c@{\ } 
r@{\ }r@{\ }r@{\ \ \ }
r@{\ }r@{\ }r@{\ \ \ }
r@{\ }r@{\ }r@{\ \ \ }
r@{\ }r@{\ }r} \hline \hline \\ [-10pt]
 & \multicolumn{3}{c}{$U$} 
 & \multicolumn{3}{c}{$U'$} 
 & \multicolumn{3}{c}{$J$} 
 & \multicolumn{3}{c}{$V$} \\ \hline \\ [-10pt] 
 $k^3$ 
 & bare & cRPA  & RPA    
 & bare  & cRPA & RPA 
 & bare  & cRPA & RPA 
 & bare  & cRPA & RPA \\ \hline \\ [-10pt] 
   $5^3$   & 15.12  & 3.37 & 1.25  & 13.83 & 2.26  & 0.27  
           &  0.60  & 0.53 & 0.48  &  3.56 & 0.65  & 0.01 \\ 
   $6^3$   & 15.16  & 3.38 & 1.30  & 13.87 & 2.26  & 0.30 
           & 0.60   & 0.53 & 0.48  &  3.57 & 0.65  & 0.01 \\ 
   $7^3$   & 15.20  & 3.42 & 1.30  & 13.91 & 2.30  & 0.31 
           & 0.60   & 0.53 & 0.48  & 3.59  & 0.67  & 0.02 \\ 
   $8^3$   & 15.22  & 3.50 & 1.21  & 13.93 & 2.38  & 0.24 
           & 0.60   & 0.53 & 0.47  &  3.60 & 0.71  & 0.02 \\ 
   $9^3$   & 15.23  & 3.47 & 1.24  & 13.94 & 2.35  & 0.26  
           & 0.60   & 0.53 & 0.47  &  3.61 & 0.70  & 0.02 \\ 
   $10^3$  & 15.24  & 3.46 & 1.26  & 13.96 & 2.35  & 0.28 
           &  0.60  & 0.53 & 0.48  &  3.62 & 0.70  & 0.02 \\ 
   $11^3$  & 15.26  & 3.47 & 1.26  & 13.97 & 2.35  & 0.28 
           &  0.60  & 0.53 & 0.48  &  3.63 & 0.70  & 0.02 \\ 
   $12^3$  & 15.27  & 3.49 & 1.25  & 13.98 & 2.37  & 0.27 
           &  0.60  & 0.53 & 0.48  &  3.64 & 0.71  & 0.02 \\ \hline \hline
\end{tabular} 
} 
\end{center} 
\label{k-dependence} 
\end{table}

Table~\ref{band-dependence} is a dependence of the static interaction parameters of SrVO$_3$ on the total number of bands. We see that the 50 bands is enough to obtain the converged results. 
\begin{table}[htpb] 
\caption{Dependence of the static ($\omega$=0) interaction parameters of SrVO$_3$ on the total number of calculated bands $N_{band}$. The view of the table is the same as Table~\ref{k-dependence}. Other conditions are as follows:  $8\times8\times8$ $k$-grid, $E_{cut}^{\epsilon}=10$ Ry, $\delta=0.1$ eV. The unit of the interaction parameter is eV.} 
\begin{center} 
\scalebox{0.85}{
\begin{tabular}{c 
r@{\ \ }r@{\ \ \ }
r@{\ \ }r@{\ \ \ }
r@{\ \ }r@{\ \ \ }
r@{\ \ }r} \hline \hline \\ [-10pt] 
 & \multicolumn{2}{c}{$U$} 
 & \multicolumn{2}{c}{$U'$} 
 & \multicolumn{2}{c}{$J$} 
 & \multicolumn{2}{c}{$V$} \\ \hline \\ [-10pt] 
 $N_{band}$ 
 & cRPA & RPA    
 & cRPA & RPA 
 & cRPA & RPA 
 & cRPA & RPA \\ \hline \\ [-10pt]
   $30$  & 3.59 & 1.22  & 2.45  & 0.24  
         & 0.53 & 0.47  & 0.73  & 0.02 \\ 
   $50$  & 3.50 & 1.21  & 2.38  & 0.24  
         & 0.53 & 0.47  & 0.71  & 0.02 \\ 
   $100$ & 3.47 & 1.21  & 2.36  & 0.24 
         & 0.53 & 0.47  & 0.70  & 0.02 \\ 
   $150$ & 3.46 & 1.21  & 2.36  & 0.24 
         & 0.52 & 0.47  & 0.70  & 0.02 \\ 
   $200$ & 3.46 & 1.21  & 2.36  & 0.24 
         & 0.52 & 0.47  & 0.70  & 0.02 \\ \hline \hline
\end{tabular} 
} 
\end{center} 
\label{band-dependence} 
\end{table}

Table~\ref{ecut-dependence} gives a convergence behavior of the static interaction parameters with respect to the cutoff energy $E_{cut}^{\epsilon}$ of the polarization function. This parameter is important for the convergence and it is desirable to take large enough. The convergence within 0.01 eV of the onsite constrained-RPA interaction parameters requires about 30 Ry, and in the case of the RPA parameter, it is about 40 Ry. However, the large $E_{cut}^{\epsilon}$ needs the large computational time because of the double-loop calculation on the {\bf G} and {\bf G}' vectors in Eq.~(\ref{eq:chi}). By default, $E_{cut}^{\epsilon}$ is set to  $E_{cut}^{\psi}/10$.   
\begin{table}[htpb] 
\caption{Dependence of the static ($\omega$=0) interaction parameters of SrVO$_3$ on the cutoff energy $E_{cut}^{\epsilon}$ of the polarization function. The view of the table is the same as Table~\ref{k-dependence}. Other conditions are as follows: $8\times8\times8$ $k$-grid, $N_{band}=50$, $\delta=0.1$ eV. The units are in Rydberg for $E_{cut}^{\epsilon}$ and eV for interaction parameters.} 
\begin{center} 
\scalebox{0.85}{
\begin{tabular}{c 
r@{\ \ }r@{\ \ \ }
r@{\ \ }r@{\ \ \ }
r@{\ \ }r@{\ \ \ }
r@{\ \ }r} \hline \hline \\ [-10pt] 
 & \multicolumn{2}{c}{$U$} 
 & \multicolumn{2}{c}{$U'$} 
 & \multicolumn{2}{c}{$J$} 
 & \multicolumn{2}{c}{$V$} \\ \hline \\ [-10pt] 
 $E_{cut}^{\epsilon}$ 
 & cRPA & RPA    
 & cRPA & RPA 
 & cRPA & RPA 
 & cRPA & RPA \\ \hline  \\ [-10pt] 
   $5$   & 3.64 & 1.47  & 2.45  & 0.32  
         & 0.57 & 0.56  & 0.70  & 0.02 \\ 
   $10$  & 3.50 & 1.21  & 2.38  & 0.24  
         & 0.53 & 0.47  & 0.71  & 0.02 \\ 
   $15$  & 3.50 & 1.13  & 2.40  & 0.26 
         & 0.50 & 0.41  & 0.71  & 0.02 \\ 
   $20$  & 3.46 & 1.04  & 2.39  & 0.25 
         & 0.49 & 0.37  & 0.71  & 0.02 \\ 
   $25$  & 3.44 & 0.98  & 2.38  & 0.24 
         & 0.48 & 0.35  & 0.71  & 0.02 \\ 
   $30$  & 3.43 & 0.94  & 2.38  & 0.23 
         & 0.48 & 0.33  & 0.71  & 0.02 \\ 
   $35$  & 3.42 & 0.91  & 2.37  & 0.23 
         & 0.48 & 0.32  & 0.71  & 0.02 \\        
   $40$  & 3.42 & 0.89  & 2.37  & 0.23 
         & 0.48 & 0.31  & 0.71  & 0.02 \\         
         
\hline \hline
\end{tabular} 
} 
\end{center} 
\label{ecut-dependence} 
\end{table}

Table~\ref{delta-dependence} is a dependence of the static onsite interaction parameter on the broadening factor $\delta$ introduced in the polarization-function calculation of Eqs.~(\ref{eq:chi}), (\ref{Xakqbk}), and (\ref{tetrahedron}). We see that the $\delta$ does not affect the static cRPA and RPA results. 
\begin{table}[htpb] 
\caption{Dependence of static ($\omega$=0) interaction parameters of SrVO$_3$ on the broadening factor $\delta$ in the polarization-function calculation in Eqs.~(\ref{eq:chi}), (\ref{Xakqbk}), and (\ref{tetrahedron}). The view of the table is the same as Table~\ref{k-dependence}. Other conditions are as follows: $8\times8\times8$ $k$-grid, $N_{band}=50$, $E_{cut}^{\epsilon}=10$ Ry. The units of $\delta$ and the interaction parameter are eV.} 
\begin{center} 
\scalebox{0.85}{
\begin{tabular}{c 
r@{\ \ }r@{\ \ \ }
r@{\ \ }r@{\ \ \ }
r@{\ \ }r@{\ \ \ }
r@{\ \ }r} \hline \hline \\ [-10pt] 
 & \multicolumn{2}{c}{$U$} 
 & \multicolumn{2}{c}{$U'$} 
 & \multicolumn{2}{c}{$J$} 
 & \multicolumn{2}{c}{$V$} \\ \hline \\ [-10pt] 
 $\delta$ 
 & cRPA & RPA    
 & cRPA & RPA 
 & cRPA & RPA 
 & cRPA & RPA \\ 
 \hline  \\ [-10pt] 
   $0.001$  & 3.50 & 1.19  & 2.38  & 0.22  
            & 0.53 & 0.47  & 0.71  & 0.02 \\ 
   $0.005$  & 3.50 & 1.19  & 2.38  & 0.22  
            & 0.53 & 0.47  & 0.71  & 0.02 \\ 
   $0.01$   & 3.50 & 1.19  & 2.38  & 0.23  
            & 0.53 & 0.47  & 0.71  & 0.02 \\ 
   $0.05$   & 3.50 & 1.20  & 2.38  & 0.23 
            & 0.53 & 0.47  & 0.71  & 0.02 \\ 
   $0.1$    & 3.50 & 1.21  & 2.38  & 0.24 
            & 0.53 & 0.47  & 0.71  & 0.02 \\ 
   $0.2$    & 3.50 & 1.23  & 2.38  & 0.25 
            & 0.53 & 0.47  & 0.71  & 0.02 \\      
\hline \hline
\end{tabular} 
} 
\end{center} 
\label{delta-dependence} 
\end{table}

\begin{figure}[htpb] 
\centering
\includegraphics[width=0.4\textwidth]{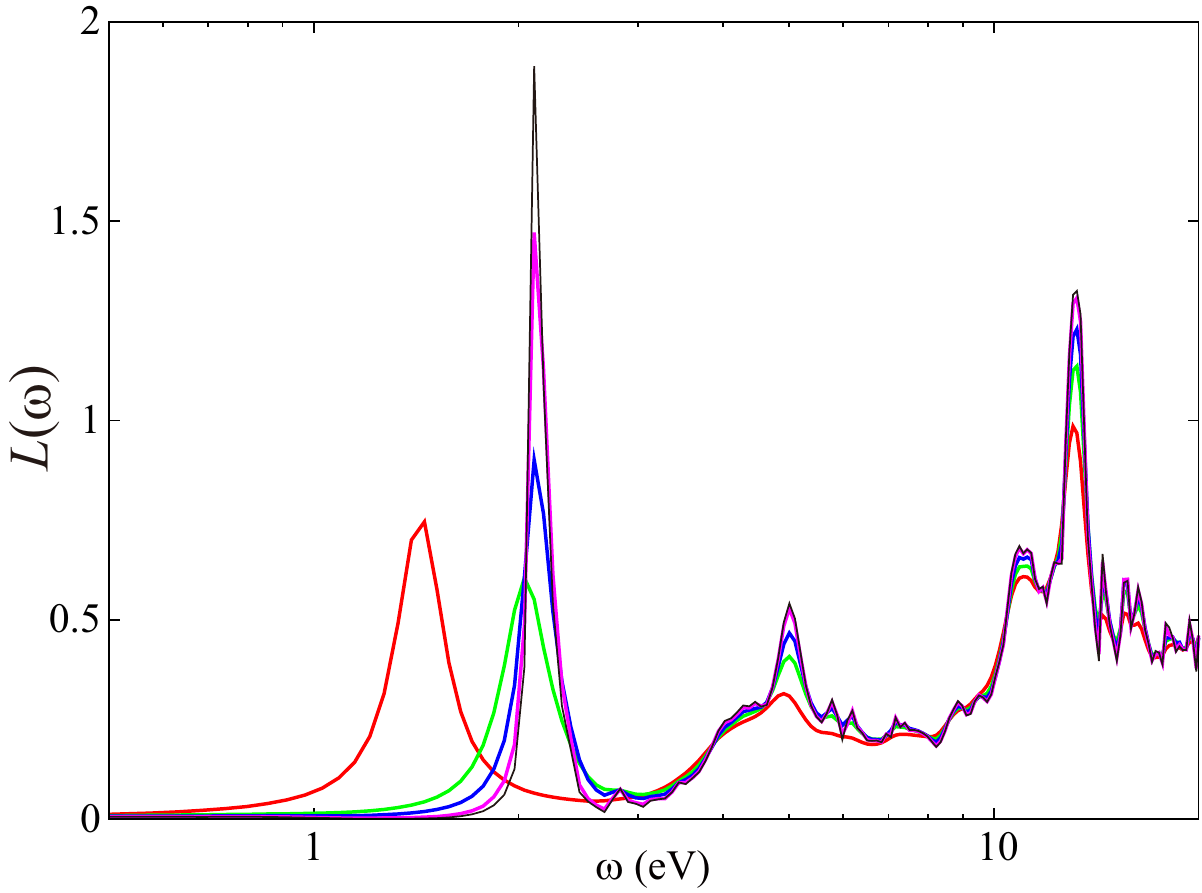}
\caption{Dependence of EELS function $L(\omega)$ in Eq.~(\ref{eq:eels}) of SrVO$_3$ on the broadening factor $\delta$ introduced in the polarization-function calculation [Eqs.~(\ref{eq:chi}), (\ref{Xakqbk}), and (\ref{tetrahedron})]. The calculation is based on the usual RPA. The spectra drawn by black, purple, blue, green, and red indicates the spectra with $\delta$ = 0.001 eV, 0.01 eV, 0.05 eV, 0.1 eV, and 0.2 eV, respectively. A spectral behavior around 1-2 eV results from the plasmon excitation within the $t_{2g}$ band. As the $\delta$ increases, the peak intensity decreases and the peak position shifts to the lower-energy side eventually.} 
\label{delta-dep-eels}
\end{figure}

We note that the $\delta$ may affect dynamical properties, especially in the low-energy collective excitation. Figure~\ref{delta-dep-eels} is the $\delta$ dependence of EELS function $L(\omega)$ in Eq.~(\ref{eq:eels}). As the $\delta$ value increases [0.001 eV (black curves), 0.01 eV (purple curves), 0.05 eV (blue curves), 0.1 eV (green curves), and 0.2 eV (red curves)], the intensity of the plasmon peak around 1-2 eV decreases and eventually the peak position shifts to the lower energy. The plasma excitation is also sensitive to the $k$-point density~\cite{Draxl_2006}. Therefore, one should be careful about computational conditions for the quantitative discussion of the dynamical properties.   

\subsection{Pseudopotential dependence} \label{pseudopotntial-dependence} 
Here, we discuss the pseudopotential dependence of the interaction parameters in details. The pseudopotential depends mainly on the cutoff radius $r_{loc}$ for local pseudopotential. The pseudopotential with a small $r_{loc}$ parameter makes deeper potential, and the resulting pseudo wavefunction tends to be more localized near the ion core. We constructed the five pseudopotentials with the different $r_{loc}$ values (0.8, 1.0, 1.5, 1.8 and 2.1 bohr) for a vanadium atom. The pseudopotential with $r_{loc}$ = 0.8, 1.0, and 1.5 bohr were constructed for an ionic semicore configuration of $(3s)^{2}(3p)^{6}(3d)^{3}$. The pseudopotentials with $r_{loc}$ = 1.8 and 2.1 bohr were constructed with an ionic valence configuration of $(3d)^{3}(4s)^{0}(4p)^{0}$. The Troullier-Martins (TM) type was adopted as a function form of the pseudo wavefunction~\cite{Troullier_1991}. $8\times8\times8$ $k$-point sampling and wavefunction cutoff $E_{cut}^{\psi}$ of 196 Ry are employed. The total number of bands $N_{band}$ is 50, and the broadening factor $\delta$ was set to be 0.1 eV. A polarization-function cutoff $E_{cut}^{\epsilon}$ is important for the effective interaction parameters, so the convergence behavior are discussed for this parameter. We note that the $t_{2g}$-band structures obtained with the above 5 pseudopotentials are in almost perfect agreements.  

\begin{figure}[htpb]  
\centering
\includegraphics[width=0.47\textwidth]{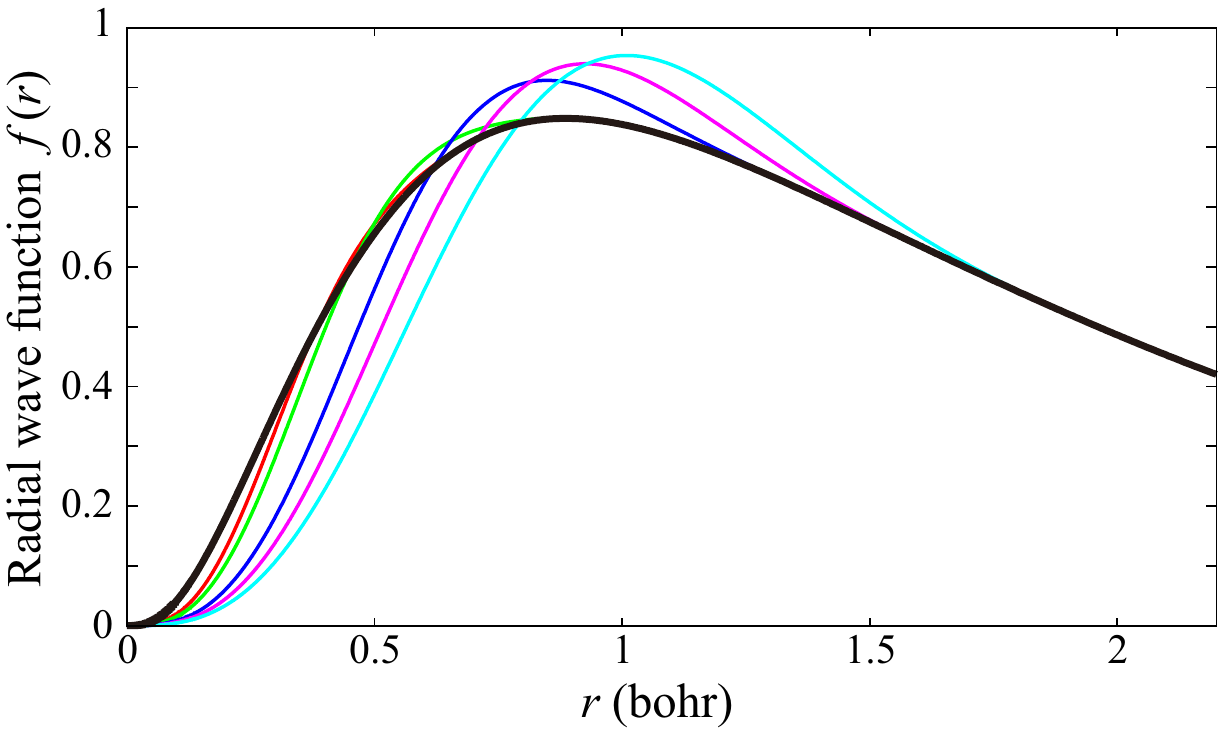}
\caption{Comparison of radial part of pseudo wavefunction and all-electron wavefunction (thick black curve) of the $3d$ orbital of a vanadium atom. The figure compares the pseudo wavefunctions obtained with different pseudopotentials, where a cutoff radius of the local pseudopotential $r_{loc}$ is changed. Red, green, blue, purple, and light-blue curves are the results with $r_{loc}$ = 0.8, 1.0, 1.5, 1.8, and 2.1 bohr, respectively. By reducing the $r_{loc}$ parameter, the maximum-amplitude position of the pseudo wavefunction is shifted to the ion-core side ($r=0$).} 
\label{pwn-vs-aewn}
\end{figure}

Figure~\ref{pwn-vs-aewn} compares the atomic pseudo wavefunctions obtained with the different $r_{loc}$ parameters with the atomic all-electron wavefunction (black curve). As the $r_{loc}$ parameter is reduced from 2.1 bohr (light blue) $\to$ 1.8 bohr (purple) $\to$ 1.5 bohr (blue) $\to$ 1.0 bohr (green) $\to$ 0.8 bohr (red), the maximum amplitude position of the pseudo wavefunction is shifted to the ion-core side. The pseudo wavefunctions with the $r_{loc}=0.8$ and 1.0 bohr are almost the same as the all-electron wave function. 

This trend can affect the localization of the Wannier function; the Wannier function generated with a pseudopotential with a small $r_{loc}$ cutoff tends to be more localized. As a result, it can give a large bare (unscreened) direct-Coulomb integral. Table~\ref{rc-vs-interaction-bare} shows the $r_{loc}$ dependence of the bare interaction parameters for the $t_{2g}$ Wannier function of SrVO$_3$. We see from the table that the $r_{loc}$ and the Wannier spread $S_{i{\bf 0}}$ in Eq.~(\ref{wannier-spread}) has clear positive correlation; a smaller $r_{loc}$ leads to a smaller Wannier spread. As a result, the $r_{loc}$ and bare onsite direct-Coulomb integrals $U_{bare}$ and $U'_{bare}$ correlate negatively (i.e., the small $r_{loc}$ brings about the large $U_{bare}$ and $U'_{bare}$). There are no discernible effects on the bare exchange $J_{bare}$ and the bare nearest-neighbor direct integrals $V_{bare}$. 
\begin{table}[htpb] 
\caption{Dependence of bare (unscreened) interaction parameters of SrVO$_3$ on pseudopotential, where we change the $r_{loc}$ parameter which is a cutoff radius for the local pseudopotential. In this table, we also list the Wannier spread in Eq.~(\ref{wannier-spread}). Note that the pseudopotentials with $r_{loc}$ = 0.8, 1.0, and 1.5 bohr are constructed for the semicore configuration and those with $r_{loc}$ = 1.8 and 2.1 bohr are constructed for the valence configuration (see the text). Calculation condition is $8\times8\times8$ $k$-point sampling and 196-Ry wavefunction cutoff.  The unit of the interaction parameter is eV.} 
\begin{center} 
\scalebox{0.9}{
\begin{tabular}{c@{\ \ \ }c@{\ \ \ }c@{\ \ \ }c@{\ \ \ }c@{\ \ }c} \hline \hline \\ [-10pt] 
 $r_{loc}$ [bohr] & $S$ [\AA$^2$] & $U_{bare}$ & $U_{bare}'$ & $J_{bare}$ & $V_{bare}$ \\ 
 \hline \\ [-10pt] 
 0.8 & 1.88490 & 16.13 & 14.92 &  0.57 & 3.60  \\ 
 1.0 & 1.88735 & 16.13 & 14.91 &  0.58 & 3.60  \\ 
 1.5 & 1.89448 & 15.91 & 14.65 &  0.60 & 3.60  \\ 
 1.8 & 1.96969 & 15.59 & 14.31 &  0.60 & 3.60  \\ 
 2.1 & 1.97853 & 15.21 & 13.92 &  0.60 & 3.60  \\  
\hline \hline
\end{tabular} 
} 
\end{center} 
\label{rc-vs-interaction-bare} 
\end{table} 

Tables~\ref{rc-vs-interaction-rpa} is the pseudopotential dependence of constrained-RPA and usual-RPA interaction parameters. An interesting trend can be seen in the table; details of the pseudopotential hardly affect the interaction values in contrast to the bare (unscreened) interaction parameters. In general, however, sufficiently large $E_{cut}^{\epsilon}$ would be desirable for a safer quantitative discussion about effective interaction parameters, especially for the usual RPA case. 
\begin{table}[htpb] 
\caption{Dependence of static ($\omega$=0) interaction parameters of SrVO$_3$ on pseudopotential, where we change the $r_{loc}$ parameter which is a cutoff radius for the local pseudopotential. Note that the pseudopotentials with $r_{loc}$ = 0.8, 1.0, and 1.5 bohr are constructed for the semicore configuration and those with $r_{loc}$ = 1.8 and 2.1 bohr are constructed for the valence configuration (see the text). Under each pseudopotential, the $E_{cut}^{\epsilon}$ cutoff dependence on the interaction parameters is investigated. Other calculation condition is as follows: $8\times8\times8$ $k$-point sampling, $N_{band}=50$, $E_{cut}^{\psi}=196$ Ry, and $\delta=0.1$ eV. The unit of the interaction parameter is eV.} 
\begin{center} 
\scalebox{0.75}{
\begin{tabular}{c@{\ \ \ }c@{\ \ }c@{\ \ \ }c@{\ \ }c@{\ \ \ }c@{\ \ }c@{\ \ \ }c@{\ \ }c}
\hline \hline \\ [-10pt] 
  $r_{loc}=0.8$ bohr 
   & \multicolumn{2}{c}{$U$} 
   & \multicolumn{2}{c}{$U'$} 
   & \multicolumn{2}{c}{$J$} 
   & \multicolumn{2}{c}{$V$} \\ \hline \\ [-10pt] 
 $E_{cut}^{\epsilon}$ [Ry]
 & cRPA & RPA 
 & cRPA & RPA 
 & cRPA & RPA 
 & cRPA & RPA \\ \hline \\ [-10pt]
  5 & 3.85 & 1.62 & 2.73 & 0.52 & 0.54 & 0.53 & 0.66 & 0.02 \\
 10 & 3.53 & 1.19 & 2.46 & 0.23 & 0.52 & 0.47 & 0.67 & 0.02 \\
 15 & 3.51 & 1.09 & 2.46 & 0.22 & 0.50 & 0.42 & 0.68 & 0.02 \\
 20 & 3.49 & 1.03 & 2.47 & 0.23 & 0.48 & 0.38 & 0.68 & 0.02 \\
 25 & 3.48 & 0.99 & 2.48 & 0.24 & 0.47 & 0.36 & 0.68 & 0.02 \\ 
 30 & 3.47 & 0.96 & 2.48 & 0.24 & 0.47 & 0.34 & 0.68 & 0.02 \\ 
 35 & 3.47 & 0.94 & 2.48 & 0.25 & 0.47 & 0.33 & 0.68 & 0.02 \\
 40 & 3.46 & 0.92 & 2.48 & 0.25 & 0.46 & 0.32 & 0.68 & 0.02 \\
\hline \hline \\ [-10pt] 
   $r_{loc}=1.0$ bohr
   & \multicolumn{2}{c}{$U$} 
   & \multicolumn{2}{c}{$U'$} 
   & \multicolumn{2}{c}{$J$} 
   & \multicolumn{2}{c}{$V$} \\ \hline \\ [-10pt] 
 $E_{cut}^{\epsilon}$ [Ry]
 & cRPA & RPA 
 & cRPA & RPA 
 & cRPA & RPA 
 & cRPA & RPA \\ \hline \\ [-10pt]
  5 & 3.84 & 1.62 & 2.71 & 0.52 & 0.55 & 0.54 & 0.66 & 0.02 \\ 
 10 & 3.53 & 1.20 & 2.45 & 0.23 & 0.52 & 0.47 & 0.67 & 0.02 \\ 
 15 & 3.50 & 1.10 & 2.45 & 0.22 & 0.50 & 0.42 & 0.68 & 0.02 \\ 
 20 & 3.48 & 1.04 & 2.46 & 0.24 & 0.49 & 0.39 & 0.68 & 0.02 \\  
 25 & 3.47 & 1.00 & 2.46 & 0.24 & 0.48 & 0.36 & 0.68 & 0.02 \\   
 30 & 3.47 & 0.96 & 2.47 & 0.25 & 0.47 & 0.35 & 0.68 & 0.02 \\ 
 35 & 3.46 & 0.94 & 2.47 & 0.25 & 0.47 & 0.33 & 0.68 & 0.02 \\
 40 & 3.45 & 0.92 & 2.47 & 0.25 & 0.47 & 0.32 & 0.68 & 0.02 \\ 
\hline \hline \\ [-10pt] 
  $r_{loc}=1.5$ bohr
  & \multicolumn{2}{c}{$U$} 
  & \multicolumn{2}{c}{$U'$} 
  & \multicolumn{2}{c}{$J$} 
  & \multicolumn{2}{c}{$V$} \\ \hline \\ [-10pt] 
 $E_{cut}^{\epsilon}$ [Ry]
 & cRPA & RPA 
 & cRPA & RPA 
 & cRPA & RPA 
 & cRPA & RPA \\ \hline \\ [-10pt] 
  5 & 3.80 & 1.59 & 2.62 & 0.44 & 0.57 & 0.56 & 0.66 & 0.02 \\ 
 10 & 3.55 & 1.22 & 2.43 & 0.22 & 0.54 & 0.49 & 0.67 & 0.02 \\ 
 15 & 3.54 & 1.13 & 2.45 & 0.24 & 0.51 & 0.43 & 0.68 & 0.02 \\ 
 20 & 3.51 & 1.07 & 2.46 & 0.25 & 0.50 & 0.39 & 0.68 & 0.02 \\ 
 25 & 3.50 & 1.01 & 2.46 & 0.25 & 0.49 & 0.37 & 0.68 & 0.02 \\   
 30 & 3.49 & 0.98 & 2.46 & 0.25 & 0.49 & 0.35 & 0.68 & 0.02 \\ 
 35 & 3.48 & 0.95 & 2.46 & 0.25 & 0.48 & 0.34 & 0.68 & 0.02 \\ 
 40 & 3.47 & 0.92 & 2.45 & 0.24 & 0.48 & 0.33 & 0.68 & 0.02 \\ 
\hline \hline \\ [-10pt] 
 $r_{loc}=1.8$ bohr
 & \multicolumn{2}{c}{$U$} 
 & \multicolumn{2}{c}{$U'$} 
 & \multicolumn{2}{c}{$J$} 
 & \multicolumn{2}{c}{$V$} \\ \hline \\ [-10pt] 
 $E_{cut}^{\epsilon}$ [Ry]
 & cRPA & RPA 
 & cRPA & RPA 
 & cRPA & RPA 
 & cRPA & RPA \\ \hline \\ [-10pt] 
  5 & 3.68 & 1.53 & 2.49 & 0.38 & 0.57 & 0.56 & 0.69 & 0.02 \\ 
 10 & 3.47 & 1.22 & 2.35 & 0.22 & 0.53 & 0.47 & 0.70 & 0.02 \\ 
 15 & 3.47 & 1.14 & 2.37 & 0.25 & 0.51 & 0.43 & 0.70 & 0.02 \\ 
 20 & 3.44 & 1.06 & 2.37 & 0.25 & 0.49 & 0.38 & 0.70 & 0.02 \\ 
 25 & 3.42 & 1.00 & 2.36 & 0.24 & 0.49 & 0.36 & 0.70 & 0.02 \\  
 30 & 3.41 & 0.96 & 2.35 & 0.24 & 0.48 & 0.34 & 0.70 & 0.02 \\
 35 & 3.40 & 0.93 & 2.35 & 0.23 & 0.48 & 0.33 & 0.70 & 0.02 \\ 
 40 & 3.39 & 0.91 & 2.35 & 0.23 & 0.48 & 0.32 & 0.70 & 0.02 \\ 
\hline \hline \\ [-10pt] 
  $r_{loc}=2.1$ bohr
  & \multicolumn{2}{c}{$U$} 
  & \multicolumn{2}{c}{$U'$} 
  & \multicolumn{2}{c}{$J$} 
  & \multicolumn{2}{c}{$V$} \\ \hline \\ [-10pt] 
 $E_{cut}^{\epsilon}$ [Ry]
 & cRPA & RPA 
 & cRPA & RPA 
 & cRPA & RPA 
 & cRPA & RPA \\ \hline \\ [-10pt] 
  5 & 3.64 & 1.47 & 2.45 & 0.32 & 0.57 & 0.56 & 0.70 & 0.02 \\ 
 10 & 3.50 & 1.21 & 2.38 & 0.24 & 0.53 & 0.47 & 0.71 & 0.02 \\
 15 & 3.49 & 1.13 & 2.40 & 0.26 & 0.51 & 0.41 & 0.71 & 0.02 \\ 
 20 & 3.46 & 1.04 & 2.38 & 0.25 & 0.49 & 0.37 & 0.71 & 0.02 \\  
 25 & 3.44 & 0.98 & 2.37 & 0.24 & 0.49 & 0.35 & 0.71 & 0.02 \\  
 30 & 3.43 & 0.94 & 2.37 & 0.23 & 0.49 & 0.33 & 0.71 & 0.02 \\
 35 & 3.42 & 0.91 & 2.37 & 0.22 & 0.49 & 0.32 & 0.71 & 0.02 \\ 
 40 & 3.42 & 0.90 & 2.36 & 0.23 & 0.48 & 0.31 & 0.71 & 0.02 \\ 
\hline \hline
\end{tabular} 
} 
\end{center} 
\label{rc-vs-interaction-rpa} 
\end{table}

Finally, we mention the effects of the pseudopotential type. As the famous pseudopotential types, besides the TM type, there are the ONCV (Optimized Norm-Conserving Vanderbilt) type~\cite{Hamann_2013} and the RRKJ (Rappe-Rabe-Kaxiras-Joannopoulos) type~\cite{Rappe_1990}. Even if the same cutoff $r_{loc}$ is employed, the results can be quantitatively different due to the difference in the functional form of the pseudopotential, so the user should be careful about this point. Table~\ref{xTAPP-vs-QE} compares the calculated interaction parameters based on the TM-type, ONCV-type~\cite{about-ONCV-QE} and RRKJ-type~\cite{about-RRKJ-QE} pseudopotentials. The calculations with the TM-type pseudopotential were performed with xTAPP, which is referred to as TM-xTAPP. The calculations with the ONCV-type and RRKJ-type pseudopotentials were performed with {\sc Quantum Espresso}, which are referred to as ONCV-QE and RRKJ-QE, respectively. For the TM-xTAPP, we give two results TM(v)-xTAPP and TM(s)-xTAPP, for which the former is the results based on the pseudopotential constructed with the valence-electron configuration, and the latter is the results based on the pseudopotential with the semicore configuration for V and Sr. The ONCV and RRKJ pseudopotentials are also constructed for the semicore configurations for V and Sr. We found that, for the screened direct-Coulomb interaction $U$ and $U'$, RRKJ-QE gives significantly smaller values than others.  
\begin{table}[htpb] 
\caption{Comparison of the static ($\omega$=0) interaction parameters among the pseudopotential type, where the results based on the TM-type, ONCV-type, and RRKJ-type pseudopotentials are compared. The calculation results with the TM-type pseudopotential are obtained via xTAPP, which are denoted by TM-xTAPP, while the results with the ONCV and RRKJ pseudopotential are obtained via {\sc Quantum Espresso}, which are abbreviated by ONCV-QE and RRKJ-QE, respectively. The symbol in parentheses attached to TM, i.e., (v) or (s), indicates whether the pseudopotential was made for the valence or semicore configuration, respectively. The calculation condition is as follows: $8\times8\times8$ $k$-point sampling, $N_{band}=50$, $E_{cut}^{\psi}=100$ Ry, $E_{cut}^{\epsilon}=20$ Ry, and $\delta=0.1$ eV. The unit of the interaction parameter is eV.
} 
\begin{center} 
\scalebox{0.8}{
\begin{tabular}{c@{\ \ \ }c@{\ \ \ }c@{\ \ \ }c@{\ \ \ }c@{\ \ \ }c@{\ \ }c} \hline \hline \\ [-10pt] 
      &      & TM(v)-xTAPP & TM(s)-xTAPP & ONCV-QE & RRKJ-QE  \\  \hline \\ [-10pt] 
      & bare & 15.22  & 15.93 &  15.73  &  15.72 \\ 
 $U$  & cRPA & 3.46   &  3.50 &  3.50   &   3.19 \\ 
      &  RPA & 1.04   &  1.05 &  1.06  &   0.99 \\ 
      \hline \\ [-10pt] 
      & bare & 13.93  & 14.71 & 14.44   &  14.35 \\ 
 $U'$ & cRPA & 2.39   &  2.48 &  2.43  &   2.08 \\ 
      &  RPA & 0.25   &  0.25 &  0.25  &   0.20 \\ 
      \hline \\ [-10pt] 
      & bare & 0.60   & 0.57  & 0.61   &   0.64 \\ 
 $J$  & cRPA & 0.49   & 0.48  & 0.51   &   0.52 \\ 
      &  RPA & 0.37   & 0.38  & 0.39   &   0.38 \\ 
      \hline \\ [-10pt]       
      & bare & 3.60   & 3.60  & 3.60   &  3.60  \\ 
 $V$  & cRPA & 0.71   & 0.66  & 0.65   &  0.56  \\ 
      &  RPA & 0.02   & 0.02  & 0.02   &  0.02  \\       
\hline \hline
\end{tabular} 
} 
\end{center} 
\label{xTAPP-vs-QE} 
\end{table} 

\section{Conclusion}\label{sec_conclusion} 
In conclusion, we present a new software RESPACK for deriving effective low-energy models of materials from first principles. The software contains programs for computing maximally-localized Wannier functions, RPA response functions, and matrix elements for screened interaction with respect to the Wannier functions. RESPACK is freely available under the GNU General Public Licence~\cite{GPL_URL}. RESPACK is written in {\sc Fortran90} and supports plane-wave DFT codes xTAPP~\cite{xTAPP_URL} and {\sc Quantum Espresso}~\cite{QE-2017,QE-2009} , for which an interface script that converts the band-calculation results to the RESPACK inputs is provided. As an important notice, RESPACK currently supports {\it ab initio} codes with the norm-conserving pseudopotential. 
The present paper focuses on the derivation of the effective model, but it has currently been extended to the GW calculation, effective-model derivation with spin-orbit interaction, and electron-lattice coupling evaluation. We will report on these additional features in  near future. 

\section*{Acknowledgments}
We thank Yoshiro Nohara for providing a module for generalized tetrahedron calculation. We also acknowledge Maxime Charlebois and Jean-Baptiste Mor\'{e}e for useful discussions about the code development. We thank Masatoshi Imada, Ryotaro Arita, and Takashi Miyake for the helpful discussions for the development of the RESPACK program and applications to various materials. A part of RESPACK is developed under the support of "Project for advancement of software usability in materials science" in fiscal year 2018 by the Institute for Solid State Physics, The University of Tokyo. In connection with this project, we acknowledge Taisuke Ozaki for his kind support and useful discussions during this project. We acknowledge the financial support of JSPS Kakenhi Grant No. 16H06345 (YN, TT, YY, TM, and KN), No. 17K14336 (YN), No. 18H01158 (YN), 16K17746 (TM), No. 19K03739 (TM), No. 16K05452 (KN), No. 17H03393 (KN), No. 17H03379 (KN), and No. 19K03673 (KN). MK and TM were supported by Priority Issue (creation of new functional devices and high-performance materials to support next-generation industries) to be tackled by using Post `K' Computer from the MEXT of Japan. KY, TM, and YM were supported by Building of Consortia for the Development of Human Resources in Science and Technology from MEXT of Japan. 

\appendix
\section{Input files}\label{sec_input}
In this appendix, we describe details of the input file {\tt input.in} for specifying the calculation condition of RESPACK. The {\tt input.in} is given in the {\sc Fortran} namelist format. Table~\ref{namelist} lists six namelists available in {\tt input.in}. The namelist {\tt \&param\_wannier} contains variables for the Wannier-function calculation. The namelists {\tt \&param\_interpolation} and {\tt \&param\_visualization} describe variables for the Wannier-interpolation band and the visualization of the realspace Wannier functions, respectively. These three namelists are read by a common executable file {\tt calc\_wannier}. This executable file is generated by compilation of source code in the directory {\tt RESPACK/src/wannier}. The namelist {\tt \&param\_chiqw} contains parameters for calculations of polarization and dielectric functions and are read by an executable file {\tt calc\_chiqw} which is created by compilation of source code in {\tt RESPACK/src/chiqw}. The namelist {\tt \&param\_calc\_int} describes variables for direct-Coulomb and exchange-integral calculation and are read by executable files {\tt calc\_w3d} and {\tt calc\_j3d}. These are made by compilation in the source code in {\tt RESPACK/src/calc\_int}. 
\begin{table}[htpb] 
\caption{Namelists available in an input file ({\tt input.in}) of RESPACK. The second column gives an executable file to which each namelist relates. The third column is a brief explanation of each namelist.} 
\vspace{-0.3cm} 
\begin{center} 
\scalebox{0.7}{
\begin{tabular}{l@{\ \ \ }l@{\ \ \ }l} \hline \hline  \\ [-10pt] 
  namelist                    & executable file     & explanation  \\  \hline \\ [-10pt] 
 {\tt \&param\_wannier}       & {\tt calc\_wannier} & Wannier-function calculation \\   
 {\tt \&param\_interpolation} & {\tt calc\_wannier} & Wannier-interpolated band calculation \\ 
 {\tt \&param\_visualization} & {\tt calc\_wannier} & visualization of Wannier function \\ 
 {\tt \&param\_chiqw}         & {\tt calc\_chiqw}   & polarization and dielectric calculations \\ 
 {\tt \&param\_calc\_int}     & {\tt calc\_w3d}     & direct-Coulomb integral calculation \\ 
 {\tt \&param\_calc\_int}     & {\tt calc\_j3d}     & exchange integral calculation \\ 
 \hline \hline
\end{tabular} 
} 
\end{center}
\label{namelist} 
\end{table} 

\subsection{{\tt \&param\_wannier}} \label{param-wannier}
Table~\ref{param_wannier} shows main four variables in the namelist {\tt \&param\_wannier}. {\tt N\_wannier} is the total number of the Wannier function to be calculated. {\tt Lower\_energy\_window} and {\tt Upper\_energy\_window} specify the energy window and the Bloch bands in the energy window are used for the Wannier-function calculation. The user has to set proper values by seeing the band dispersion obtained with DFT band calculations. {\tt N\_initial\_guess} is the number of initial Gaussian for the Wannier-function calculation. 
The user must enter initial-guess information just below this namelist (see Table~\ref{input-SrVO3}). {\tt vec\_ini(1:N\_initial\_guess)} in Table~\ref{param_wannier} are variables to specify the initial guess and 
is treated as {\it type} in {\sc Fortran}. {\tt vec\_ini(}$i${\tt )\%orb} specifies the orbital type of the $i$th initial guess, where {\tt \%orb} can take a character of {\tt s, px, py, pz, dxy, dyz, dzx, dx2, dz2}. The initial guess is treated with the Gaussian function, $\exp(-\alpha_i|{\bf r}-{\bf R}_i|^2$), where $\alpha_i$={\tt \%a} is orbital exponent of the $i$th Gaussian. The Gaussian position ${\bf R}_i=x{\bf a}_1+y{\bf a}_2+z{\bf a}_3$ with $x$={\tt \%x}, $y$={\tt \%y}, and $z$={\tt \%z} are given in the fractional coordinates, where ${\bf a}_1$, ${\bf a}_2$, and ${\bf a}_3$ are the lattice vectors in the calculation cell. Note that one has to enter {\tt N\_initial\_guess} pieces of the {\tt vec\_init} information. Specific descriptions are given in Table~\ref{input-SrVO3}.   
\begin{table}[htpb] 
\caption{Main variables in namelist {\tt \&param\_wannier} to construct maximally-localized Wannier functions. The variable type and its brief explanation are shown in the second and third columns, respectively. The lower five variables in {\tt vec\_ini} are information about initial guesses. Note that {\tt vec\_ini} is treated as {\it type} of {\sc Fortran}. The initial guess is specified by the orbital type ``{\tt s, px, py, pz, dxy, dyz, dzx, dx2, dz2}" and parameters in the Gaussian function, $\exp(-\alpha_i|{\bf r}-{\bf R}_i|^2$). Here, $\alpha_i$ is an orbital exponent of the $i$th Gaussian, and ${\bf R}_i=x{\bf a}_1+y{\bf a}_2+z{\bf a}_3$ is the Gaussian position in the fractional coordinates, where ${\bf a}_1$, ${\bf a}_2$, and ${\bf a}_3$ are the lattice vectors in the calculation cell.}  
\begin{center} 
\scalebox{0.76}{
\begin{tabular}{l@{\ \ \ }l@{\ \ \ }l} \hline \hline  \\ [-10pt] 
 \multicolumn{3}{l}{namelist {\tt \&param\_wannier}}  \\ \hline \\ [-10pt] 
  variable                   & type    & explanation   \\ \hline \\ [-10pt] 
 {\tt N\_wannier}            & integer & number of the Wannier functions \\   
 {\tt Lower\_energy\_window} & real    & lower limit of energy window in eV \\ 
 {\tt Upper\_energy\_window} & real    & upper limit of energy window in eV \\ 
 {\tt N\_initial\_guess}     & integer & number of initial Gaussians  \\  \hline \\ [-10pt] 
 {\tt vec\_ini(}$i${\tt )\%orb} & character & orbital type of $i$th Gaussian \\   
 {\tt vec\_ini(}$i${\tt )\%a}   & real      & orbital exponent of $i$th Gaussian \\ 
 {\tt vec\_ini(}$i${\tt )\%x}   & real      & $x$ component of $i$th Gaussian position \\ 
 {\tt vec\_ini(}$i${\tt )\%y}   & real      & $y$ component of $i$th Gaussian position \\ 
 {\tt vec\_ini(}$i${\tt )\%z}   & real      & $z$ component of $i$th Gaussian position \\  
\hline \hline 
\end{tabular}
} 
\end{center}
\label{param_wannier} 
\end{table} 

\subsection{{\tt \&param\_interpolation}} \label{param-interpolation}
Tables~\ref{interpolation1} summarize main variables in the namelist {\tt \&param\_interpolation} to draw the Wannier-interpolated band. The user specifies the total number of symmetric $k$ points in the dispersion line as {\tt N\_sym\_points}, and, just below this namelist, describes the coordinates \{($s_1, s_2, s_3$)\} of the symmetric $k$ vectors (see the Table~\ref{input-SrVO3} for concrete descriptions). Here, ${\bf k}_i=s_1 {\bf b}_1+s_2{\bf b}_2+s_3{\bf b}_3$ with ${\bf b}_1$, ${\bf b}_2$, and ${\bf b}_3$ being the basic reciprocal lattice vectors. These variables are read as {\it array} {\tt SK\_sym\_pts(1:3,1:N\_sym\_points)}. Usually, the same $k$ points as those employed in the DFT band calculation are adopted. 
\begin{table}[htpb] 
\caption{Main variables in namelist {\tt \&param\_interpolation} to draw the Wannier-interpolated band. The type and brief explanation of the variable are shown in the second and third columns, respectively. Lower three variables specify the symmetric $k$ points \{${\bf k}_i$\} in the dispersion line. These \{${\bf k}_i$\} information are treated as {\it array} {\tt SK\_sym\_pts(1:3,1:N\_sym\_points)} in {\sc Fortran}, where ${\bf k}_i=s_1 {\bf b}_1+s_2{\bf b}_2+s_3{\bf b}_3$ is the $i$th symmetric $k$ point with ${\bf b}_1$, ${\bf b}_2$, and ${\bf b}_3$ being the basic reciprocal lattice vectors. .}   
\begin{center} 
\scalebox{0.8}{
\begin{tabular}{l@{\ \ \ }l@{\ \ \ }l} \hline \hline  \\ [-10pt] 
 \multicolumn{3}{l}{namelist {\tt \&param\_interpolation}}  \\ \hline \\ [-10pt] 
  variable         & type    & explanation   \\  \hline \\ [-10pt] 
 {\tt N\_sym\_points}  & integer & number of symmetric $k$ points \\ \hline \\ [-10pt] 
 {\tt SK\_sym\_pts(}$1,i${\tt )} & real & $s1$ component of $i$th symmetric $k$ point \\   
 {\tt SK\_sym\_pts(}$2,i${\tt )} & real & $s2$ component of $i$th symmetric $k$ point \\   
 {\tt SK\_sym\_pts(}$3,i${\tt )} & real & $s3$ component of $i$th symmetric $k$ point \\ 
 \hline \hline
\end{tabular}
}
\end{center}
\label{interpolation1} 
\end{table} 

\subsection{{\tt \&param\_visualization}} \label{param-visalization} 
We show in Table~\ref{visalization} parameters in the namelist {\tt \&param\_visualization}. By default, RESPACK skips the visualization calculation of the realspace Wannier function. So, the user must set {\tt Flg\_vis\_wannier} to 1 when calculating visualization data. Structural data of the same spatial range as the Wannier-function data is also output as cif (Crystallographic Information File) format. All the results can be drawn by software VESTA~\cite{Momma_2011}. 
\begin{table}[htpb] 
\caption{Main parameter in namelist {\tt \&param\_visualization} to calculate realspace Wannier functions. When the variable {\tt Flg\_vis\_wannier} is set to 1, RESPACK calculates the visualization data of the Wannier function. Outputs can be drawn by software VESTA~\cite{Momma_2011}. } 
\begin{center} 
\scalebox{0.7}{
\begin{tabular}{l@{\ \ \ }l@{\ \ \ }l} \hline \hline  \\ [-10pt] 
 \multicolumn{3}{l}{namelist {\tt \&param\_visualization}}  \\ \hline \\ [-10pt] 
  variable               & type    & explanation   \\  \hline \\ [-10pt] 
 {\tt Flg\_vis\_wannier} & integer & visualization of Wannier function (do not: 0, do: 1) \\ 
 \hline \hline
\end{tabular} 
}
\end{center}
\label{visalization} 
\end{table} 

\subsection{{\tt \&param\_chiqw}} \label{param-chiqw}
Next, we list in Table~\ref{param_chiqw} main parameters of the namelist {\tt \&param\_chiqw}. In this namelist, the user specifies the calculation condition of the polarization and dielectric functions. {\tt Ecut\_for\_eps} is the cutoff energy for the polarization function. Default value is set to 1/10 of wave-function cutoff $E_{cut}^{\psi}$. {\tt Num\_freq\_grid} is the total number of the frequency grid. By default, a log grid of 70 points is generated. The maximum excitation energy $E_{max}$ is estimated from the total bands considered in the polarization function, and the range of the frequency grid is set to $3E_{max}$ to consider the Lorentz tail of the polarization function. See Section~\ref{sec_frequency_grid} for details. {\tt Green\_func\_delt} is smearing value used in the generalized tetrahedron calculation (0.1 eV by default). In RESPACK, the polarization calculation is performed in parallel. To this end, two variables {\tt MPI\_num\_qcomm} and {\tt MPI\_num\_proc\_per\_qcomm} are prepared. {\tt MPI\_num\_qcomm} is a variable for parallel computation on the $q$ points and defines the number of $q$-community. {\tt MPI\_num\_proc\_per\_qcomm} specifies the total number of MPI processes in each $q$-community. Note that when {\tt mod(N\_MPI,MPI\_num\_qcomm)/=0}, the {\it chiqw} program stops, where {\tt N\_MPI} is the total number of the MPI processes and this value is set in the job script. By default, {\tt MPI\_num\_qcomm=1} and {\tt MPI\_num\_proc\_per\_qcomm=N\_MPI}. Details of the parallel calculation can be found in Section.~\ref{sec_parallel_calculation}. Lastly, {\tt Flg\_cRPA} is a variable to specify the constrained RPA calculation.  In the default setting, {\tt Flg\_cRPA=0} and the program performs a usual RPA calculation. 
\begin{table}[htpb] 
\caption{Main variables in namelist {\tt \&param\_chiqw} to calculate polarization and dielectric functions. The variable type and its brief explanation are shown in the second and third columns, respectively. See the text for more detailed explanation of each variable.} 
\begin{center} 
\scalebox{0.71}{
\begin{tabular}{l@{\ \ \ }l@{\ \ \ }l} \hline \hline  \\ [-10pt] 
 \multicolumn{3}{l}{namelist {\tt \&param\_chiqw}}  \\ \hline \\ [-10pt] 
  variable             & type    & explanation   \\  \hline \\ [-10pt] 
 {\tt Ecut\_for\_eps}  & real    & cutoff energy for polarization in Rydberg \\ 
 {\tt Num\_freq\_grid} & integer & number of frequency grid \\ 
 {\tt Green\_func\_delt} & real  & smearing value in eV \\ 
 {\tt MPI\_num\_qcomm} & integer & degree of parallelism for $q$-point calculation \\ 
 {\tt MPI\_num\_proc\_per\_qcomm} & integer & number of MPI processes per $q$-community \\ 
 {\tt Flg\_cRPA}       & integer & constrained RPA (do not: 0, do: 1) \\
 \hline \hline
\end{tabular} 
} 
\end{center}
\label{param_chiqw} 
\end{table} 

\subsection{{\tt \&param\_calc\_int}} \label{param-calcint}
Table~\ref{param_calc_int} gives a main variable in the namelist {\tt \&param\_calc\_int}. Evaluations of the direct-Coulomb and exchange integrals are executed in separate executable files (the direct-Coulomb integral for {\tt calc\_w3d} and the exchange integral for {\tt calc\_j3d}), but these calculations are performed with the common namelist {\tt \&param\_calc\_int}. RESPACK calculates the frequency-dependent effective interaction but its data size is huge. So, the user can set an output frequency number of the effective interaction matrix as {\tt Calc\_ifreq}. By default, the results of $\omega=0$ are output ({\tt Calc\_ifreq=1}). 
\begin{table}[htpb] 
\caption{Main parameter in namelist {\tt \&param\_calc\_int} to evaluate the direct-Coulomb and exchange integrals. A common namelist is used for the direct-Coulomb and exchange calculations. The variable {\tt Calc\_ifreq} specifies the frequency number of the output interaction integrals, and by default, the $\omega=0$ results are output ({\tt Calc\_ifreq=1}).} 
\begin{center} 
\scalebox{0.8}{
\begin{tabular}{l@{\ \ \ }l@{\ \ \ }l} \hline \hline  \\ [-10pt] 
 \multicolumn{3}{l}{namelist {\tt \&param\_calc\_int}}  \\ \hline \\ [-10pt] 
  variable         & type    & explanation   \\  \hline \\ [-10pt] 
 {\tt Calc\_ifreq} & integer & frequency number of output interaction matrix \\ 
 \hline \hline
\end{tabular} 
} 
\end{center}
\label{param_calc_int} 
\end{table} 

\section{transfer analysis} \label{sec_tr}
When the RESPACK job ({\it wannier}, {\it chiqw}, {\it calc\_w3d}, {\it calc\_j3d}) is completed, a directory {\tt dir-model} is created under the calculation directory, and the following 9 files are generated in {\tt dir-model}: 
\vspace{3mm}\hrule
\begin{enumerate}
\item {\tt zvo\_hr.dat} (transfer integrals)
\item {\tt zvo\_dr.dat} (density matrix)
\item {\tt zvo\_ur.dat} (direct Coulomb integrals)
\item {\tt zvo\_jr.dat} (exchange integrals)
\item {\tt zvo\_geom.dat} (Wannier centers: text format)
\item {\tt zvo\_geom.xsf} (Wannier centers: xsf format)
\item {\tt zvo\_bandkpts.dat} [$k$-grid in Eq.~(\ref{k-disp-grid})]
\item {\tt zvo\_mkkpts.dat} [$k$-grid in Eq.~(\ref{k-dos-grid})]
\item {\tt zvo\_ef.dat} (Fermi energy)
\end{enumerate}
\hrule\vspace{3mm}
These {\tt zvo} files are used for the connection to model-solver software mVMC~\cite{mVMC_URL} and $\mathcal{H}\Phi$~\cite{HPHI_URL}. The {\tt zvo\_hr.dat}, {\tt zvo\_dr.dat}, {\tt zvo\_ur.dat}, and {\tt zvo\_jr.dat} are output in the {\sc Waniier90} format~\cite{Wannier90}. From these {\tt zvo} files, the interface code provided from mVMC and $\mathcal{H}\Phi$ generate the input files of mVMC and $\mathcal{H}\Phi$ automatically. Thus, the user can easily make the inputs for the post-RESPACK calculations. Refer to the manual sites of mVMC and $\mathcal{H}\Phi$ for the interface program and usage. 

RESPACK provides a utility tool to analyze transfer data, and the above 9 files are also used for this purpose. Using a {\sc python} script {\tt tr.py} in the directory {\tt RESPACK/util/transfer\_analysis} and an executable file {\tt calc\_tr} by compiling source code in the directory {\tt RESPACK/src/transfer\_analysis}, one can execute the transfer analysis. 
Preparation for the calculations so far is as follows: 
\vspace{3mm}\hrule\vspace{3mm}
\texttt{> cd RESPACK/src/transfer\_analysis}

\texttt{> make} 

\texttt{> cp calc\_tr calculation\_directory/.} 

\texttt{> cd RESPACK/util/transfer\_analysis} 

\begin{figure}[htpb] 
\begin{center} 
\includegraphics[width=0.30\textwidth]{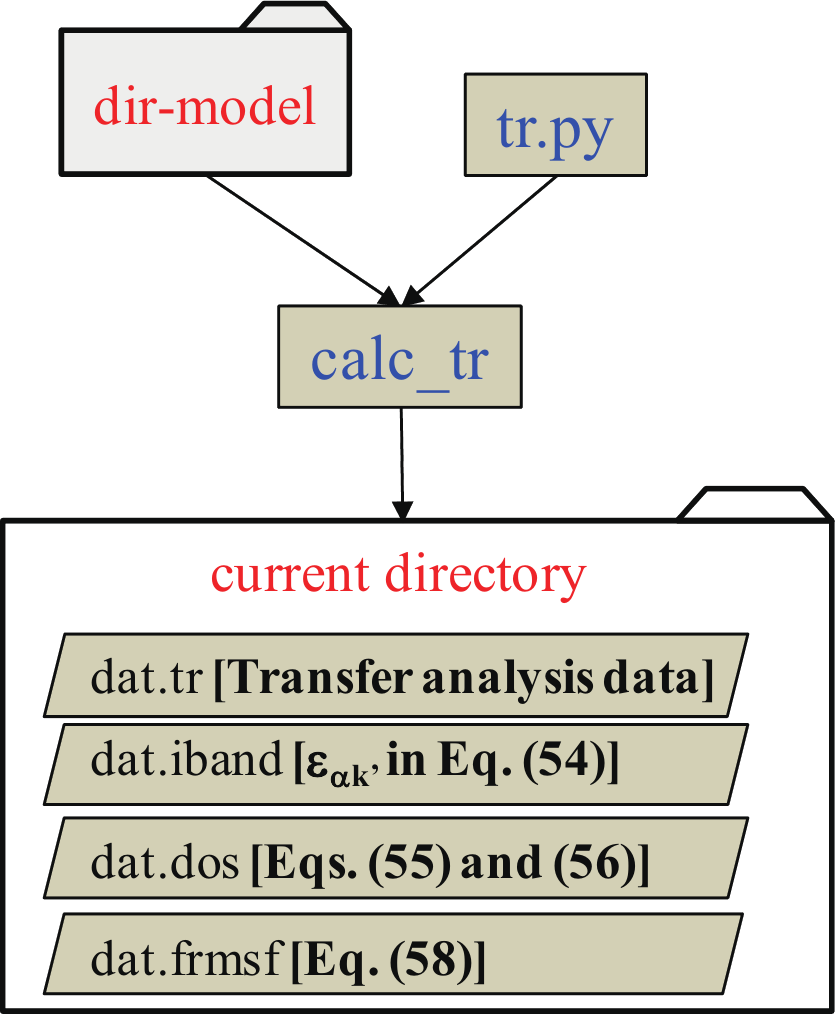}
\end{center} 
\caption{Flow diagram of transfer analysis. With the data in {\tt dir-model}, an executable file {\tt calc\_tr}, a {\sc python} script {\tt tr.py}, the transfer analysis is performed. Note that {\tt tr.py} calls the executable file {\tt calc\_tr} internally. Calculation results are output in the current directory.} 
\label{fig-tr}
\end{figure}

\texttt{> cp tr.py calculation\_directory/.} 
\vspace{3mm}\hrule\vspace{3mm} 
\noindent 
Figure~\ref{fig-tr} is a flow diagram of the transfer analysis. With the data files in {\tt dir-model}, the executable file {\tt calc\_tr}, and the {\sc python} script {\tt tr.py}, the transfer analysis is performed.  

This {\tt tr.py} code is basically a sorting code for the transfer data. The user can classify and sort the transfer data with the same absolute value. Besides the sorting function, the band-dispersion calculation, the density-of-states calculation, and the Fermi-surface calculation, etc. can be executed. Table~\ref{param_tr} summarizes the command lines available in {\tt tr.py}. With this program, the user can change the calculation condition such as energy threshold ({\tt --ecut}) and spatial range ({\tt --rcut}) of transfers, and $k$-grid size ({\tt --kgrd}) considered in the band-dispersion and density-of-states calculations. 
\begin{table}[htpb] 
\caption{Command line options available in the {\sc python} script {\tt tr.py} in Fig.~\ref{fig-tr}. A brief explanation is given in the right column. The part written in bold font indicates the default setting. See the text for the usage of {\tt tr.py}.} 
\begin{center} 
\scalebox{0.72}{
\begin{tabular}{l@{\ \ \ }l} \hline \hline  \\ [-10pt] 
 command  & explanation   \\  \hline \\ [-10pt] 
 {\tt -h} & display argument list\\ 
 {\tt --bnd} & Wannier-interpolated band calculation\\ 
 {\tt --dos} & density of state (DOS) calculation\\ 
 {\tt --frm} & Fermi-surface data generation for Fermisurfer\\ 
 {\tt --diff} $D$ & threshold for detecting equivalent transfers ({\bf 0.01 eV}) \\ 
 {\tt --ecut} $E$& cut transfers below $E$ ({\bf 0.00 eV})\\ 
 {\tt --rcut} $R$& cut transfers over $R$ ({\bf 100 \AA})\\ 
 {\tt --elec} $N$& set electron numbers in unit cell to $N$ ({\bf no setting})\\ 
 {\tt --delta} $\delta$& set broadening for DOS calculation to $\delta$ ({\bf 0.01 eV})\\ 
 {\tt --kgrd=}$`n\ m\ l'$& set $k$-grid to $n\times m\times l$ ({\bf the same as the band calculation})\\ 
 \hline \hline
\end{tabular} 
} 
\end{center}
\label{param_tr} 
\end{table} 

\begin{figure}[htpb] 
\centering
\includegraphics[width=0.43\textwidth]{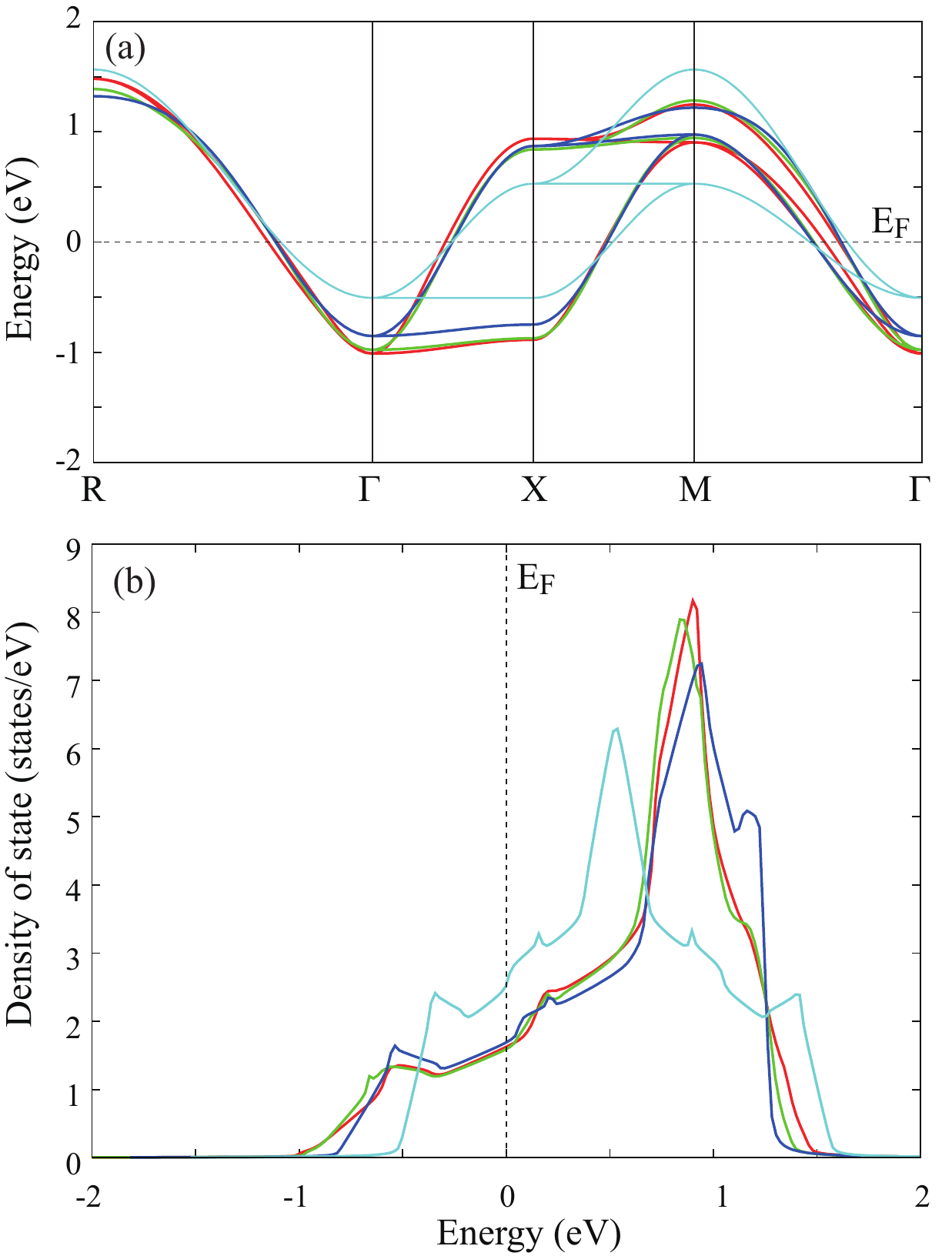}
\caption{Dependence of the $t_{2g}$-band properties of SrVO$_3$ on energy cutoff of transfer integrals: (a) The Wannier-interpolated band: {\tt python tr.py --bnd --ecut=0.00} (red), {\tt 0.01} (green), {\tt 0.05} (blue), {\tt 0.10} (right blue). (b) Density of states: {\tt python tr.py --dos --ecut=0.00} (red), {\tt 0.01} (green), {\tt 0.05} (blue), {\tt 0.10} (right blue). The Fermi energy is zero.
} 
\label{tr-iband-dos}
\end{figure}

Figure~\ref{tr-iband-dos} is examples of the transfer analysis for SrVO$_3$. The panel (a) and (b) show the energy-cutoff dependence of the Wannier-interpolated band and density of states, respectively. These results were obtained by performing the following commands  
\begin{itemize}
    \item Band-dispersion calculation [Fig.~\ref{tr-iband-dos}(a)]:
    
    {\tt > python tr.py --bnd --ecut=}$E$
    
    \item Density-of-states calculation [Fig.~\ref{tr-iband-dos}(b)]: 
    
    {\tt > python tr.py --dos --ecut=}$E$
\end{itemize}
with $E$ being a value of the energy threshold in unit of eV. In the figures, we show the results of $E=0.00$ (red), 0.01 (green), 0.05 (blue), 0.1 (right-blue) eV. 
With {\tt --ecut=0.1}, only the nearest neighbor transfers are considered in the calculations (see Table~\ref{param_transfer}), and therefore the band dispersion and density of states are appreciably changed from others. 

\begin{figure}[htpb] 
\centering
\includegraphics[width=0.475\textwidth]{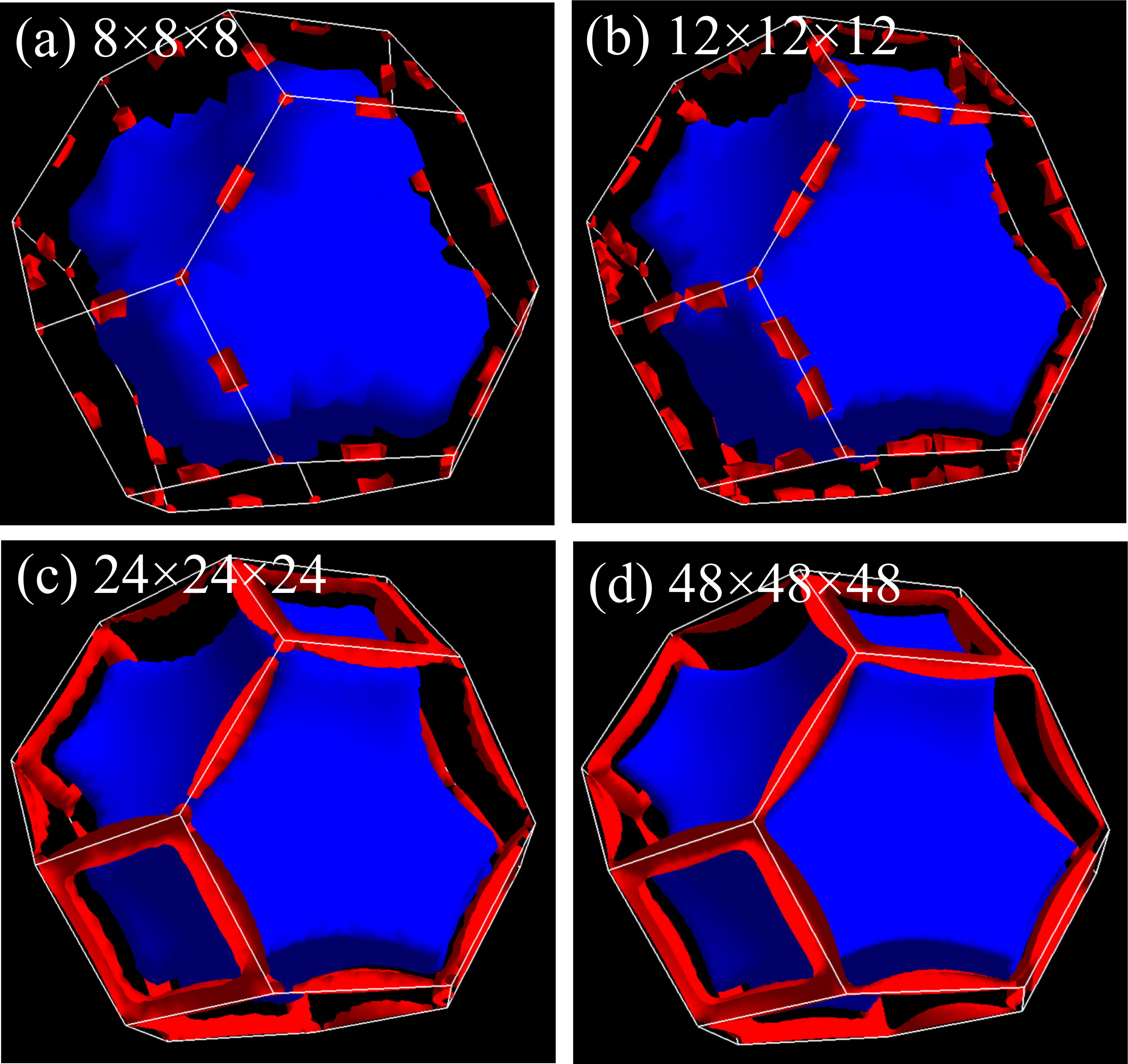}
\caption{Dependence of the Fermi surface of Al on the $k$-grid density (drawn by {\sc FermiSurfer}~\cite{fermisurfer,KAWAMURA2019197}): {\tt python tr.py --frm --kgrd=}`8 8 8' (a), `12 12 12' (b), `24 24 24' (c), `48 48 48' (d).}
\label{fermi-surface}
\end{figure}

As another example, we show in Fig.~\ref{fermi-surface} the $k$-grid density dependence of the Fermi surface. This is an Al result. Accurate description for a small Fermi surface colored by red needs dense $k$-grid density. The calculation is performed as   
\vspace{3mm}\hrule\vspace{3mm}
{\tt > python tr.py --frm --kgrd='}$n\ m\ l${\tt '}  
\vspace{3mm}\hrule\vspace{3mm}
\noindent 
with $n$, $m$, and $l$ being numbers of $k$-grid along the basic reciprocal lattice vectors ${\bf b}_1$, ${\bf b}_2$, and ${\bf b}_3$, respectively. 

\bibliographystyle{elsarticle-num}

\begin{thebibliography}{10}
\expandafter\ifx\csname url\endcsname\relax
  \def\url#1{\texttt{#1}}\fi
\expandafter\ifx\csname urlprefix\endcsname\relax\def\urlprefix{URL }\fi
\expandafter\ifx\csname href\endcsname\relax
  \def\href#1#2{#2} \def\path#1{#1}\fi

\bibitem{Hohenverg_1964}
P.~Hohenberg, W.~Kohn,
  \href{https://link.aps.org/doi/10.1103/PhysRev.136.B864}{Inhomogeneous
  electron gas}, Phys. Rev. 136 (1964) B864--B871.
\newblock \href {http://dx.doi.org/10.1103/PhysRev.136.B864}
  {\path{doi:10.1103/PhysRev.136.B864}}.
\newline\urlprefix\url{https://link.aps.org/doi/10.1103/PhysRev.136.B864}

\bibitem{Kohn_1965}
W.~Kohn, L.~J. Sham,
  \href{https://link.aps.org/doi/10.1103/PhysRev.140.A1133}{Self-consistent
  equations including exchange and correlation effects}, Phys. Rev. 140 (1965)
  A1133--A1138.
\newblock \href {http://dx.doi.org/10.1103/PhysRev.140.A1133}
  {\path{doi:10.1103/PhysRev.140.A1133}}.
\newline\urlprefix\url{https://link.aps.org/doi/10.1103/PhysRev.140.A1133}

\bibitem{Iwata_2010}
J.-I. Iwata, D.~Takahashi, A.~Oshiyama, T.~Boku, K.~Shiraishi, S.~Okada,
  K.~Yabana,
  \href{http://www.sciencedirect.com/science/article/pii/S0021999109006676}{A
  massively-parallel electronic-structure calculations based on real-space
  density functional theory}, Journal of Computational Physics 229~(6) (2010)
  2339 -- 2363.
\newblock \href {http://dx.doi.org/https://doi.org/10.1016/j.jcp.2009.11.038}
  {\path{doi:https://doi.org/10.1016/j.jcp.2009.11.038}}.
\newline\urlprefix\url{http://www.sciencedirect.com/science/article/pii/S0021999109006676}

\bibitem{Hine_2009}
N.~Hine, P.~Haynes, A.~Mostofi, C.-K. Skylaris, M.~Payne,
  \href{http://www.sciencedirect.com/science/article/pii/S0010465508004414}{Linear-scaling
  density-functional theory with tens of thousands of atoms: Expanding the
  scope and scale of calculations with onetep}, Computer Physics Communications
  180~(7) (2009) 1041 -- 1053.
\newblock \href {http://dx.doi.org/https://doi.org/10.1016/j.cpc.2008.12.023}
  {\path{doi:https://doi.org/10.1016/j.cpc.2008.12.023}}.
\newline\urlprefix\url{http://www.sciencedirect.com/science/article/pii/S0010465508004414}

\bibitem{Duy_2014-1}
T.~V.~T. Duy, T.~Ozaki,
  \href{http://www.sciencedirect.com/science/article/pii/S0010465513003020}{A
  decomposition method with minimum communication amount for parallelization of
  multi-dimensional ffts}, Computer Physics Communications 185~(1) (2014) 153
  -- 164.
\newblock \href {http://dx.doi.org/https://doi.org/10.1016/j.cpc.2013.08.028}
  {\path{doi:https://doi.org/10.1016/j.cpc.2013.08.028}}.
\newline\urlprefix\url{http://www.sciencedirect.com/science/article/pii/S0010465513003020}

\bibitem{Duy_2014-2}
T.~V.~T. Duy, T.~Ozaki,
  \href{http://www.sciencedirect.com/science/article/pii/S0010465513004013}{A
  three-dimensional domain decomposition method for large-scale dft electronic
  structure calculations}, Computer Physics Communications 185~(3) (2014) 777
  -- 789.
\newblock \href {http://dx.doi.org/https://doi.org/10.1016/j.cpc.2013.11.008}
  {\path{doi:https://doi.org/10.1016/j.cpc.2013.11.008}}.
\newline\urlprefix\url{http://www.sciencedirect.com/science/article/pii/S0010465513004013}

\bibitem{Bowler_2006}
D.~R. Bowler, R.~Choudhury, M.~J. Gillan, T.~Miyazaki,
  \href{https://onlinelibrary.wiley.com/doi/abs/10.1002/pssb.200541386}{Recent
  progress with large-scale ab initio calculations: the conquest code}, physica
  status solidi (b) 243~(5) (2006) 989--1000.
\newblock \href
  {http://arxiv.org/abs/https://onlinelibrary.wiley.com/doi/pdf/10.1002/pssb.200541386}
  {\path{arXiv:https://onlinelibrary.wiley.com/doi/pdf/10.1002/pssb.200541386}},
  \href {http://dx.doi.org/10.1002/pssb.200541386}
  {\path{doi:10.1002/pssb.200541386}}.
\newline\urlprefix\url{https://onlinelibrary.wiley.com/doi/abs/10.1002/pssb.200541386}

\bibitem{vandeWalle_2002}
A.~van~de Walle, M.~Asta, G.~Ceder,
  \href{http://www.sciencedirect.com/science/article/pii/S0364591602800062}{The
  alloy theoretic automated toolkit: A user guide}, Calphad 26~(4) (2002) 539
  -- 553.
\newblock \href
  {http://dx.doi.org/https://doi.org/10.1016/S0364-5916(02)80006-2}
  {\path{doi:https://doi.org/10.1016/S0364-5916(02)80006-2}}.
\newline\urlprefix\url{http://www.sciencedirect.com/science/article/pii/S0364591602800062}

\bibitem{Zhang_2013}
W.~Zhang, A.~R. Oganov, A.~F. Goncharov, Q.~Zhu, S.~E. Boulfelfel, A.~O.
  Lyakhov, E.~Stavrou, M.~Somayazulu, V.~B. Prakapenka, Z.~Kon{\^o}pkov{\'a},
  \href{https://science.sciencemag.org/content/342/6165/1502}{Unexpected stable
  stoichiometries of sodium chlorides}, Science 342~(6165) (2013) 1502--1505.
\newblock \href
  {http://arxiv.org/abs/https://science.sciencemag.org/content/342/6165/1502.full.pdf}
  {\path{arXiv:https://science.sciencemag.org/content/342/6165/1502.full.pdf}},
  \href {http://dx.doi.org/10.1126/science.1244989}
  {\path{doi:10.1126/science.1244989}}.
\newline\urlprefix\url{https://science.sciencemag.org/content/342/6165/1502}

\bibitem{Imada_2010}
M.~Imada, T.~Miyake, \href{https://doi.org/10.1143/JPSJ.79.112001}{Electronic
  structure calculation by first principles for strongly correlated electron
  systems}, Journal of the Physical Society of Japan 79~(11) (2010) 112001.
\newblock \href {http://arxiv.org/abs/https://doi.org/10.1143/JPSJ.79.112001}
  {\path{arXiv:https://doi.org/10.1143/JPSJ.79.112001}}, \href
  {http://dx.doi.org/10.1143/JPSJ.79.112001}
  {\path{doi:10.1143/JPSJ.79.112001}}.
\newline\urlprefix\url{https://doi.org/10.1143/JPSJ.79.112001}

\bibitem{Nakamura_2008}
K.~Nakamura, R.~Arita, M.~Imada,
  \href{https://doi.org/10.1143/JPSJ.77.093711}{Ab initio derivation of
  low-energy model for iron-based superconductors lafeaso and lafepo}, Journal
  of the Physical Society of Japan 77~(9) (2008) 093711.
\newblock \href {http://arxiv.org/abs/https://doi.org/10.1143/JPSJ.77.093711}
  {\path{arXiv:https://doi.org/10.1143/JPSJ.77.093711}}, \href
  {http://dx.doi.org/10.1143/JPSJ.77.093711}
  {\path{doi:10.1143/JPSJ.77.093711}}.
\newline\urlprefix\url{https://doi.org/10.1143/JPSJ.77.093711}

\bibitem{Nakamura_2009}
K.~Nakamura, Y.~Yoshimoto, T.~Kosugi, R.~Arita, M.~Imada,
  \href{https://doi.org/10.1143/JPSJ.78.083710}{Ab initio derivation of
  low-energy model for $\kappa$-et type organic conductors}, Journal of the
  Physical Society of Japan 78~(8) (2009) 083710.
\newblock \href {http://arxiv.org/abs/https://doi.org/10.1143/JPSJ.78.083710}
  {\path{arXiv:https://doi.org/10.1143/JPSJ.78.083710}}, \href
  {http://dx.doi.org/10.1143/JPSJ.78.083710}
  {\path{doi:10.1143/JPSJ.78.083710}}.
\newline\urlprefix\url{https://doi.org/10.1143/JPSJ.78.083710}

\bibitem{Nakamura_2010}
K.~Nakamura, Y.~Yoshimoto, Y.~Nohara, M.~Imada,
  \href{https://doi.org/10.1143/JPSJ.79.123708}{Ab initio low-dimensional
  physics opened up by dimensional downfolding: Application to lafeaso},
  Journal of the Physical Society of Japan 79~(12) (2010) 123708.
\newblock \href {http://arxiv.org/abs/https://doi.org/10.1143/JPSJ.79.123708}
  {\path{arXiv:https://doi.org/10.1143/JPSJ.79.123708}}, \href
  {http://dx.doi.org/10.1143/JPSJ.79.123708}
  {\path{doi:10.1143/JPSJ.79.123708}}.
\newline\urlprefix\url{https://doi.org/10.1143/JPSJ.79.123708}

\bibitem{Nakamura_2016}
K.~Nakamura, Y.~Nohara, Y.~Yosimoto, Y.~Nomura,
  \href{https://link.aps.org/doi/10.1103/PhysRevB.93.085124}{{Ab initio $GW$
  plus cumulant calculation for isolated band systems: Application to organic
  conductor ${(\mathrm{TMTSF})}_{2}{\mathrm{PF}}_{6}$ and transition-metal
  oxide ${\mathrm{SrVO}}_{3}$}}, Phys. Rev. B 93 (2016) 085124.
\newblock \href {http://dx.doi.org/10.1103/PhysRevB.93.085124}
  {\path{doi:10.1103/PhysRevB.93.085124}}.
\newline\urlprefix\url{https://link.aps.org/doi/10.1103/PhysRevB.93.085124}

\bibitem{cRPA_Miyake_2009}
T.~Miyake, F.~Aryasetiawan, M.~Imada,
  \href{https://link.aps.org/doi/10.1103/PhysRevB.80.155134}{Ab initio
  procedure for constructing effective models of correlated materials with
  entangled band structure}, Phys. Rev. B 80 (2009) 155134.
\newblock \href {http://dx.doi.org/10.1103/PhysRevB.80.155134}
  {\path{doi:10.1103/PhysRevB.80.155134}}.
\newline\urlprefix\url{https://link.aps.org/doi/10.1103/PhysRevB.80.155134}

\bibitem{cRPA_Wehling_2011}
T.~O. Wehling, E.~\ifmmode \mbox{\c{S}}\else \c{S}\fi{}a\ifmmode
  \mbox{\c{s}}\else \c{s}\fi{}\ifmmode \imath \else \i
  \fi{}o\ifmmode~\breve{g}\else \u{g}\fi{}lu, C.~Friedrich, A.~I. Lichtenstein,
  M.~I. Katsnelson, S.~Bl\"ugel,
  \href{https://link.aps.org/doi/10.1103/PhysRevLett.106.236805}{Strength of
  effective coulomb interactions in graphene and graphite}, Phys. Rev. Lett.
  106 (2011) 236805.
\newblock \href {http://dx.doi.org/10.1103/PhysRevLett.106.236805}
  {\path{doi:10.1103/PhysRevLett.106.236805}}.
\newline\urlprefix\url{https://link.aps.org/doi/10.1103/PhysRevLett.106.236805}

\bibitem{cRPA_Nomura_2012}
Y.~Nomura, K.~Nakamura, R.~Arita,
  \href{https://link.aps.org/doi/10.1103/PhysRevB.85.155452}{Ab initio
  derivation of electronic low-energy models for c${}_{60}$ and aromatic
  compounds}, Phys. Rev. B 85 (2012) 155452.
\newblock \href {http://dx.doi.org/10.1103/PhysRevB.85.155452}
  {\path{doi:10.1103/PhysRevB.85.155452}}.
\newline\urlprefix\url{https://link.aps.org/doi/10.1103/PhysRevB.85.155452}

\bibitem{cRPA_Nomura_2013}
Y.~Nomura, M.~Kaltak, K.~Nakamura, C.~Taranto, S.~Sakai, A.~Toschi, R.~Arita,
  K.~Held, G.~Kresse, M.~Imada,
  \href{https://link.aps.org/doi/10.1103/PhysRevB.86.085117}{Effective on-site
  interaction for dynamical mean-field theory}, Phys. Rev. B 86 (2012) 085117.
\newblock \href {http://dx.doi.org/10.1103/PhysRevB.86.085117}
  {\path{doi:10.1103/PhysRevB.86.085117}}.
\newline\urlprefix\url{https://link.aps.org/doi/10.1103/PhysRevB.86.085117}

\bibitem{cRPA_Ersoy_2012}
E.~\ifmmode \mbox{\c{S}}\else \c{S}\fi{}a\ifmmode \mbox{\c{s}}\else
  \c{s}\fi{}\ifmmode \imath \else \i \fi{}o\ifmmode~\breve{g}\else
  \u{g}\fi{}lu, C.~Friedrich, S.~Bl\"ugel,
  \href{https://link.aps.org/doi/10.1103/PhysRevLett.109.146401}{Strength of
  the effective coulomb interaction at metal and insulator surfaces}, Phys.
  Rev. Lett. 109 (2012) 146401.
\newblock \href {http://dx.doi.org/10.1103/PhysRevLett.109.146401}
  {\path{doi:10.1103/PhysRevLett.109.146401}}.
\newline\urlprefix\url{https://link.aps.org/doi/10.1103/PhysRevLett.109.146401}

\bibitem{cRPA_Vaugier_2012}
L.~Vaugier, H.~Jiang, S.~Biermann,
  \href{https://link.aps.org/doi/10.1103/PhysRevB.86.165105}{Hubbard $u$ and
  hund exchange $j$ in transition metal oxides: Screening versus localization
  trends from constrained random phase approximation}, Phys. Rev. B 86 (2012)
  165105.
\newblock \href {http://dx.doi.org/10.1103/PhysRevB.86.165105}
  {\path{doi:10.1103/PhysRevB.86.165105}}.
\newline\urlprefix\url{https://link.aps.org/doi/10.1103/PhysRevB.86.165105}

\bibitem{cRPA_Hansmann_2013}
P.~Hansmann, L.~Vaugier, H.~Jiang, S.~Biermann,
  \href{https://doi.org/10.1088%2F0953-8984%2F25%2F9%2F094005}{What about u on
  surfaces? extended hubbard models for adatom systems from first principles},
  Journal of Physics: Condensed Matter 25~(9) (2013) 094005.
\newblock \href {http://dx.doi.org/10.1088/0953-8984/25/9/094005}
  {\path{doi:10.1088/0953-8984/25/9/094005}}.
\newline\urlprefix\url{https://doi.org/10.1088%2F0953-8984%2F25%2F9%2F094005}

\bibitem{cRPA_Nilsson_2013}
F.~Nilsson, R.~Sakuma, F.~Aryasetiawan,
  \href{https://link.aps.org/doi/10.1103/PhysRevB.88.125123}{Ab initio
  calculations of the hubbard $u$ for the early lanthanides using the
  constrained random-phase approximation}, Phys. Rev. B 88 (2013) 125123.
\newblock \href {http://dx.doi.org/10.1103/PhysRevB.88.125123}
  {\path{doi:10.1103/PhysRevB.88.125123}}.
\newline\urlprefix\url{https://link.aps.org/doi/10.1103/PhysRevB.88.125123}

\bibitem{cRPA_Amadon_2014}
B.~Amadon, T.~Applencourt, F.~Bruneval,
  \href{https://link.aps.org/doi/10.1103/PhysRevB.89.125110}{Screened coulomb
  interaction calculations: crpa implementation and applications to dynamical
  screening and self-consistency in uranium dioxide and cerium}, Phys. Rev. B
  89 (2014) 125110.
\newblock \href {http://dx.doi.org/10.1103/PhysRevB.89.125110}
  {\path{doi:10.1103/PhysRevB.89.125110}}.
\newline\urlprefix\url{https://link.aps.org/doi/10.1103/PhysRevB.89.125110}

\bibitem{cRPA_Yamaji_2014}
Y.~Yamaji, Y.~Nomura, M.~Kurita, R.~Arita, M.~Imada,
  \href{https://link.aps.org/doi/10.1103/PhysRevLett.113.107201}{First-principles
  study of the honeycomb-lattice iridates ${\mathrm{na}}_{2}{\mathrm{iro}}_{3}$
  in the presence of strong spin-orbit interaction and electron correlations},
  Phys. Rev. Lett. 113 (2014) 107201.
\newblock \href {http://dx.doi.org/10.1103/PhysRevLett.113.107201}
  {\path{doi:10.1103/PhysRevLett.113.107201}}.
\newline\urlprefix\url{https://link.aps.org/doi/10.1103/PhysRevLett.113.107201}

\bibitem{cRPA_Okamoto_2014}
S.~Okamoto, W.~Zhu, Y.~Nomura, R.~Arita, D.~Xiao, N.~Nagaosa,
  \href{https://link.aps.org/doi/10.1103/PhysRevB.89.195121}{Correlation
  effects in (111) bilayers of perovskite transition-metal oxides}, Phys. Rev.
  B 89 (2014) 195121.
\newblock \href {http://dx.doi.org/10.1103/PhysRevB.89.195121}
  {\path{doi:10.1103/PhysRevB.89.195121}}.
\newline\urlprefix\url{https://link.aps.org/doi/10.1103/PhysRevB.89.195121}

\bibitem{cRPA_Kim_2016}
M.~Kim, Y.~Nomura, M.~Ferrero, P.~Seth, O.~Parcollet, A.~Georges,
  \href{https://link.aps.org/doi/10.1103/PhysRevB.94.155152}{Enhancing
  superconductivity in ${A}_{3}{\mathrm{c}}_{60}$ fullerides}, Phys. Rev. B 94
  (2016) 155152.
\newblock \href {http://dx.doi.org/10.1103/PhysRevB.94.155152}
  {\path{doi:10.1103/PhysRevB.94.155152}}.
\newline\urlprefix\url{https://link.aps.org/doi/10.1103/PhysRevB.94.155152}

\bibitem{cRPA_Hirayama_2018}
M.~Hirayama, Y.~Yamaji, T.~Misawa, M.~Imada,
  \href{https://link.aps.org/doi/10.1103/PhysRevB.98.134501}{{Ab initio
  effective Hamiltonians for cuprate superconductors}}, Phys. Rev. B 98 (2018)
  134501.
\newblock \href {http://dx.doi.org/10.1103/PhysRevB.98.134501}
  {\path{doi:10.1103/PhysRevB.98.134501}}.
\newline\urlprefix\url{https://link.aps.org/doi/10.1103/PhysRevB.98.134501}

\bibitem{cRPA_Hirayama_2019}
M.~Hirayama, T.~Misawa, T.~Ohgoe, Y.~Yamaji, M.~Imada,
  \href{https://link.aps.org/doi/10.1103/PhysRevB.99.245155}{{Effective
  Hamiltonian for cuprate superconductors derived from multiscale ab initio
  scheme with level renormalization}}, Phys. Rev. B 99 (2019) 245155.
\newblock \href {http://dx.doi.org/10.1103/PhysRevB.99.245155}
  {\path{doi:10.1103/PhysRevB.99.245155}}.
\newline\urlprefix\url{https://link.aps.org/doi/10.1103/PhysRevB.99.245155}

\bibitem{cRPA_Tadano_2019}
T.~Tadano, Y.~Nomura, M.~Imada,
  \href{https://link.aps.org/doi/10.1103/PhysRevB.99.155148}{{Ab initio
  derivation of an effective Hamiltonian for the
  ${\mathrm{La}}_{2}{\mathrm{CuO}}_{4}/{\mathrm{La}}_{1.55}{\mathrm{Sr}}_{0.45}{\mathrm{CuO}}_{4}$
  heterostructure}}, Phys. Rev. B 99 (2019) 155148.
\newblock \href {http://dx.doi.org/10.1103/PhysRevB.99.155148}
  {\path{doi:10.1103/PhysRevB.99.155148}}.
\newline\urlprefix\url{https://link.aps.org/doi/10.1103/PhysRevB.99.155148}

\bibitem{cRPA_Nomura_arXiv}
Y.~Nomura, M.~Hirayama, T.~Tadano, Y.~Yoshimoto, K.~Nakamura, R.~Arita,
  \href{https://link.aps.org/doi/10.1103/PhysRevB.100.205138}{Formation of a
  two-dimensional single-component correlated electron system and band
  engineering in the nickelate superconductor ${\mathrm{ndnio}}_{2}$}, Phys.
  Rev. B 100 (2019) 205138.
\newblock \href {http://dx.doi.org/10.1103/PhysRevB.100.205138}
  {\path{doi:10.1103/PhysRevB.100.205138}}.
\newline\urlprefix\url{https://link.aps.org/doi/10.1103/PhysRevB.100.205138}

\bibitem{cRPA_Hirayama_arXiv}
M.~Hirayama, T.~Tadano, Y.~Nomura, R.~Arita, Materials design of dynamically
  stable d$^9$ layered nickelates.
\newblock \href {http://arxiv.org/abs/1910.03974} {\path{arXiv:1910.03974}}.

\bibitem{Aryasetiawan_2004}
F.~Aryasetiawan, M.~Imada, A.~Georges, G.~Kotliar, S.~Biermann, A.~I.
  Lichtenstein,
  \href{https://link.aps.org/doi/10.1103/PhysRevB.70.195104}{Frequency-dependent
  local interactions and low-energy effective models from electronic structure
  calculations}, Phys. Rev. B 70 (2004) 195104.
\newblock \href {http://dx.doi.org/10.1103/PhysRevB.70.195104}
  {\path{doi:10.1103/PhysRevB.70.195104}}.
\newline\urlprefix\url{https://link.aps.org/doi/10.1103/PhysRevB.70.195104}

\bibitem{Kawamura_2017}
M.~Kawamura, K.~Yoshimi, T.~Misawa, Y.~Yamaji, S.~Todo, N.~Kawashima,
  \href{http://www.sciencedirect.com/science/article/pii/S0010465517301200}{Quantum
  lattice model solver h$\phi$}, Computer Physics Communications 217 (2017) 180
  -- 192.
\newblock \href {http://dx.doi.org/https://doi.org/10.1016/j.cpc.2017.04.006}
  {\path{doi:https://doi.org/10.1016/j.cpc.2017.04.006}}.
\newline\urlprefix\url{http://www.sciencedirect.com/science/article/pii/S0010465517301200}

\bibitem{Georges_1996}
A.~Georges, G.~Kotliar, W.~Krauth, M.~J. Rozenberg,
  \href{https://link.aps.org/doi/10.1103/RevModPhys.68.13}{Dynamical mean-field
  theory of strongly correlated fermion systems and the limit of infinite
  dimensions}, Rev. Mod. Phys. 68 (1996) 13--125.
\newblock \href {http://dx.doi.org/10.1103/RevModPhys.68.13}
  {\path{doi:10.1103/RevModPhys.68.13}}.
\newline\urlprefix\url{https://link.aps.org/doi/10.1103/RevModPhys.68.13}

\bibitem{Kotliar_2006}
G.~Kotliar, S.~Y. Savrasov, K.~Haule, V.~S. Oudovenko, O.~Parcollet, C.~A.
  Marianetti,
  \href{https://link.aps.org/doi/10.1103/RevModPhys.78.865}{Electronic
  structure calculations with dynamical mean-field theory}, Rev. Mod. Phys. 78
  (2006) 865--951.
\newblock \href {http://dx.doi.org/10.1103/RevModPhys.78.865}
  {\path{doi:10.1103/RevModPhys.78.865}}.
\newline\urlprefix\url{https://link.aps.org/doi/10.1103/RevModPhys.78.865}

\bibitem{Tahara_2008_1}
D.~Tahara, M.~Imada, \href{https://doi.org/10.1143/JPSJ.77.093703}{Variational
  monte carlo study of electron differentiation around mott transition},
  Journal of the Physical Society of Japan 77~(9) (2008) 093703.
\newblock \href {http://arxiv.org/abs/https://doi.org/10.1143/JPSJ.77.093703}
  {\path{arXiv:https://doi.org/10.1143/JPSJ.77.093703}}, \href
  {http://dx.doi.org/10.1143/JPSJ.77.093703}
  {\path{doi:10.1143/JPSJ.77.093703}}.
\newline\urlprefix\url{https://doi.org/10.1143/JPSJ.77.093703}

\bibitem{Tahara_2008_2}
D.~Tahara, M.~Imada, \href{https://doi.org/10.1143/JPSJ.77.114701}{Variational
  monte carlo method combined with quantum-number projection and multi-variable
  optimization}, Journal of the Physical Society of Japan 77~(11) (2008)
  114701.
\newblock \href {http://arxiv.org/abs/https://doi.org/10.1143/JPSJ.77.114701}
  {\path{arXiv:https://doi.org/10.1143/JPSJ.77.114701}}, \href
  {http://dx.doi.org/10.1143/JPSJ.77.114701}
  {\path{doi:10.1143/JPSJ.77.114701}}.
\newline\urlprefix\url{https://doi.org/10.1143/JPSJ.77.114701}

\bibitem{Misawa_2019}
T.~Misawa, S.~Morita, K.~Yoshimi, M.~Kawamura, Y.~Motoyama, K.~Ido, T.~Ohgoe,
  M.~Imada, T.~Kato,
  \href{http://www.sciencedirect.com/science/article/pii/S0010465518303102}{mvmc$-$open-source
  software for many-variable variational monte carlo method}, Computer Physics
  Communications 235 (2019) 447 -- 462.
\newblock \href {http://dx.doi.org/https://doi.org/10.1016/j.cpc.2018.08.014}
  {\path{doi:https://doi.org/10.1016/j.cpc.2018.08.014}}.
\newline\urlprefix\url{http://www.sciencedirect.com/science/article/pii/S0010465518303102}

\bibitem{Nakamura_CI_2008}
K.~Nakamura, Y.~Yoshimoto, R.~Arita, S.~Tsuneyuki, M.~Imada,
  \href{https://link.aps.org/doi/10.1103/PhysRevB.77.195126}{Optical absorption
  study by ab initio downfolding approach: Application to gaas}, Phys. Rev. B
  77 (2008) 195126.
\newblock \href {http://dx.doi.org/10.1103/PhysRevB.77.195126}
  {\path{doi:10.1103/PhysRevB.77.195126}}.
\newline\urlprefix\url{https://link.aps.org/doi/10.1103/PhysRevB.77.195126}

\bibitem{RESPACK_URL}
\href{{https://sites.google.com/view/kazuma7k6r}}{{RESPACK Web page}} [online,
  cited {https://sites.google.com/view/kazuma7k6r}].

\bibitem{Marzari_1997}
N.~Marzari, D.~Vanderbilt,
  \href{https://link.aps.org/doi/10.1103/PhysRevB.56.12847}{Maximally localized
  generalized wannier functions for composite energy bands}, Phys. Rev. B 56
  (1997) 12847--12865.
\newblock \href {http://dx.doi.org/10.1103/PhysRevB.56.12847}
  {\path{doi:10.1103/PhysRevB.56.12847}}.
\newline\urlprefix\url{https://link.aps.org/doi/10.1103/PhysRevB.56.12847}

\bibitem{Souza_2001}
I.~Souza, N.~Marzari, D.~Vanderbilt,
  \href{https://link.aps.org/doi/10.1103/PhysRevB.65.035109}{Maximally
  localized wannier functions for entangled energy bands}, Phys. Rev. B 65
  (2001) 035109.
\newblock \href {http://dx.doi.org/10.1103/PhysRevB.65.035109}
  {\path{doi:10.1103/PhysRevB.65.035109}}.
\newline\urlprefix\url{https://link.aps.org/doi/10.1103/PhysRevB.65.035109}

\bibitem{Fujiwara_2003}
T.~Fujiwara, S.~Yamamoto, Y.~Ishii,
  \href{https://doi.org/10.1143/JPSJ.72.777}{{Generalization of the Iterative
  Perturbation Theory and Metal--Insulator Transition in Multi-Orbital Hubbard
  Bands}}, J. Phys. Soc. Jpn. 72~(4) (2003) 777--780.
\newblock \href {http://dx.doi.org/10.1143/JPSJ.72.777}
  {\path{doi:10.1143/JPSJ.72.777}}.
\newline\urlprefix\url{https://doi.org/10.1143/JPSJ.72.777}

\bibitem{Nohara_2009}
Y.~Nohara, S.~Yamamoto, T.~Fujiwara,
  \href{https://link.aps.org/doi/10.1103/PhysRevB.79.195110}{{Electronic
  structure of perovskite-type transition metal oxides
  $\text{La}M{\text{O}}_{3}$ $(M=\text{Ti}\ensuremath{\sim}\text{Cu})$ by
  $\text{U}+\text{GW}$ approximation}}, Phys. Rev. B 79 (2009) 195110.
\newblock \href {http://dx.doi.org/10.1103/PhysRevB.79.195110}
  {\path{doi:10.1103/PhysRevB.79.195110}}.
\newline\urlprefix\url{https://link.aps.org/doi/10.1103/PhysRevB.79.195110}

\bibitem{Sasioglu_2011}
E.~\ifmmode \mbox{\c{S}}\else \c{S}\fi{}a\ifmmode \mbox{\c{s}}\else
  \c{s}\fi{}\ifmmode \imath \else \i \fi{}o\ifmmode~\breve{g}\else
  \u{g}\fi{}lu, C.~Friedrich, S.~Bl\"ugel,
  \href{https://link.aps.org/doi/10.1103/PhysRevB.83.121101}{Effective coulomb
  interaction in transition metals from constrained random-phase
  approximation}, Phys. Rev. B 83 (2011) 121101.
\newblock \href {http://dx.doi.org/10.1103/PhysRevB.83.121101}
  {\path{doi:10.1103/PhysRevB.83.121101}}.
\newline\urlprefix\url{https://link.aps.org/doi/10.1103/PhysRevB.83.121101}

\bibitem{Hyvertsen_chi_1987}
M.~S. Hybertsen, S.~G. Louie,
  \href{https://link.aps.org/doi/10.1103/PhysRevB.35.5585}{Ab initio static
  dielectric matrices from the density-functional approach. i. formulation and
  application to semiconductors and insulators}, Phys. Rev. B 35 (1987)
  5585--5601.
\newblock \href {http://dx.doi.org/10.1103/PhysRevB.35.5585}
  {\path{doi:10.1103/PhysRevB.35.5585}}.
\newline\urlprefix\url{https://link.aps.org/doi/10.1103/PhysRevB.35.5585}

\bibitem{Draxl_2006}
C.~Ambrosch-Draxl, J.~O. Sofo,
  \href{http://www.sciencedirect.com/science/article/pii/S0010465506001299}{Linear
  optical properties of solids within the full-potential linearized augmented
  planewave method}, Computer Physics Communications 175~(1) (2006) 1 -- 14.
\newblock \href {http://dx.doi.org/https://doi.org/10.1016/j.cpc.2006.03.005}
  {\path{doi:https://doi.org/10.1016/j.cpc.2006.03.005}}.
\newline\urlprefix\url{http://www.sciencedirect.com/science/article/pii/S0010465506001299}

\bibitem{Marini2001}
A.~Marini, G.~Onida, R.~Del~Sole,
  \href{https://link.aps.org/doi/10.1103/PhysRevB.64.195125}{Plane-wave dft-lda
  calculation of the electronic structure and absorption spectrum of copper},
  Phys. Rev. B 64 (2001) 195125.
\newblock \href {http://dx.doi.org/10.1103/PhysRevB.64.195125}
  {\path{doi:10.1103/PhysRevB.64.195125}}.
\newline\urlprefix\url{https://link.aps.org/doi/10.1103/PhysRevB.64.195125}

\bibitem{Hyvertsen_GW_1986}
M.~S. Hybertsen, S.~G. Louie,
  \href{https://link.aps.org/doi/10.1103/PhysRevB.34.5390}{Electron correlation
  in semiconductors and insulators: Band gaps and quasiparticle energies},
  Phys. Rev. B 34 (1986) 5390--5413.
\newblock \href {http://dx.doi.org/10.1103/PhysRevB.34.5390}
  {\path{doi:10.1103/PhysRevB.34.5390}}.
\newline\urlprefix\url{https://link.aps.org/doi/10.1103/PhysRevB.34.5390}

\bibitem{fermisurfer}
\href{{http://fermisurfer.osdn.jp/}}{{FermiSurfer Web page}} [online, cited
  {http://fermisurfer.osdn.jp/}].

\bibitem{Yamauchi_1996}
J.~Yamauchi, M.~Tsukada, S.~Watanabe, O.~Sugino,
  \href{https://link.aps.org/doi/10.1103/PhysRevB.54.5586}{First-principles
  study on energetics of c-bn(001) reconstructed surfaces}, Phys. Rev. B 54
  (1996) 5586--5603.
\newblock \href {http://dx.doi.org/10.1103/PhysRevB.54.5586}
  {\path{doi:10.1103/PhysRevB.54.5586}}.
\newline\urlprefix\url{https://link.aps.org/doi/10.1103/PhysRevB.54.5586}

\bibitem{QE-2009}
P.~Giannozzi, S.~Baroni, N.~Bonini, M.~Calandra, R.~Car, C.~Cavazzoni,
  D.~Ceresoli, G.~L. Chiarotti, M.~Cococcioni, I.~Dabo, A.~{Dal Corso},
  S.~de~Gironcoli, S.~Fabris, G.~Fratesi, R.~Gebauer, U.~Gerstmann,
  C.~Gougoussis, A.~Kokalj, M.~Lazzeri, L.~Martin-Samos, N.~Marzari, F.~Mauri,
  R.~Mazzarello, S.~Paolini, A.~Pasquarello, L.~Paulatto, C.~Sbraccia,
  S.~Scandolo, G.~Sclauzero, A.~P. Seitsonen, A.~Smogunov, P.~Umari, R.~M.
  Wentzcovitch, \href{http://www.quantum-espresso.org}{Quantum espresso: a
  modular and open-source software project for quantum simulations of
  materials}, Journal of Physics: Condensed Matter 21~(39) (2009) 395502
  (19pp).
\newline\urlprefix\url{http://www.quantum-espresso.org}

\bibitem{QE-2017}
P.~Giannozzi, O.~Andreussi, T.~Brumme, O.~Bunau, M.~B. Nardelli, M.~Calandra,
  R.~Car, C.~Cavazzoni, D.~Ceresoli, M.~Cococcioni, N.~Colonna, I.~Carnimeo,
  A.~D. Corso, S.~de~Gironcoli, P.~Delugas, R.~A.~D. Jr, A.~Ferretti,
  A.~Floris, G.~Fratesi, G.~Fugallo, R.~Gebauer, U.~Gerstmann, F.~Giustino,
  T.~Gorni, J.~Jia, M.~Kawamura, H.-Y. Ko, A.~Kokalj,
  E.~K\"{u}\c{c}\"{u}kbenli, M.~Lazzeri, M.~Marsili, N.~Marzari, F.~Mauri,
  N.~L. Nguyen, H.-V. Nguyen, A.~O. de-la Roza, L.~Paulatto, S.~Ponc\'{e},
  D.~Rocca, R.~Sabatini, B.~Santra, M.~Schlipf, A.~P. Seitsonen, A.~Smogunov,
  I.~Timrov, T.~Thonhauser, P.~Umari, N.~Vast, X.~Wu, S.~Baroni,
  \href{http://stacks.iop.org/0953-8984/29/i=46/a=465901}{{Advanced
  capabilities for materials modelling with QUANTUM ESPRESSO}}, Journal of
  Physics: Condensed Matter 29~(46) (2017) 465901.
\newline\urlprefix\url{http://stacks.iop.org/0953-8984/29/i=46/a=465901}

\bibitem{Kleinman_1982}
L.~Kleinman, D.~M. Bylander,
  \href{https://link.aps.org/doi/10.1103/PhysRevLett.48.1425}{Efficacious form
  for model pseudopotentials}, Phys. Rev. Lett. 48 (1982) 1425--1428.
\newblock \href {http://dx.doi.org/10.1103/PhysRevLett.48.1425}
  {\path{doi:10.1103/PhysRevLett.48.1425}}.
\newline\urlprefix\url{https://link.aps.org/doi/10.1103/PhysRevLett.48.1425}

\bibitem{Troullier_1991}
N.~Troullier, J.~L. Martins,
  \href{https://link.aps.org/doi/10.1103/PhysRevB.43.1993}{Efficient
  pseudopotentials for plane-wave calculations}, Phys. Rev. B 43 (1991)
  1993--2006.
\newblock \href {http://dx.doi.org/10.1103/PhysRevB.43.1993}
  {\path{doi:10.1103/PhysRevB.43.1993}}.
\newline\urlprefix\url{https://link.aps.org/doi/10.1103/PhysRevB.43.1993}

\bibitem{Perdew_1996}
J.~P. Perdew, K.~Burke, M.~Ernzerhof,
  \href{https://link.aps.org/doi/10.1103/PhysRevLett.77.3865}{{Generalized
  Gradient Approximation Made Simple}}, Phys. Rev. Lett. 77 (1996) 3865--3868.
\newblock \href {http://dx.doi.org/10.1103/PhysRevLett.77.3865}
  {\path{doi:10.1103/PhysRevLett.77.3865}}.
\newline\urlprefix\url{https://link.aps.org/doi/10.1103/PhysRevLett.77.3865}

\bibitem{Momma_2011}
K.~Momma, F.~Izumi, {{\it VESTA3} for three-dimensional visualization of
  crystal, volumetric and morphology data}, Journal of Applied Crystallography
  44 (2011) 1272.
\newblock \href {http://dx.doi.org/{https://doi.org/10.1107/S0021889811038970}}
  {\path{doi:{https://doi.org/10.1107/S0021889811038970}}}.

\bibitem{about-Re-eM-w-0}
In this calculation, since the broadening factor $\delta$ is introduced, the
  real part of the macroscopic dielectric function does not actually diverge
  and takes a value $\sim -\omega_{pl}^2/\delta^2$ at the $\omega \to 0$
  limit~\cite{Draxl_2006}.

\bibitem{Makino-1998}
H.~Makino, I.~H. Inoue, M.~J. Rozenberg, I.~Hase, Y.~Aiura, S.~Onari,
  \href{https://link.aps.org/doi/10.1103/PhysRevB.58.4384}{Bandwidth control in
  a perovskite-type ${3d}^{1}$-correlated metal
  ${\mathrm{ca}}_{1\ensuremath{-}x}{\mathrm{sr}}_{x}{\mathrm{vo}}_{3}.$ ii.
  optical spectroscopy}, Phys. Rev. B 58 (1998) 4384--4393.
\newblock \href {http://dx.doi.org/10.1103/PhysRevB.58.4384}
  {\path{doi:10.1103/PhysRevB.58.4384}}.
\newline\urlprefix\url{https://link.aps.org/doi/10.1103/PhysRevB.58.4384}

\bibitem{Hamann_2013}
D.~R. Hamann,
  \href{https://link.aps.org/doi/10.1103/PhysRevB.88.085117}{Optimized
  norm-conserving vanderbilt pseudopotentials}, Phys. Rev. B 88 (2013) 085117.
\newblock \href {http://dx.doi.org/10.1103/PhysRevB.88.085117}
  {\path{doi:10.1103/PhysRevB.88.085117}}.
\newline\urlprefix\url{https://link.aps.org/doi/10.1103/PhysRevB.88.085117}

\bibitem{Rappe_1990}
A.~M. Rappe, K.~M. Rabe, E.~Kaxiras, J.~D. Joannopoulos,
  \href{https://link.aps.org/doi/10.1103/PhysRevB.41.1227}{Optimized
  pseudopotentials}, Phys. Rev. B 41 (1990) 1227--1230.
\newblock \href {http://dx.doi.org/10.1103/PhysRevB.41.1227}
  {\path{doi:10.1103/PhysRevB.41.1227}}.
\newline\urlprefix\url{https://link.aps.org/doi/10.1103/PhysRevB.41.1227}

\bibitem{about-ONCV-QE}
For ONCV-QE, we employ Perdew-Burke-Ernzerhof (PBE)~\cite{Perdew_1996}
  norm-conserving pseudopotentials generated by the code ONCVPSP (Optimized
  Norm-Conserving Vanderbilt PSeudopotential)~\cite{Hamann_2013}, which are
  downloaded from the PseudoDojo~\cite{Setten_2018}.

\bibitem{about-RRKJ-QE}
The RRKJ pseudopotentials are generated by using the Opium code with the input
  files provided by the developers at
  \url{https://www.sas.upenn.edu/rappegroup/research/pseudo-potential-gga.html}.

\bibitem{GPL_URL}
\href{{https://www.gnu.org/}}{{GPL Web page}} [online, cited
  {https://www.gnu.org/}].

\bibitem{xTAPP_URL}
\href{{http://xtapp.cp.is.s.u-tokyo.ac.jp}}{{xTAPP Web page}} [online, cited
  {http://xtapp.cp.is.s.u-tokyo.ac.jp}].

\bibitem{mVMC_URL}
\href{{https://issp-center-dev.github.io/mVMC/docs/index.html}}{{mVMC Web
  page}} [online, cited
  {https://issp-center-dev.github.io/mVMC/docs/index.html}].

\bibitem{HPHI_URL}
\href{{http://issp-center-dev.github.io/HPhi/index.html}}{{H$\Phi$ Web page}}
  [online, cited {http://issp-center-dev.github.io/HPhi/index.html}].

\bibitem{Wannier90}
A.~A. Mostofi, J.~R. Yates, Y.-S. Lee, I.~Souza, D.~Vanderbilt, N.~Marzari,
  \href{http://www.sciencedirect.com/science/article/pii/S0010465507004936}{wannier90:
  A tool for obtaining maximally-localised wannier functions}, Computer Physics
  Communications 178~(9) (2008) 685 -- 699.
\newblock \href {http://dx.doi.org/https://doi.org/10.1016/j.cpc.2007.11.016}
  {\path{doi:https://doi.org/10.1016/j.cpc.2007.11.016}}.
\newline\urlprefix\url{http://www.sciencedirect.com/science/article/pii/S0010465507004936}

\bibitem{KAWAMURA2019197}
M.~Kawamura,
  \href{http://www.sciencedirect.com/science/article/pii/S0010465519300347}{Fermisurfer:
  Fermi-surface viewer providing multiple representation schemes}, Computer
  Physics Communications 239 (2019) 197 -- 203.
\newblock \href {http://dx.doi.org/https://doi.org/10.1016/j.cpc.2019.01.017}
  {\path{doi:https://doi.org/10.1016/j.cpc.2019.01.017}}.
\newline\urlprefix\url{http://www.sciencedirect.com/science/article/pii/S0010465519300347}

\bibitem{Setten_2018}
M.~van Setten, M.~Giantomassi, E.~Bousquet, M.~Verstraete, D.~Hamann, X.~Gonze,
  G.-M. Rignanese,
  \href{http://www.sciencedirect.com/science/article/pii/S0010465518300250}{{The
  PseudoDojo: Training and grading a 85 element optimized norm-conserving
  pseudopotential table}}, Computer Physics Communications 226 (2018) 39 -- 54.
\newblock \href {http://dx.doi.org/https://doi.org/10.1016/j.cpc.2018.01.012}
  {\path{doi:https://doi.org/10.1016/j.cpc.2018.01.012}}.
\newline\urlprefix\url{http://www.sciencedirect.com/science/article/pii/S0010465518300250}

\end{thebibliography}
\providecommand{\noopsort}[1]{}\providecommand{\singleletter}[1]{#1}%

\end{document}